%% file: main.tex
\documentclass[leqno]{amsart}

\usepackage[foot]{amsaddr}

\usepackage{color}
\usepackage{geometry}
\usepackage{subfiles}
\usepackage{graphicx}
\usepackage[section]{placeins}
\usepackage{hyperref}
\usepackage{stmaryrd}
\usepackage{amsmath,amssymb,amsfonts,amsthm,textcomp}
\usepackage{mathtools}
\usepackage{tensor}
\usepackage{multicol}
\usepackage{subfig}
\usepackage{array}
\usepackage{accents}
\usepackage[colorinlistoftodos]{todonotes}
\usepackage[ruled,vlined]{algorithm2e}
\input{commands}
\newcommand{\SFdf}{{\texttt{SF\scriptsize{D-F}}}}

\begin{document}

\title[\SFdf: Sparse Fourier divergence-free approximations]{Statistical Learning for Fluid Flows: Sparse Fourier divergence-free approximations}
\author{Luis Espath$^1$, Dmitry Kabanov$^1$, Jonas Kiessling$^{2,3}$ \& Ra\'{u}l Tempone$^{1,4,5}$}
\address{$^1$Department of Mathematics, RWTH Aachen University, Geb\"{a}ude-1953 1.OG, Pontdriesch 14-16, 161, 52062 Aachen, Germany.}
\address{$^2$H-Ai AB is: Box 5216, 102 45 Stockholm, Sweden}
\address{$^3$Institutionen för Matematik, Kungl. Tekniska Högskolan, 100 44 Stockholm, Sweden.}
\address{$^4$Alexander von Humboldt Professor in Mathematics for Uncertainty Quantification, RWTH Aachen University, Germany.}
\address{$^5$King Abdullah University of Science \& Technology (KAUST), Computer, Electrical and Mathematical Sciences \& Engineering Division (CEMSE), Thuwal 23955-6900, Saudi Arabia.}
\email{espath@gmail.com}

\date{\today}

\begin{abstract}
\noindent
\subfile{./abstract}
\end{abstract}

\maketitle

\tableofcontents                        % Print table of contents

%-------------------------------------------------------------------------------%

\subfile{./section-1}

\subfile{./section-2}

\subfile{./section-3}

\subfile{./section-4}

\subfile{./section-5}

\subfile{./section-6}

%-------------------------------------------------------------------------------%

\appendix

\subfile{./appendix}

%-------------------------------------------------------------------------------%

% \bibliographystyle{unsrt}
% \bibliography{bib_section-1,bib_section-2,bib_section-3,bib_section-4,bib_section-5}
\end{document}

%% file: commands.tex
\newfont{\tenbfsl}{cmbxti9 scaled 1200}
\newfont{\tenbbb}{msbm10}
\newfont{\svnbbb}{msbm8}

\newcommand{\rbull}{\red{\bullet}}
\newcommand{\bbull}{\blue{\bullet}}

\newcommand{\bs}[1]{\boldsymbol{#1}}
\newcommand{\cl}[1]{\mathcal{#1}}

\newcommand{\bb}[1]{\mathbb{#1}}

\newcommand{\red}[1]{{\color{red} #1}}
\newcommand{\blue}[1]{{\color{blue} #1}}

\newcommand{\id}{\mbox{\tenbfsl 1\/}}

\newcommand{\norm}[2]{\left\lVert {#1} \right\rVert_{#2}}

\newcommand{\fr}[2]{\textstyle{\frac{{#1}}{{#2}}}}

\newcommand{\di}[1]{\,\mathrm{d}{#1}}

\newcommand{\dv}{\,\mathrm{d}v}
\newcommand{\da}{\,\mathrm{d}a}

\newcommand{\trans}{{\scriptscriptstyle\mskip-1mu\top\mskip-2mu}}

\newcommand{\sym}{\mathrm{sym}\mskip2mu}
\newcommand{\skw}{\mathrm{skw}\mskip2mu}

\newcommand{\Grad}{\mathrm{grad}\mskip2mu}
\newcommand{\Div}{\mathrm{div}\mskip2mu}

\newcommand{\dd}[2]{\frac{\mathrm{d}{#1}}{\mathrm{d}{#2}}}

\newtheorem{rmk}{Remark}

\newcommand{\R}{\mathbb R}
\newcommand{\T}{\trans}
\newcommand{\diag}{\mathrm{diag}}

%% file: abstract.tex
%auto-ignore
We reconstruct the velocity field of incompressible flows given a finite set of measurements. For the spatial approximation, we introduce the Sparse Fourier divergence-free (\SFdf) approximation based on a discrete $L^2$ projection. Within this physics-informed type of statistical learning framework, we adaptively build a sparse set of Fourier basis functions with corresponding coefficients by solving a sequence of minimization problems where the set of basis functions is augmented greedily at each optimization problem. We regularize our minimization problems with the seminorm of the fractional Sobolev space in a Tikhonov fashion. In the Fourier setting, the incompressibility (divergence-free) constraint becomes a finite set of linear algebraic equations. We couple our spatial approximation with the truncated Singular Value Decomposition (SVD) of the flow measurements for temporal compression. Our computational framework thus combines supervised and unsupervised learning techniques. We assess the capabilities of our method in various numerical examples arising in fluid mechanics.
\\
\textbf{keywords:} Sparse approximation; Fourier series; Statistical learning; Fluid mechanics.
\\
\textbf{AMS subject classifications:}
$\cdot$
76-10 % Mathematical modeling or simulation for problems pertaining to fluid mechanics
$\cdot$
68T05 %	Learning and adaptive systems in artificial intelligence
$\cdot$
30E10 % Approximation in the complex plane
$\cdot$
35L65 % Partial differential equations - Conservation laws
$\cdot$

%% file: section-1.tex
%auto-ignore
\section{Introduction}

Machine learning strategies for fluid flows have been extensively developed in recent years. Particular attention has been paid to physics-informed deep neural networks \cite{Ram19,Maz20} in a statistical learning context. Such models combine measurements with physical properties to improve the reconstruction quality, especially when there are not enough velocity measurements. A comprehensive review on Machine Learning for Fluid Mechanics is presented in \cite{Bru20}. However, the idea of incorporating physical constraints may be combined with other mathematical models instead of neural networks to achieve good reconstruction results. Particularly, the constraint of divergence-free velocity field may be used to ensure that the flow is incompressible. For instance, Tempone \cite{Tem99} uses Fourier basis functions coupled with the divergence-free constraint to approximate wind velocity fields. In that setting, the differential divergence-free constraint is written as a set of linear algebraic constraints. Similarly, to approximate solutions of the incompressible Navier--Stokes equation, Lowitzsch \cite{Low04} proposes a new class of radial basis functions that by construction are divergence free. Thus, any linear combination of these basis functions will render a divergence-free flow. Also, Cervantes Cabrera et al.\ \cite{Cab13} also employs radial basis functions; however, the divergence-free approximation arises from penalizing the target energy functional with the divergence of the vector field. Wang et al.\ \cite{Wan16} construct a divergence-free smoothing for particle image velocimetry. They use a divergence corrective scheme constructed upon finite-difference approximations to render divergence-free field while dealing with measurement errors. Last, Busch et al.\ \cite{Bus13} construct divergence-free velocity fields from cine 3D phase-contrast flow measurements.

This work presents a novel computational framework combining supervised and unsupervised learning techniques. Given a finite set of velocity measurements, we aim to reconstruct the velocity field of incompressible flows. For the spatial approximation, we introduce the Sparse Fourier divergence-free (\SFdf) approximation based on a discrete $L^2$ projection. Our supervised learning technique is a type of physically informed statistical learning framework. Moreover, to make the approximation sparse, we proceed as follows. We start with a minimal set of Fourier basis functions and construct, in a greedy manner, a larger sparse set of basis functions along a sequence of optimization problems. In each optimization problem, we increase the set of Fourier basis functions and measure their relative energy. Only the most energetic high-wavenumber modes are kept and the optimization sequence continues. To regularize these ill-posed optimization problems, we use a Tikhonov regularizer based on the seminorm of the fractional Sobolev space. Focusing on incompressible flows, we impose the incompressibility constraint, that is, the reconstructed vector field is divergence free. From a numerical standpoint, it is often intricate to impose the divergence-free constraint to a vector-valued approximation. However, when using Fourier approximation, this differential constraint becomes algebraic and pointwise divergence-free fields can be easily constructed. Last, to emulate solid-wall types of boundary conditions, we penalize the normal velocity on the boundary of an immersed body. For the temporal approximation, we employ truncated Singular Value Decomposition (SVD) to find a low-rank approximation of time-evolving data, which then is combined with an ensemble of \SFdf\ approximations to reconstruct such data. We conclude this work presenting various numerical examples to assess the capabilities of our method in fluid mechanics.

Although our method can accurately and adaptively reconstruct divergence-free fields, one would ideally wish to incorporate the Navier--Stokes equations as additional constraints to be satisfied. However, these equations require the acceleration and pressure measurements or at least one would need to devise an additional model for these quantities. Our computational framework assumes that we can access only a finite set of velocity measurements. Moreover, although any acceleration, velocity, and pressure fields arising from the incompressible Navier--Stokes equations automatically satisfy the mechanical version of the second law of thermodynamics, whether this property holds for the discretized versions of these equations within a statistical learning framework is unclear. Thus, one may argue that if the incompressible Navier–Stokes equations are to be included in the computational framework, so is the mechanical version of the second law of thermodynamics.

The remainder of this work is organized as follows. \S\ref{sc:notation.problem} presents the mathematical notation and problem statement and \S\ref{sc:sfdf} introduces our Sparse Fourier divergence-free approximation (\SFdf) for spatial data. \S\ref{sc:numerics.spatial} assesses the accuracy of our method in numerical examples. \S\ref{sc:compress.sensing} describes how a set of such spatial approximations can be applied to the time series of velocity measurements in an economical manner by exploiting the low-rank structure of the time series. \S\ref{sc:numerics.spatio.temporal} demonstrates the approach developed in \S\ref{sc:sfdf} and \S\ref{sc:compress.sensing} by recovering the shear flow in the Kelvin--Helmholtz instability phenomenon. Conclusions are drawn in \S\ref{sc:conclusions}.

%% file: section-2.tex
%auto-ignore
\section{Notation and problem statement}
\label{sc:notation.problem}

Recalling that an incompressible flow is endowed with a divergence-free velocity field, we let $\bs{\upsilon}\colon\cl{D}\subset\bb{R}^n\mapsto\bb{R}^n$ denote a real-valued divergence-free velocity field, where $\cl{D}\coloneqq\Pi_{\iota=1}^n[0, D_\iota]$ is a physical domain of length $D_\iota$ per direction $\iota$. Consider a set $\{\bs{u}_i\}_{i=1}^P$ of observed pointwise velocity measurements of an unknown divergence-free vector field at locations $\{\bs{x}_i\}_{i=1}^P\subset\cl{D}$. Consider further an immersed body $\cl{B}$ in $\cl{D}$ with boundary $\partial\cl{B}$. We then seek to reconstruct the unknown divergence-free vector field out of the measurements $\{\bs{u}_i\}_{i=1}^P$ and the immersed boundary $\partial\cl{B}$.

We construct our approximation on the fractional Sobolev space of all periodic functions that are square integrable on a toroidal $\cl{D}$. Thus, letting
\begin{equation}\label{eq:alpha}
\hat{\bs{\alpha}}=(\alpha_1/D_1,\ldots,\alpha_n/D_n)\qquad\text{with}\qquad\bs{\alpha}\in\bb{Z}^n,
\end{equation}
with $k\in(1,\infty)$, we respectively define the $L^2$ and $H^k\coloneqq{W}^{k,2}$ spaces of vector-valued functions $\bs{\upsilon}\colon\cl{D}\mapsto\bb{R}^n$ as
\begin{equation}\label{eq:L2}
L^2(\cl{D})\coloneqq\left\{\bs{\upsilon}=\sum\limits_{\bs{\alpha} \in \bb{Z}^n} \bs{\upsilon}_{\bs{\alpha}} \exp(2\pi\jmath\,\hat{\bs{\alpha}}\cdot\bs{x})\,\bigg\vert\,\dfrac{1}{|\cl{D}|}\int_\cl{D}\|\bs{\upsilon}\|^2\dv=\sum\limits_{\bs{\alpha} \in \bb{Z}^n}\|\bs{\upsilon}_{\bs{\alpha}}\|^2<\infty\right\},
\end{equation}
and
\begin{equation}\label{eq:Hp}
H^{k}(\cl{D})\coloneqq\left\{\bs{\upsilon}\in L^2(\cl{D})\,\bigg\vert\,\sum_{\bs{\alpha}\in\bb{Z}^n}(2\pi\|\hat{\bs{\alpha}}\|)^{2k}\|\bs{\upsilon}_{\bs{\alpha}}\|^2<\infty\right\},
\end{equation}
where $\jmath\coloneqq\sqrt{-1}$ is the imaginary unit and $\bs{\upsilon}_{\bs{\alpha}}$ the $\bs{\alpha}$th component of the Fourier transform of $\bs{\upsilon}$, that is,
\begin{equation}
\bs{\upsilon}_{\bs{\alpha}}\coloneqq\dfrac{1}{|\cl{D}|}\int_{\cl{D}}\bs{\upsilon}(\bs{x})\exp(-2\pi\jmath\mskip2mu\hat{\bs{\alpha}}\cdot\bs{x})\dv\qquad\forall\,\bs{\alpha}\in\bb{Z}^n.
\end{equation}
Note that in expression \eqref{eq:Hp}, we only defined the seminorm of $H^k$ for the vector-valued $\bs{\upsilon}$. Thus, the full norm that induces the space $H^k$ is
\begin{equation}
\norm{\bs{\upsilon}}{H^k(\cl{D})}^2\coloneqq\sum_{\bs{\alpha}\in\bb{Z}^n}(1+(2\pi\|\hat{\bs{\alpha}}\|)^{2k})\|\bs{\upsilon}_{\bs{\alpha}}\|^2,
\end{equation}
where $\Grad^k\bs{\upsilon}$ is the $k$th gradient of $\bs{\upsilon}$. In what follows, we use $\mathring{H}^k$ to indicate the seminorm
\begin{equation}\label{eq:regularization}
\norm{\bs{\upsilon}}{\mathring{H}^k(\cl{D})}^2\coloneqq\sum_{\bs{\alpha}\in\bb{Z}^n}(2\pi\|\hat{\bs{\alpha}}\|)^{2k}\|\bs{\upsilon}_{\bs{\alpha}}\|^2=\norm{\Grad^k\bs{\upsilon}}{L^2(\cl{D})}^2.
\end{equation}
Alternatively, the $L^2$ inner product for vector-valued functions may assume the following conventional form, in an arbitray domain, $\cl{D}$, and on its boundary, $\partial\cl{D}$,
\begin{equation}
(\bs{\upsilon},\bs{\omega})_{L^2(\cl{D})}\coloneqq\int_{\cl{D}}\bs{\upsilon}\cdot\bs{\omega}^\ast\dv\qquad\text{and}\qquad(\bs{\upsilon},\bs{\omega})_{L^2(\partial\cl{D})}\coloneqq\int_{\partial\cl{D}}\bs{\upsilon}\cdot\bs{\omega}^\ast\da,
\end{equation}
where the asterisk represents the complex-conjugate pair.

We can now state the reconstruction problem as follows. Given $k>1$, $\epsilon>0$ and $\lambda_B>0$, find $\bs{\upsilon}^\ast$ such that
\begin{equation}\label{eq:optimization}
\left\{
\begin{aligned}
&\bs{\upsilon}^{\mathrm{opt}}\coloneqq\underset{\bs{\upsilon}\in H^k(\cl{D})}{\arg\min} \, \dfrac{1}{P}\sum_{i=1}^P\norm{\bs{\upsilon}({\bs{x}}_i)-{\bs{u}_i}}{}^2 + \lambda_B (\upsilon_n,\upsilon_n)_{L^2(\partial\cl{B})}+\epsilon\norm{\bs{\upsilon}}{\mathring{H}^k(\cl{D})}^2,\\[4pt]
&\text{subject to }\Div\bs{\upsilon}=0,
\end{aligned}
\right.
\end{equation}
where $\Div\bs{\upsilon}$ represents the divergence of $\bs{\upsilon}$, $\upsilon_n\coloneqq\bs{\upsilon}\cdot\bs{n}$ is the normal velocity on the immersed boundary, and $\bs{n}$ is the outward unit normal to the immersed boundary $\partial\cl{B}$.

Immersed boundaries $\partial\cl{B}$ are treated with no-penetration boundary conditions, that is, $\upsilon_n\coloneqq\bs{\upsilon}\cdot\bs{n}=0$, where $\bs{n}$ is the unit normal to $\partial\cl{B}$. This no-penetration boundary condition is penalized by the second term in \eqref{eq:optimization} in the $L^2$ sense on $\partial\cl{B}$ and reads as follows.
\begin{equation}
(\upsilon_n,\upsilon_n)_{L^2(\partial\cl{B})}\coloneqq\int_{\partial\cl{B}}\upsilon_n\upsilon_n^\ast\da=\int_{\partial\cl{B}}(\bs{\upsilon}\cdot\bs{n})(\bs{\upsilon}\cdot\bs{n})^\ast\da.
\end{equation}
Last, the third term in \eqref{eq:optimization} is a regularization term that penalizes the $\mathring{H}^k$ norm of the field $\bs{\upsilon}$.

From the embedding Sobolev theorem (see Appendix \ref{ap:sobolev.embedding}) with $k>n/p$, we have that $\bs{\upsilon}$ belongs to the H\"{o}lder space ${C}^{k-\left[\frac{n}{p}\right]-1,\gamma}(\cl{D})$, namely, H\"{o}lder continuous with some positive exponent $\gamma$. Thus, for two-dimensional problems $n=2$ with $p=2$, we have that $k>1$ whereas in three-dimensional problems $n=3$ with $p=2$, we have $k>1.5$.

\section{Spatial approximation: Sparse Fourier divergence-free (\texttt{SFd-f})}
\label{sc:sfdf}

Consider the following finite-dimensional representation $\bs{\upsilon}_{\cl{I}}$ of $\bs{\upsilon}$ in $\cl{D}$,
\begin{equation}\label{eq:fourier.approximation}
\bs{\upsilon}_{\cl{I}}(\bs{x}) \coloneqq \sum\limits_{\bs{\alpha} \in \cl{I}} \bs{\upsilon}_{\bs{\alpha}} \varphi(\bs{x},\bs{\alpha}),\qquad\text{with}\qquad\varphi(\bs{x},\bs{\alpha}) \coloneqq \exp(2\pi\jmath\,\hat{\bs{\alpha}}\cdot\bs{x})\qquad\forall\,\bs{x}\in\Pi_{k=1}^n[0,D_k]\wedge\bs{\alpha}\in\cl{I},
\end{equation}
where $\cl{I}\subset\bb{Z}^n$ is a finite index set of tuples composed of $n$ integers defining the indices of the basis functions and $\bs{\upsilon}_{\bs{\alpha}}\in\bb{C}^n$ for all $\bs{\alpha}\in\cl{I}$ are their Fourier coefficients.

\subsection{Divergence constraint}

By differentiability properties of trigonometric functions, the divergence constraint becomes a linear algebraic constraint. Thus, from
\begin{equation}
\Div\bs{\upsilon}_{\cl{I}}=0,
\end{equation}
we arrive at
\begin{equation}\label{eq:algebraic.div.constraint}
\hat{\bs{\alpha}}\cdot\bs{\upsilon}_{\bs{\alpha}} = 0\qquad\forall\,\bs{\alpha}\in\cl{I},
\end{equation}
which in turn may be written as $\bs{C}_{\!\mathrm{a}}\bs{g}=\bs{0}$ where $\bs{C}_{\!\mathrm{a}}$ is a $m_{\mathrm{a}} \times n_{\mathrm{dof}}$ matrix and $\bs{g}$ is the concatenation of the coefficients $\bs{\upsilon}_{\bs{\alpha}}$ into a vector of dimension $n_{\mathrm{dof}}$. Here, $n_{\mathrm{dof}}$ is the total number of degrees of freedom and $m_{\mathrm{a}}$ is the number of constraints arising from the divergence constraint.

\subsection{Real-valued vector field constraint}

To obtain a real-valued representation for $\bs{\upsilon}_{\cl{I}}$, we impose an additional algebraic constraint
\begin{equation}\label{eq:algebraic.real.constraint}
\bs{\upsilon}_{\bs{\alpha}}=\bs{\upsilon}_{-\bs{\alpha}}^\ast\qquad\forall\,\bs{\alpha}\in\cl{I},
\end{equation}
This constraint is also linear and may be written as $\bs{C}_{\!\mathrm{b}}\bs{u}=\bs{0}$ where $\bs{C}_{\!\mathrm{b}}$ is a $m_{\mathrm{b}} \times n_{\mathrm{dof}}$ matrix, where $m_{\mathrm{b}}$ is the number of constraints arising from the real-valued vector constraint.

\subsection{Reformulation}

In the finite Fourier representation \eqref{eq:fourier.approximation} augmented by the algebraic constraints \eqref{eq:algebraic.div.constraint} and \eqref{eq:algebraic.real.constraint} over $\cl{I}$, the optimization problem \eqref{eq:optimization} with
\begin{equation}\label{eq:f}
f(\bs{\upsilon}_{\cl{I}})\coloneqq\dfrac{1}{P}\sum_{i=1}^P\norm{\bs{\upsilon}_{\cl{I}}({\bs{x}}_i)-{\bs{u}_i}}{}^2 + \lambda_B (\bs{\upsilon}_{\cl{I}}\cdot\bs{n},\bs{\upsilon}_{\cl{I}}\cdot\bs{n})_{L^2(\partial\cl{B})}+\epsilon\norm{\bs{\upsilon}_{\cl{I}}}{\mathring{H}^k(\cl{D})}^2,
\end{equation}
becomes
\begin{equation}\label{eq:reformulated.optimization}
\left\{
\begin{aligned}
&\bs{\upsilon}_{\cl{I}}^{\mathrm{opt}}\coloneqq\underset{\bs{\upsilon}_{\cl{I}}\in H^k(\cl{D})}{\arg\min} \, {f}(\bs{\upsilon}_{\cl{I}})\\[4pt]
&\text{subject to }\hat{\bs{\alpha}}\cdot\bs{\upsilon}_{\bs{\alpha}} = 0\quad\wedge\quad\bs{\upsilon}_{\bs{\alpha}}-\bs{\upsilon}_{-\bs{\alpha}}^\ast=0\qquad\forall\,\bs{\alpha}\in\cl{I}.
\end{aligned}
\right.
\end{equation}
To understand the need of a regularization term, see the following remark.
\begin{rmk}[Data perturbation and well-posedness]\label{rk:perturbation}
First, with $\bar{m}\ge\bar{n}\ge\bar{p}$, consider that the generalized singular value decomposition of an $\bar{m}\times\bar{n}$ matrix $\bs{A}$ and a $\bar{p}\times\bar{n}$ matrix $\bs{B}$ ($\bar{q}\coloneqq\mathrm{rank}(\bs{B})\le\bar{p}$) is given by the pair of factorizations
\begin{equation}
\bs{A}=\bs{U}
\begin{bmatrix}
\bs{\Sigma} & \bs{0}\\
\bs{0} & \id_{\bar{n}-\bar{q}}
\end{bmatrix}
\bs{Q}^{-1} \qquad\text{and}\qquad \bs{B}=\bs{V}[\bs{M},\bs{0}]\bs{Q}^{-1},
\end{equation}
where $\mathrm{rank}\bigl(\bigl[\begin{smallmatrix}\bs{A}\\\bs{B}\end{smallmatrix}\bigr]\bigr)=\bar{n}$. The columns of $\bs{U}\in\bb{R}^{\bar{m}\times\bar{n}}$ and $\bs{V}\in\bb{R}^{\bar{q}\times\bar{q}}$ are orthogonal; that is, $\bs{U}^{\trans}\bs{U}=\id_{\bar{n}}$ and $\bs{V}^{\trans}\bs{V}=\id_{\bar{q}}$. Also, $\bs{Q}\in\bb{R}^{\bar{n}\times\bar{n}}$ is nonsingular, and $\bs{\Sigma}$ and $\bs{M}$ are diagonal matrices: $\bs{\Sigma}=\mathrm{diag}\{\sigma_1,\ldots,\sigma_{\bar{q}}\}$ and $\bs{M}=\mathrm{diag}\{\mu_1,\ldots,\mu_{\bar{q}}\}$, normalized such that $\sigma_i^2+\mu_i^2=1$.

For a convenient analysis, we neglect the immersed boundaries terms and focus on the terms related to the misfit and regularization. Using \eqref{eq:f}, the optimization problem given in \eqref{eq:reformulated.optimization} then becomes
\begin{equation}\label{eq:quad}
\bs{x}^{\mathrm{opt}}(\epsilon)\coloneqq\underset{\bs{x}}{\arg\min}\, \|\bs{A}\bs{x}-\bs{b}\|^2+\epsilon\|\bs{B}\bs{x}\|^2,
\end{equation}
where $\bs{x}$ is a vector containing all $\bs{\upsilon}_\alpha$, $\|\bs{A}\bs{x}-\bs{b}\|^2=\frac{1}{P}\sum_{i=1}^P\norm{\bs{\upsilon}_{\cl{I}}({\bs{x}}_i)-{\bs{u}_i}}{}^2$ is the misfit, and $\|\bs{B}\bs{x}\|^2=\norm{\bs{\upsilon}_{\cl{I}}}{\mathring{H}^k(\cl{D})}^2$ is the fractional Sobolev seminorm.

Now, let $\delta\bs{A}$ and $\delta\bs{b}$ represent perturbations arising in the data, which may arise from positioning of the measurement device and the measurements themselves, respectively. Expression \eqref{eq:quad} then assumes the following form
\begin{equation}\label{eq:perturbed.quad}
\hat{\bs{x}}^{\mathrm{opt}}(\epsilon)\coloneqq\underset{\bs{x}}{\arg\min}\, \|(\bs{A}+\delta\bs{A})\bs{x}-(\bs{b}+\delta\bs{b})\|^2+\epsilon\|\bs{B}\bs{x}\|^2.
\end{equation}
If $0<\epsilon\le{1}$, the null spaces of $\bs{A}$ and $\bs{B}$ are trivial, that is, $\bigl[\begin{smallmatrix}\bs{A}\\\sqrt{\epsilon}\bs{B}\end{smallmatrix}\bigr]$ is full rank and the perturbation in the solution is bounded from above, namely
\begin{equation}
\|\bs{x}^{\mathrm{opt}}(\epsilon)-\hat{\bs{x}}^{\mathrm{opt}}(\epsilon)\|\le\dfrac{\kappa_\epsilon}{1-\kappa_\epsilon\dfrac{\|\delta\bs{A}\|}{\|\bs{A}\|}}\left((1+\kappa(\bs{Q}))\dfrac{\|\delta\bs{A}\|}{\|\bs{A}\|}\|\bs{x}^{\mathrm{opt}}(\epsilon)\|+\dfrac{\|\delta\bs{b}\|}{\|\bs{A}\|} + \kappa_\epsilon\dfrac{\|\delta\bs{A}\|}{\|\bs{A}\|}\dfrac{\|\bs{A}\bs{x}^{\mathrm{opt}}(\epsilon)-\bs{b}\|}{\|\bs{A}\|}\right),
\end{equation}
where $\kappa(\bs{Q})$ is the condition number of $\bs{Q}$ and $\kappa_\epsilon\coloneqq\|\bs{A}\|\|\bs{Q}\|/\sqrt{\epsilon}$.

Moreover, if $\bar{q}=\bar{n}$ and $\bs{B}$ is nonsingular, we have the following bound
\begin{equation}
\|\bs{x}^{\mathrm{opt}}(\epsilon)-\hat{\bs{x}}^{\mathrm{opt}}(\epsilon)\|\le\dfrac{\hat{\kappa}_\epsilon}{1-\hat{\kappa}_\epsilon\dfrac{\|\delta\bs{A}\|}{\|\bs{A}\|}}\left((1+\kappa(\bs{B}))\dfrac{\|\delta\bs{A}\|}{\|\bs{A}\|}\|\bs{x}^{\mathrm{opt}}(\epsilon)\|+\dfrac{\|\delta\bs{b}\|}{\|\bs{A}\|} + \hat{\kappa}_\epsilon\dfrac{\|\delta\bs{A}\|}{\|\bs{A}\|}\dfrac{\|\bs{A}\bs{x}^{\mathrm{opt}}(\epsilon)-\bs{b}\|}{\|\bs{A}\|}\right),
\end{equation}
where $\kappa(\bs{B})$ is the condition number of $\bs{B}$ and $\kappa_\epsilon\coloneqq\|\bs{A}\|\|\bs{B^{-1}}\|/\sqrt{\epsilon}$. We conclude this statement by stressing that to satisfy the above bound $\epsilon$ must be strictly positive.
Interested readers are referred to \cite{Bjo96,Tem99}.
\end{rmk}

\subsection{Constraint imposition}

Concatenating $\bs{C}_{\!\mathrm{a}}$ and $\bs{C}_{\!\mathrm{b}}$ in their first dimension, we build the matrix
\begin{equation}
\bs{C}\coloneqq\left[\bs{C}_{\!\mathrm{a}}\atop\bs{C}_{\!\mathrm{b}}\right],
\end{equation}
encompassing all the constraints. Given $m$ independent components among $n_{\mathrm{dof}}$ degrees of freedom and defining the null space of $\bs{C}$ as $\bs{Z}\coloneqq\mathrm{Null}(\bs{C})$, we recover the variables $\bs{g}$ through the linear mapping
\begin{equation}
\bs{g}=\bs{Z}\bs{g}_{\mathrm{m}}.
\end{equation}
Similarly, the gradient of $f$ given in \eqref{eq:optimization} with respect to the independent variables is computed as
\begin{equation}
-\bs{d}_{\mathrm{m}}\coloneqq\bs{Z}^{\trans} \Grad_{\!\bs{u}} f,
\end{equation}
where $\bs{d}_{\mathrm{m}}$ is the search direction.

\subsection{Regularization term}

As the bases are orthogonal, the regularization term in \eqref{eq:optimization}, defined in \eqref{eq:regularization} for $k\in\bb{R}$, is given by
\begin{align}
\norm{\bs{\upsilon}_{\cl{I}}}{\mathring{H}^k(\cl{D})}^2 & = \int_{\cl{D}} \|\Grad^k\bs{\upsilon}_{\cl{I}}\|^2 \dv = \int_{\cl{D}} \Grad^k\bs{\upsilon}_{\cl{I}} \cdot (\Grad^k\bs{\upsilon}_{\cl{I}})^\ast \dv \nonumber\\
&=\int_{\cl{D}} \left(\sum\limits_{\bs{\alpha} \in \cl{I}} \bs{\upsilon}_{\bs{\alpha}} \, \Grad^k\varphi(\bs{x},\bs{\alpha})\right) \cdot \left(\sum\limits_{\bs{\beta} \in \cl{I}} \bs{\upsilon}_{\bs{\beta}}^\ast \, \Grad^k\varphi^\ast(\bs{x},\bs{\beta})\right) \dv,\nonumber\\
&=\int_{\cl{D}} \sum\limits_{\bs{\alpha} \in \cl{I}} \left(\bs{\upsilon}_{\bs{\alpha}}\cdot\bs{\upsilon}_{\bs{\alpha}}^\ast\right) \Grad^k\varphi(\bs{x},\bs{\alpha})\cdot\Grad^k\varphi^\ast(\bs{x},\bs{\alpha}) \dv,\nonumber\\
&= \sum\limits_{\bs{\alpha} \in \cl{I}} \bs{\upsilon}_{\bs{\alpha}}\cdot\bs{\upsilon}_{\bs{\alpha}}^\ast  \int_{\cl{D}} \Grad^k\varphi(\bs{x},\bs{\alpha})\cdot\Grad^k\varphi^\ast(\bs{x},\bs{\alpha}) \dv,\nonumber\\
&= (2\pi)^{2k}\sum\limits_{\bs{\alpha} \in \cl{I}} \left(\bs{\upsilon}_{\bs{\alpha}}\cdot\bs{\upsilon}_{\bs{\alpha}}^\ast \right) (\hat{\bs{\alpha}}\cdot\hat{\bs{\alpha}})^k.
\end{align}

\subsection{Gradient of the objective function}

Here, we aim to obtain the stationary point of $f$ through the optimality condition in expression \eqref{eq:reformulated.optimization} when evaluated at $\bs{\upsilon}_{\cl{I}}$ with respect to the Fourier coefficients $\bs{\upsilon}_{\bs{\alpha}}$; that is,
\begin{equation}\label{eq:optimality}
2\dfrac{\partial}{\partial\bs{\upsilon}_{\bs{\alpha}}^\ast}\left(\dfrac{1}{P}\sum_{i=1}^P\norm{\bs{\upsilon}_{\cl{I}}(\bs{x}_i)-\bs{u}_i)}{}^2 + \lambda_B(\bs{\upsilon}_{\cl{I}}\cdot\bs{n},\bs{\upsilon}_{\cl{I}}\cdot\bs{n})_{L^2(\partial\cl{B})}+\epsilon\norm{\bs{\upsilon}_{\cl{I}}}{H^k(\cl{D})}^2\right)=\bs{0}.
\end{equation}
Interested readers are referred to \eqref{eq:gradient.wirtinger} in Appendix \ref{ap:wirtinger}.

The first term in the above expression reads
\begin{equation}
2\dfrac{\partial}{\partial\bs{\upsilon}_{\bs{\alpha}}^\ast}(\norm{\bs{\upsilon}_{\cl{I}}(\bs{x}_i)-\bs{u}_i)}{}^2) = \dfrac{2}{P}\sum_{i=1}^P \varphi^\ast(\bs{x}_i,\bs{\alpha}) (\bs{\upsilon}_{\cl{I}}(\bs{x}_i)-\bs{u}_i).
\end{equation}
Approximating the second term by the trapezoidal rule with $\lambda_B=\bar{\lambda}_B\Delta{s}$, we have that
\begin{equation}
\lambda_B(\bs{\upsilon}_{\cl{I}}\cdot\bs{n},\bs{\upsilon}_{\cl{I}}\cdot\bs{n})_{L^2(\partial\cl{B})} \approx \dfrac{\bar{\lambda}_B}{B}\sum_{i=1}^B (\bs{\upsilon}_{\cl{I}}(\bs{x}_i)\cdot\bs{n}(\bs{x}_i))^2,\qquad\forall\bs{x}_i\in\partial\cl{B},
\end{equation}
where $\bs{x}_i$ are equally spaced by $\Delta{s}$. Thus, the second term reads
\begin{equation}
2\dfrac{\partial}{\partial\bs{\upsilon}_{\bs{\alpha}}^\ast}\left(\lambda_B(\bs{\upsilon}_{\cl{I}}\cdot\bs{n},\bs{\upsilon}_{\cl{I}}\cdot\bs{n})_{L^2(\partial\cl{B})}\right) \approx 2\dfrac{\bar{\lambda}_B}{B}\sum_{i=1}^B (\bs{\upsilon}_{\cl{I}}(\bs{x}_i)\cdot\bs{n}(\bs{x}_i)) \varphi^\ast(\bs{x}_i,\bs{\alpha})\bs{n}(\bs{x}_i).
\end{equation}
The last term, related to the regularization term, reads
\begin{equation}
2\dfrac{\partial}{\partial\bs{\upsilon}_{\bs{\alpha}}^\ast}\left(\epsilon\norm{\bs{\upsilon}_{\cl{I}}}{H^k(\cl{D})}^2\right) = 2\epsilon (2\pi)^{2k} \bs{\upsilon}_{\bs{\alpha}} (\hat{\bs{\alpha}}\cdot\hat{\bs{\alpha}})^k.
\end{equation}
From the optimality condition \eqref{eq:optimality}, we finally arrive at
\begin{equation}\label{eq:optimality.explicit}
\dfrac{2}{P}\sum_{i=1}^P \varphi^\ast(\bs{x}_i,\bs{\alpha}) (\bs{\upsilon}_{\cl{I}}(\bs{x}_i)-\bs{u}_i)+2\dfrac{\bar{\lambda}_B}{B}\sum_{i=1}^B (\bs{\upsilon}_{\cl{I}}(\bs{x}_i)\cdot\bs{n}(\bs{x}_i)) \varphi^\ast(\bs{x}_i,\bs{\alpha})\bs{n}(\bs{x}_i)+2\epsilon (2\pi)^{2k} \bs{\upsilon}_{\bs{\alpha}} (\hat{\bs{\alpha}}\cdot\hat{\bs{\alpha}})^k=\bs{0}.
\end{equation}

\subsection{Algorithm}

\begin{figure}[!htb]
\centering
  \includegraphics[width=0.375\textwidth]{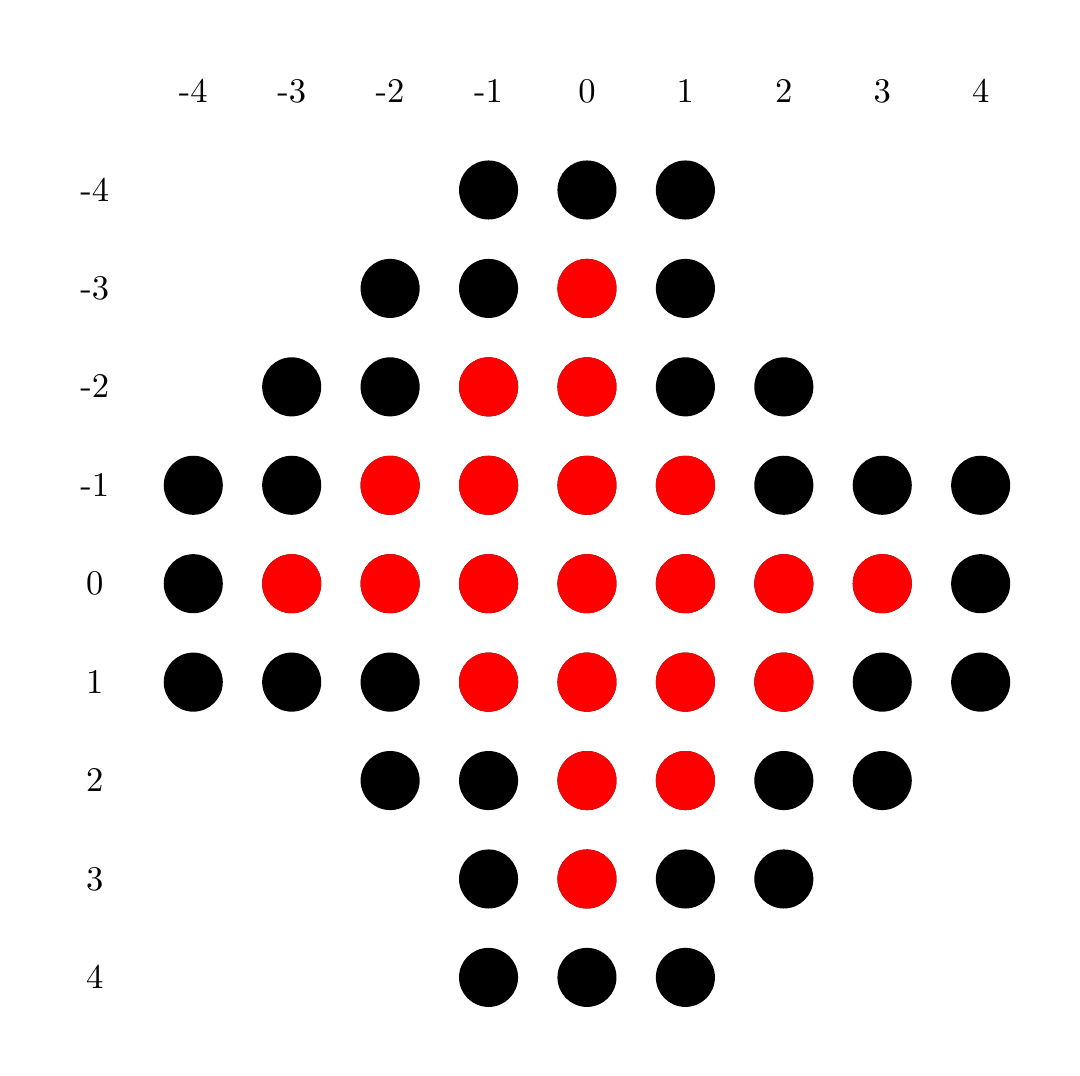}
  \caption{index-space augmentation}
\label{fg:1}
\end{figure}

In view of the \SFdf\ approximation \eqref{eq:fourier.approximation} with $\cl{I}$ being a set of $n$-tuples such that $\bs{\alpha}\in\cl{I}$, let $\delta\bs{\alpha}^\iota$ be a $n$-tuple populated with zeros and ones. The possible $\delta\bs{\alpha}^\iota$ are $\pm(1,0)$, $\pm(0,1)$, and $\pm(1,1)$ in two-dimensions and $\delta\bs{\alpha}^\iota$ are $\pm(1,0,0)$, $\pm(0,1,0)$, $\pm(0,0,1)$, $\pm(1,1,0)$, $\pm(1,0,1)$, $\pm(0,1,1)$, and $\pm(1,1,1)$ in three-dimensions. Next, let $\partial\cl{I}$ be the boundary of the index set $\cl{I}$ such that $\bs{\alpha}\in\partial\cl{I}$ if and only if $\bs{\alpha}+\delta\bs{\alpha}^\iota\not\in\cl{I}$ for some $i$. Note that, to satisfy constraint \eqref{eq:algebraic.real.constraint}, whenever an element $\bs{\beta}$ is included into (excluded from) $\cl{I}$, the element $-\bs{\beta}$ must also be included in (excluded from) $\cl{I}$. Figure \ref{fg:1} illustrates an example of $\partial\cl{I}$ (black dots) and $\cl{I}$ (complete collection of red and black dots).

The energy of the \SFdf\ approximation is defined as
\begin{equation}
\varepsilon(\cl{I})\coloneqq\sum_{\bs{\alpha}\in\cl{I}} \bs{\upsilon}_{\bs{\alpha}} \cdot \bs{\upsilon}_{\bs{\alpha}}^\ast ,
\end{equation}
and the boundary energy is
\begin{equation}
\varepsilon(\partial\cl{I})\coloneqq\sum_{\bs{\alpha}\in\partial\cl{I}} \bs{\upsilon}_{\bs{\alpha}} \cdot \bs{\upsilon}_{\bs{\alpha}}^\ast .
\end{equation}

We begin by solving \eqref{eq:reformulated.optimization} in the smallest hypercube index-space $\cl{I}\coloneqq(-1,0,1)^n$. To arrive at $\cl{I}^1$, we augment the index-space $\cl{I}^0$ to include $\bs{\alpha}+\delta\bs{\alpha}^\iota$, for all $\bs{\alpha}\in\partial\cl{I}^0$ and $1\le\iota\le3^n-1$. To obtain a sparse approximation, we retained only a percentage of these indices on the boundary $\partial\cl{I}^1$. After solving \eqref{eq:reformulated.optimization} for $\cl{I}^1$, we remove the indices on $\partial\cl{I}^1$ contributing less than a pre-established energy threshold, $\varepsilon_{\partial\cl{I}}$. This iterative procedure is considered to converge when the energy increase of the index-space augmentation is $\Delta\epsilon_{\partial\cl{I}}$ or lower. The algorithm is detailed in Algorithm \ref{al:bases.augmentation}.

\begin{algorithm}[H]\label{al:bases.augmentation}
\SetAlgoLined
\KwResult{output: $\bs{\upsilon}_{\cl{I}}$}
 data: $\{\bs{u}_i\}_{i=1}^P$, and immersed boundary $\partial\cl{B}$\;
 initialization: $\cl{I}\coloneqq(-1,0,1)^n$, $\epsilon$, $\bar{\lambda}_B$, $k$, $\varepsilon_{\partial\cl{I}}$, $\Delta\varepsilon_{\partial\cl{I}}$, \texttt{total\_it}\;
 with $\cl{I}$ construct $\cl{I}$\;
 \While{\texttt{niter} $\le$ \texttt{total\_it}}{
  $\cl{I}\gets\cl{I}\cup\left\{\bigcup_{\bs{\alpha}\in\cl{I}}\{\bs{\alpha}+\delta\bs{\alpha}^\iota\}_{\iota=1}^n\right\}$\;
  get $\{\bs{\upsilon}_{\bs{\alpha}}\}_{\bs{\alpha}\in\cl{I}}$ from solving \eqref{eq:reformulated.optimization}\;
  \eIf{$\dfrac{\varepsilon(\partial\cl{I})}{\varepsilon(\cl{I})}>\Delta\varepsilon_{\partial\cl{I}}$}{
   remove $\bs{\alpha}\in\partial\cl{I}$ corresponding to the low energy components, maintaining a fraction $1-\varepsilon_{\partial\cl{I}}$ of the relative boundary energy\;
   }{
   return $\{\bs{\upsilon}_{\bs{\alpha}}\}_{\bs{\alpha}\in\cl{I}}$\;
  }
  $\texttt{niter} +=1$\;
 }
 \caption{Sparse divergence-free discrete $L^2$ Fourier projection algorithm}
\end{algorithm}

Details about the sparse construction can be also found in \cite{Haj20}.

%% file: section-3.tex
%auto-ignore
\section{Numerical experiments on spatial approximation}
\label{sc:numerics.spatial}

\subsection{Taylor--Green vortex}

\FloatBarrier

We first consider the well-known Taylor--Green vortex described by
\begin{equation}\label{eq:taylor.green}
\bs{\upsilon}(\bs{x}) = (\cos(x_1)\sin(x_2), \, -\sin(x_1)\cos(x_2)).
\end{equation}
Naturally, this problem is exactly represented as a finite Fourier series if the discrete $L^2$-norm in \eqref{eq:optimization} is replaced by its continuous version. However, we aim to recover the divergence-free field given only $10$ velocity measurements at random points in the domain $\cl{D}=[0,2\pi]^2$.

\begin{figure}[!htb]
\centering
  \subfloat[Analytical field]{\label{fg:1.a}\includegraphics[clip,trim={1cm 0 1cm 0},width=0.25\textwidth]{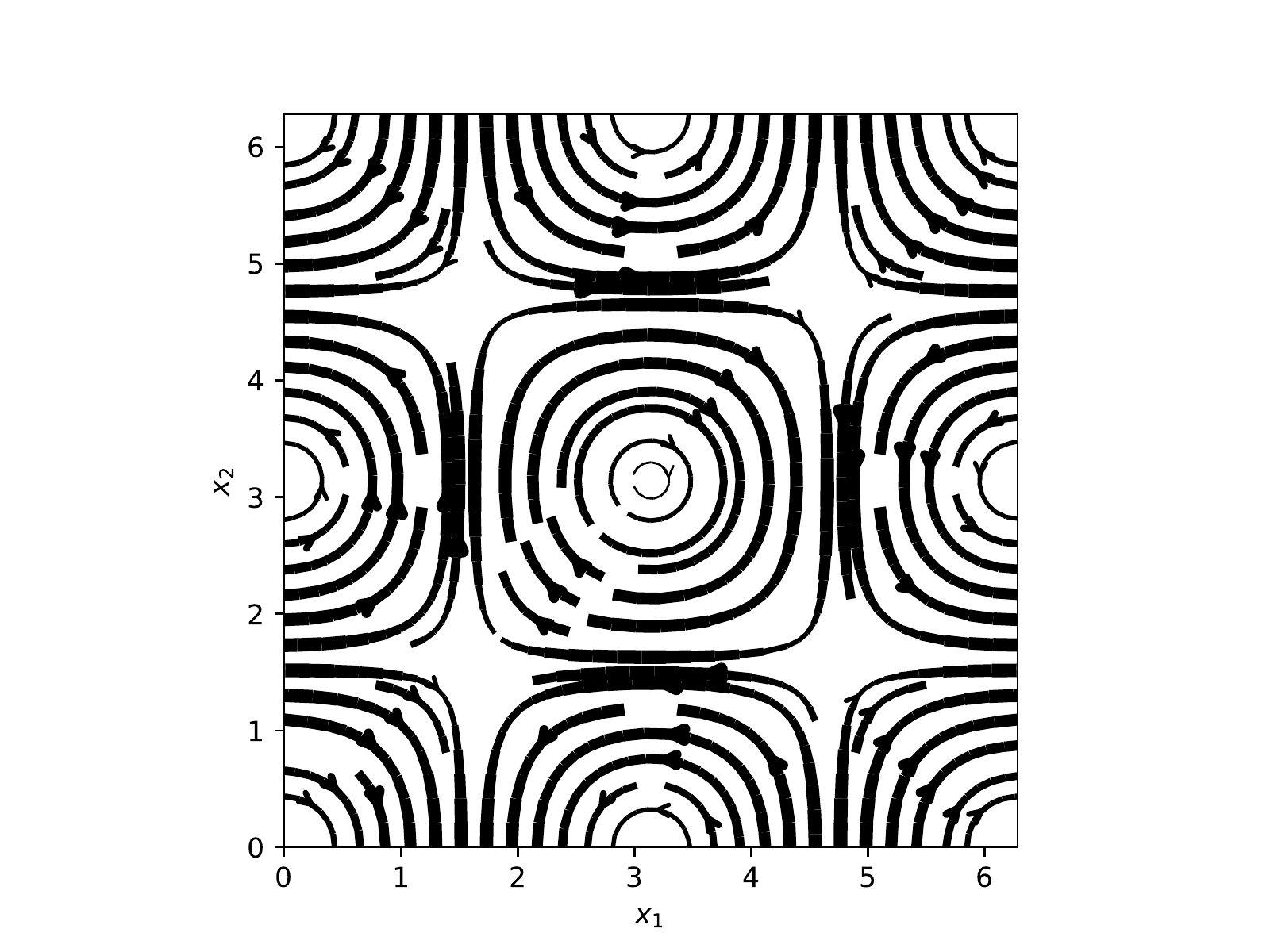}}
  \subfloat[\SFdf\ field]{\label{fg:1.b}\includegraphics[clip,trim={1cm 0 1cm 0},width=0.25\textwidth]{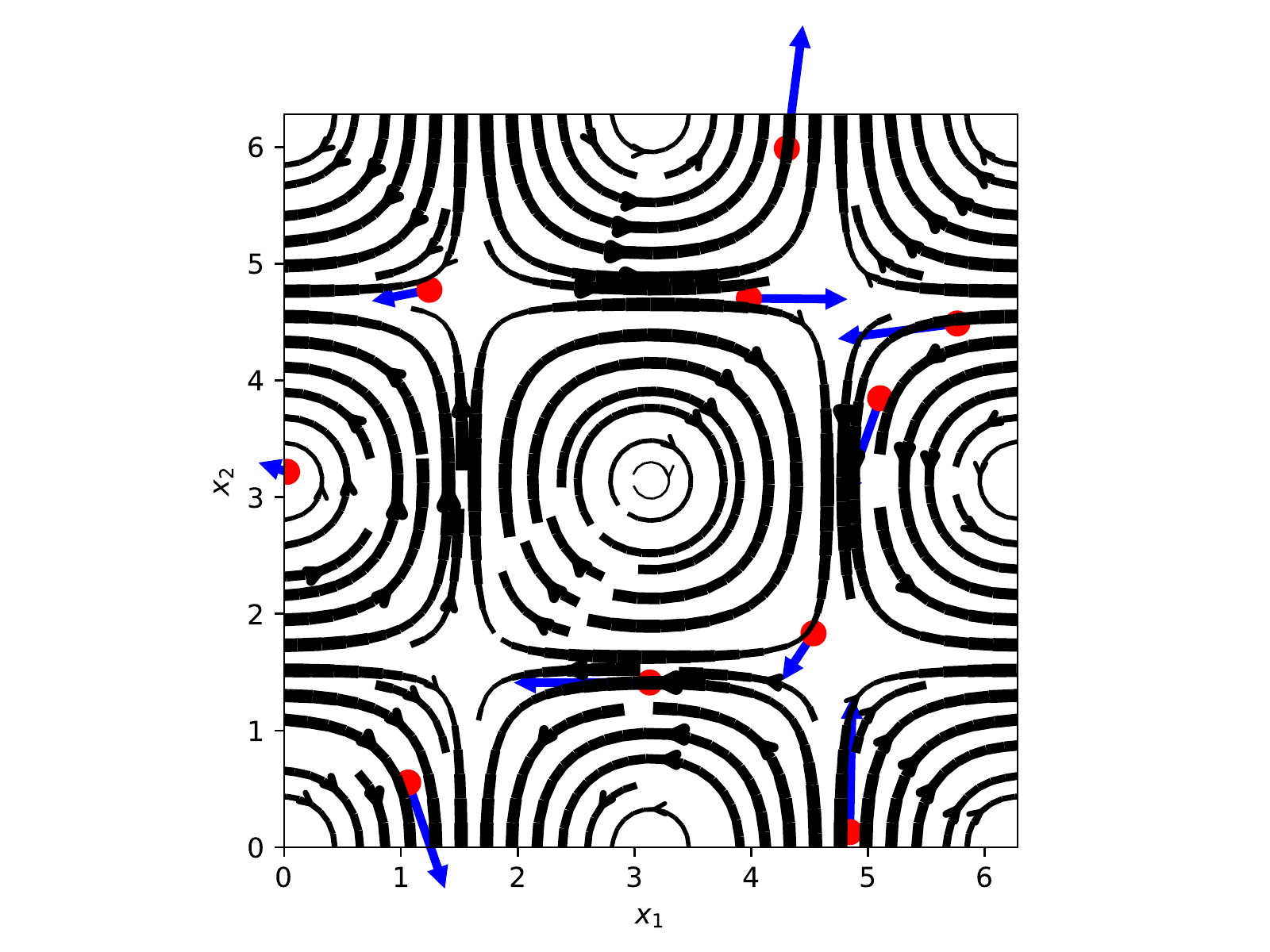}}
  \caption{Results of the Taylor--Green vortex in expression \eqref{eq:taylor.green}.}
\label{fg:1.fourier}
\end{figure}

We set the residual boundary energy $\varepsilon_{\partial\cl{I}}$ to $50\%$ and the stopping criterion $\Delta \varepsilon_{\partial\cl{I}}$ to $10^{-8}$. For the fractional Sobolev regularization, we selected $\epsilon=10^{-5}$ and $k=1.5$. After five outer iterations, we obtained the following index set with $43$ entries
\begin{equation*}
\begin{smallmatrix}
  \bbull & \bbull & \bbull & \bbull & \rbull & \rbull & \rbull\\
  \bbull & \bbull & \bbull & \bbull & \bbull & \bbull & \bbull\\
  \bbull & \bbull & \bbull & \bbull & \bbull & \bbull & \bbull\\
  \bbull & \bbull & \bbull & \bbull & \bbull & \bbull & \bbull\\
  \bbull & \bbull & \bbull & \bbull & \bbull & \bbull & \bbull\\
  \bbull & \bbull & \bbull & \bbull & \bbull & \bbull & \bbull\\
  \rbull & \rbull & \rbull & \bbull & \bbull & \bbull & \bbull\\
\end{smallmatrix}
\end{equation*}
In this example and all subsequent examples, the red and blue dots represent the dropped indices and the indices included in the approximation construction, respectively. The analytical field and the \SFdf\ approximation are presented in panels \subref{fg:1.a} and \subref{fg:1.b} of Figure \ref{fg:1.fourier}, respectively. In this figure, the red dots are the location measurements and the blue arrows are the velocity measurements.

Even though we only had access to $10$ velocity measurements, the actual field and \SFdf\ reconstructed field are indistinguishable.

\FloatBarrier

\subsection{Comparison on radial basis functions}

When the exact field representation is known, we can compute the pointwise error as
\begin{equation}\label{eq:intensive.error}
e(\bs{x})\coloneqq\norm{\bs{\upsilon}(\bs{x})-\bs{\upsilon}_{\cl{I}}^{\mathrm{opt}}(\bs{x})}{},
\end{equation}
where $\bs{\upsilon}_{\cl{I}}^{\mathrm{opt}}$ is the \SFdf\ approximation. Analogously, we define the partwise error (continuous $L^2$) error as
\begin{equation}\label{eq:extensive.error}
E\coloneqq\norm{\bs{\upsilon}(\bs{x})-\bs{\upsilon}_{\cl{I}}^{\mathrm{opt}}(\bs{x})}{L^2(\cl{D})}=\left(\int\limits_{\cl{D}}e^2(\bs{x})\dv\right)^{1/2}.
\end{equation}

To assess the efficiency of our method, we compare its results with those of standard spatial interpolators that ignore physical features. The basis functions in this assessment are listed in Table \ref{tb:rbf} (here, $r$ is the radial distance).
\begin{table}
\caption{Radial basis functions}
\begin{tabular}{ c c c c c c c }\label{tb:rbf}
 multiquadratic & inverse & Gaussian & linear & cubic & quintic & thin plate \\
 \hline\\[-10pt]
 $[(r/s)^2 + 1]^{1/2}$ & $[(r/s)^2 + 1]^{-1/2}$ & $\exp(-(r/s)^2)$ & $r$ & $r^3$ & $r^5$ & $r^2\log(r)$
\end{tabular}
\end{table}
For the three first cases, we set the parameter $s$ to $1$ and $2$, yielding the basis functions depicted in Table \ref{tb:rbf.case.1}.
\begin{table}
\caption{Radial basis functions: cases}
\begin{tabular}{ c c c c c c }\label{tb:rbf.case.1}
 $\#1$ & $\#2$ & $\#3$ & $\#4$ & $\#5$ & $\#6$ \\
 \hline\\[-10pt]
 $[r^2 + 1]^{1/2}$ & $[(r/2)^2 + 1]^{1/2}$ & $[r^2 + 1]^{-1/2}$ & $[(r/2)^2 + 1]^{-1/2}$ & $\exp(-r^2)$ & $\exp(-(r/2)^2)$
\end{tabular}
% \end{table}
% \begin{table}
\begin{tabular}{ c c c c }
 $\#7$ & $\#8$ & $\#9$ & $\#10$ \\
 \hline\\[-10pt]
 $r$ & $r^3$ & $r^5$ & $r^2\log(r)$
\end{tabular}
\end{table}

\subsubsection{First comparison}

The velocity field to be recovered is given by
\begin{equation}\label{eq:fourier.several.vortices}
\bs{\upsilon}(\bs{x})=\fr{1}{2}(\cos(x_1)\sin(x_2)+\cos(2x_1)\sin(2x_2),-\sin(x_1)\cos(x_2)-\sin(2x_1)\cos(2x_2)).
\end{equation}
As in the previous example, in a continuous $L^2$-norm, this field can be exactly computed with a finite Fourier series. However, we assume that only $36$ velocity measurements are given at randomly distributed points in a domain $[0,2\pi]^2$.

\begin{figure}[!htb]
\centering
  \subfloat[Analytical field]{\label{fg:2.a}\includegraphics[clip,trim={1cm 0 1cm 0},width=0.25\textwidth]{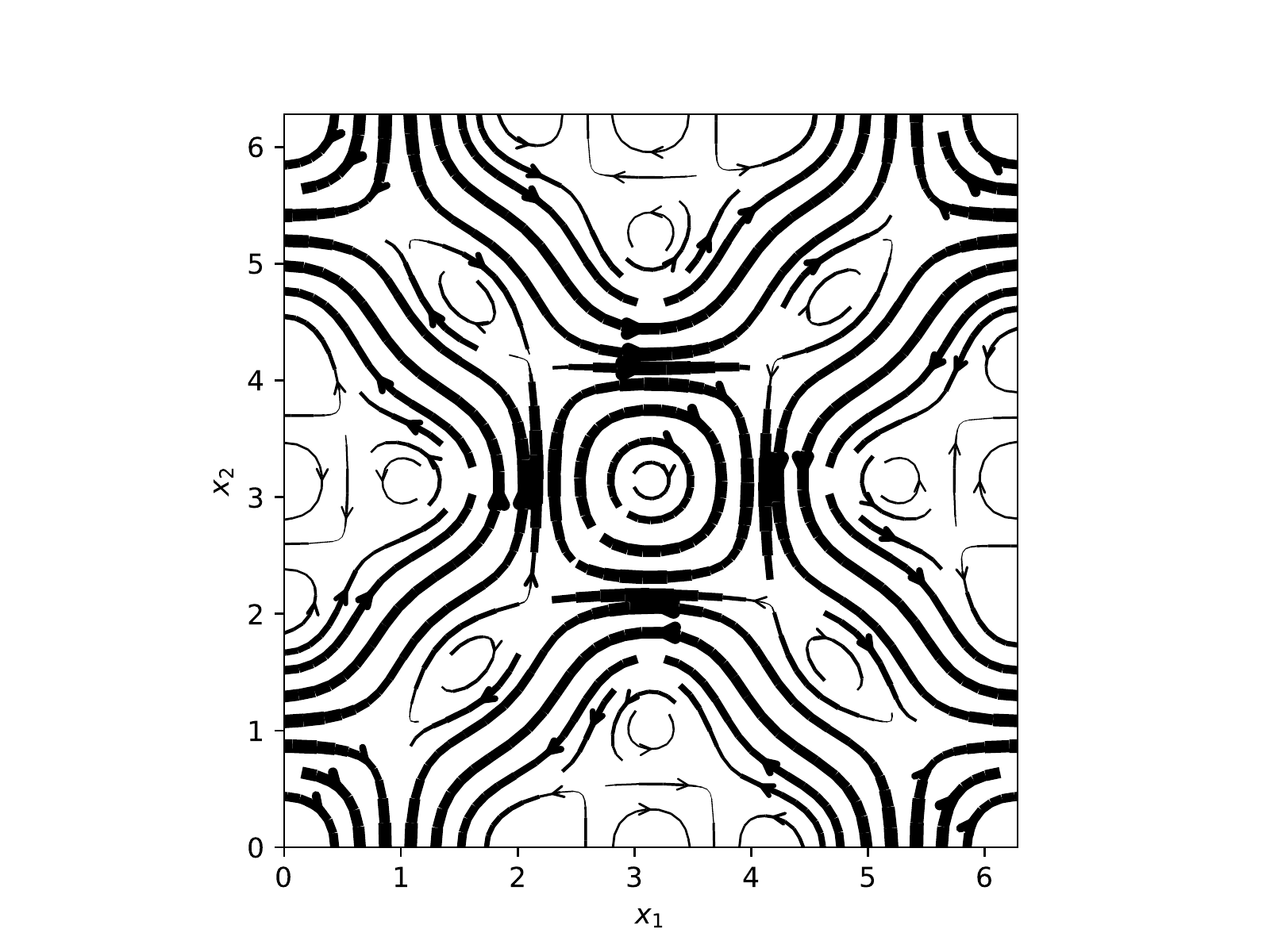}}
  \subfloat[\SFdf\ field]{\label{fg:2.b}\includegraphics[clip,trim={1cm 0 1cm 0},width=0.25\textwidth]{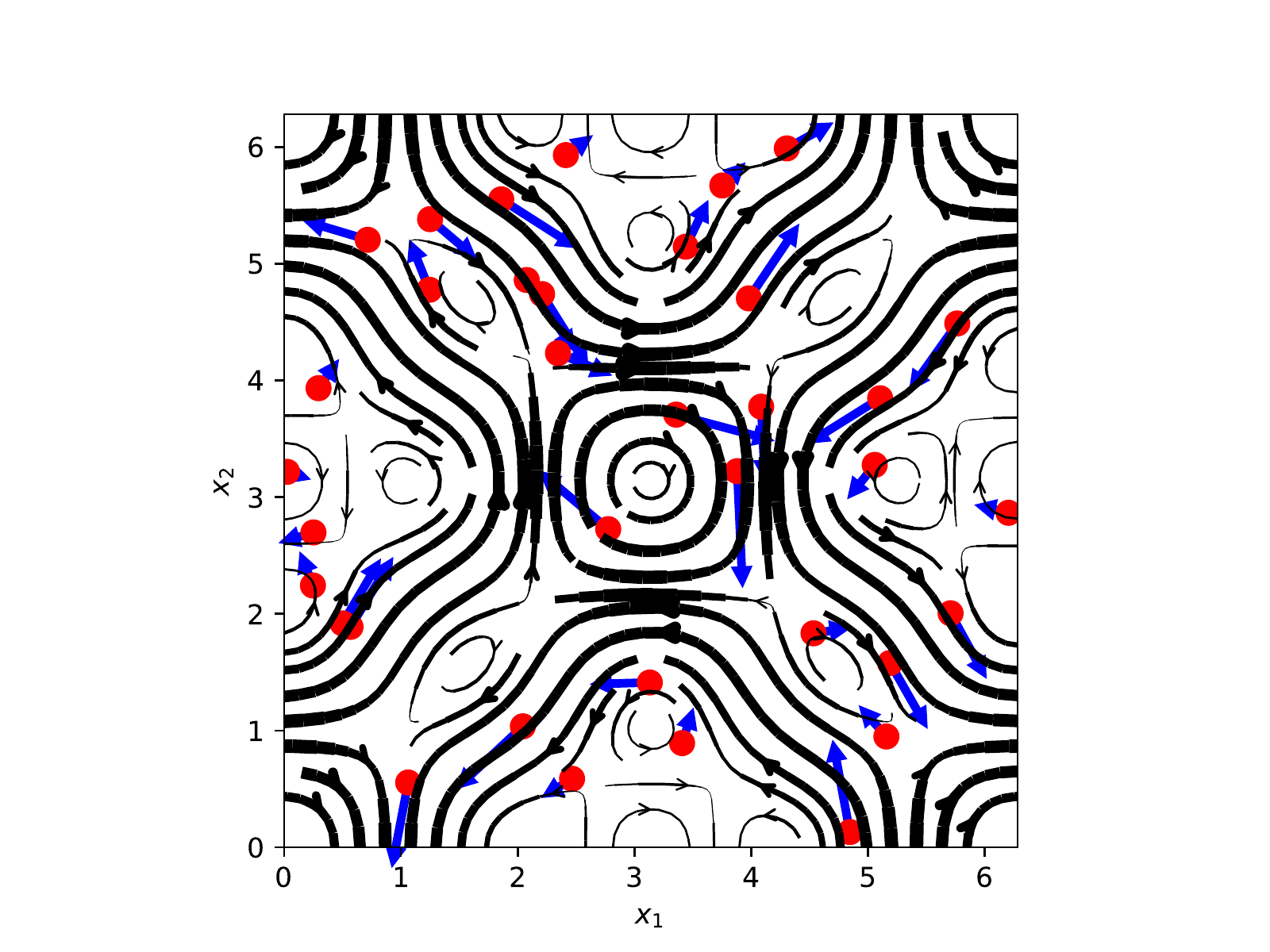}}\\
  \subfloat[Error field $e(\bs{x})$]{\label{fg:2.c}\includegraphics[clip,trim={0cm 0 0cm 0},width=0.25\textwidth]{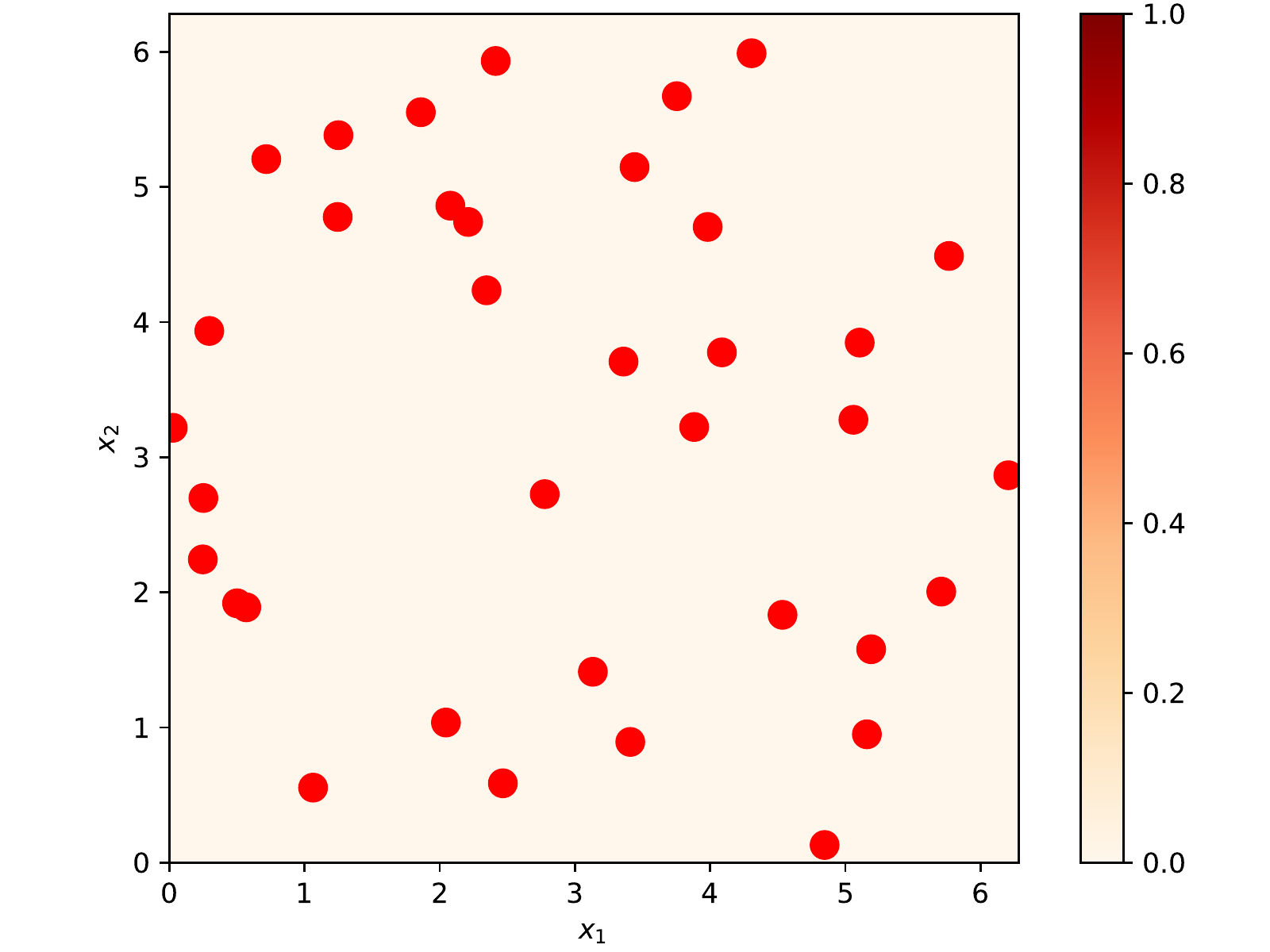}}
  \caption{Reconstruction of the analytical vortices in expression \eqref{eq:fourier.several.vortices}}
\label{fg:2.fourier}
\end{figure}

In this evaluation, the residual boundary energy was $\varepsilon_{\partial\cl{I}}=50\%$ and the stopping criterion was $\Delta \varepsilon_{\partial\cl{I}}=10^{-7}$. For the fractional Sobolev regularization, we selected $\epsilon=10^{-6}$ and $k=1.5$. After six outer iterations, we obtained the following index set with $53$ entries
\begin{equation*}
\begin{smallmatrix}
  \rbull & \rbull & \rbull & \rbull & \bbull & \bbull & \bbull\\
  \bbull & \bbull & \bbull & \rbull & \bbull & \bbull & \bbull\\
  \bbull & \bbull & \bbull & \bbull & \bbull & \bbull & \bbull\\
  \bbull & \bbull & \bbull & \bbull & \bbull & \bbull & \bbull\\
  \bbull & \bbull & \bbull & \bbull & \bbull & \bbull & \bbull\\
  \bbull & \bbull & \bbull & \bbull & \bbull & \bbull & \bbull\\
  \bbull & \bbull & \bbull & \bbull & \bbull & \bbull & \bbull\\
  \bbull & \bbull & \bbull & \rbull & \bbull & \bbull & \bbull\\
  \bbull & \bbull & \bbull & \rbull & \rbull & \rbull & \rbull\\
\end{smallmatrix}
\end{equation*}
Panels \subref{fg:2.a}, \subref{fg:2.b}, and \subref{fg:2.b} of Figure \ref{fg:2.fourier} display the analytical field, the \SFdf\ approximation, and the error field of the \SFdf\ approximation, respectively. Note that, the actual field and \SFdf\ reconstructed field are indistinguishable.

\begin{figure}[!htb]
\centering
  \subfloat[$\#1$]{\label{fg:2.d}\includegraphics[clip,trim={2cm 0 2cm 0},width=0.2\textwidth]{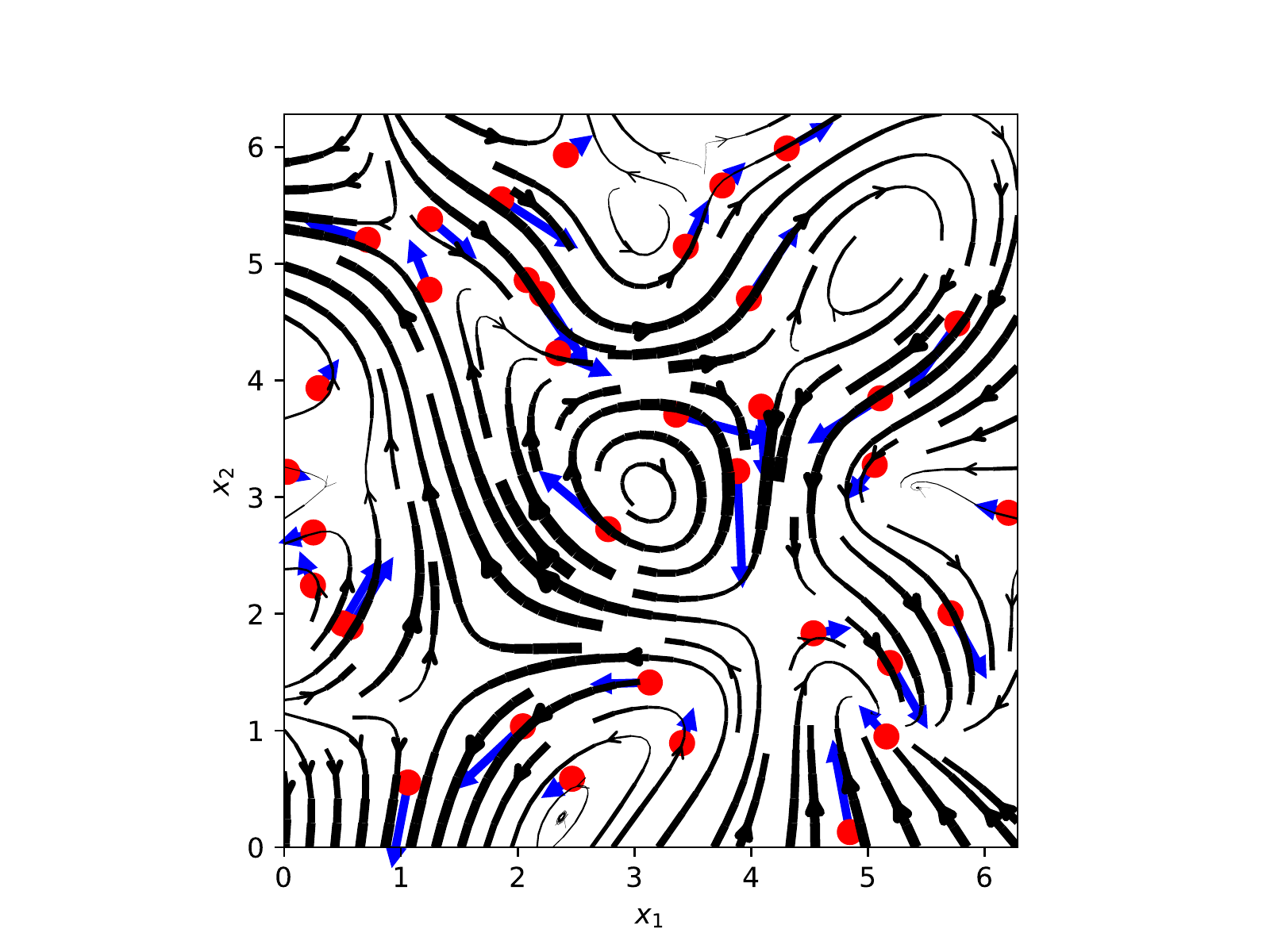}}
  \subfloat[$\#2$]{\label{fg:2.e}\includegraphics[clip,trim={2cm 0 2cm 0},width=0.2\textwidth]{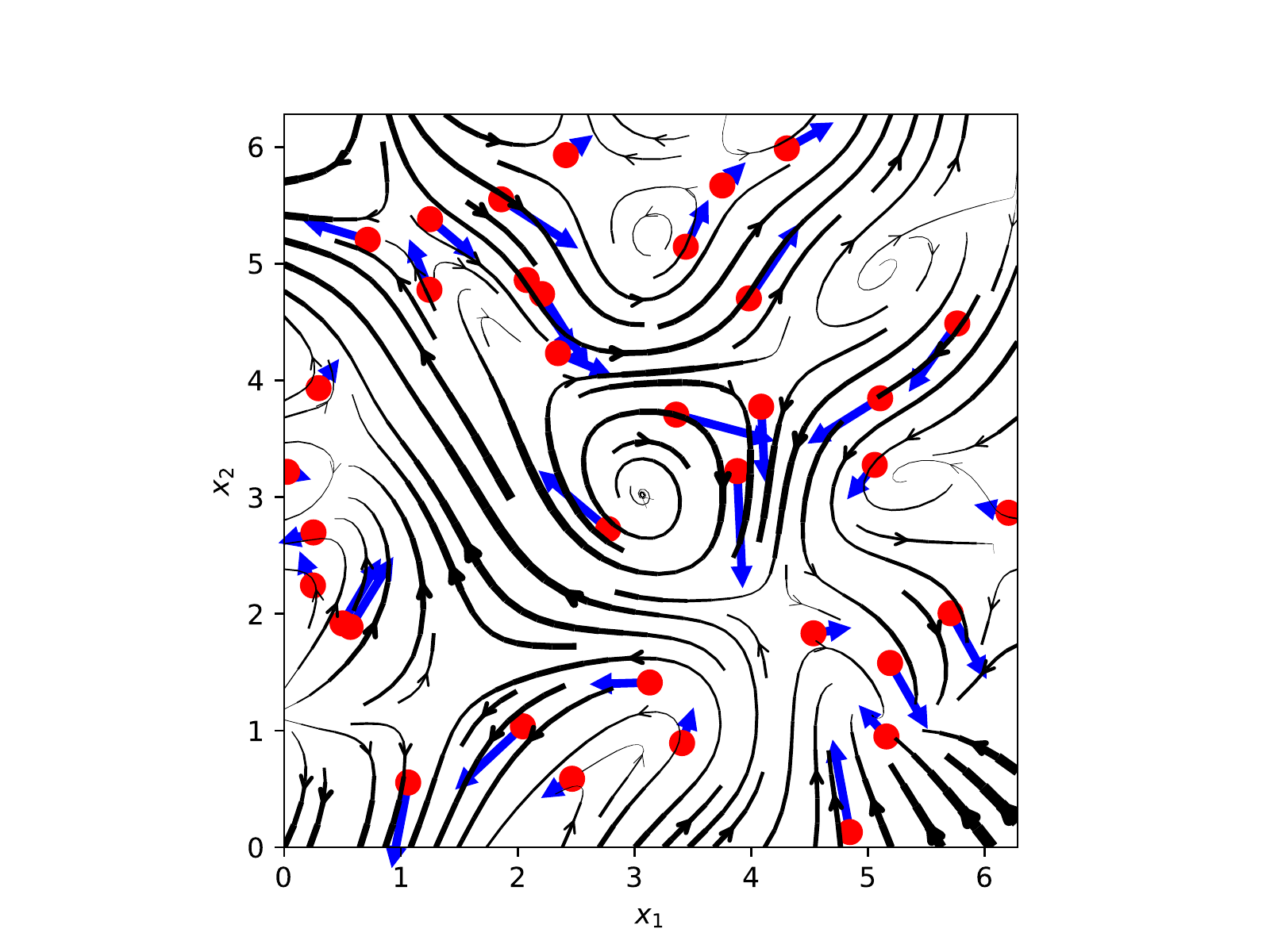}}
  \subfloat[$\#3$]{\label{fg:2.f}\includegraphics[clip,trim={2cm 0 2cm 0},width=0.2\textwidth]{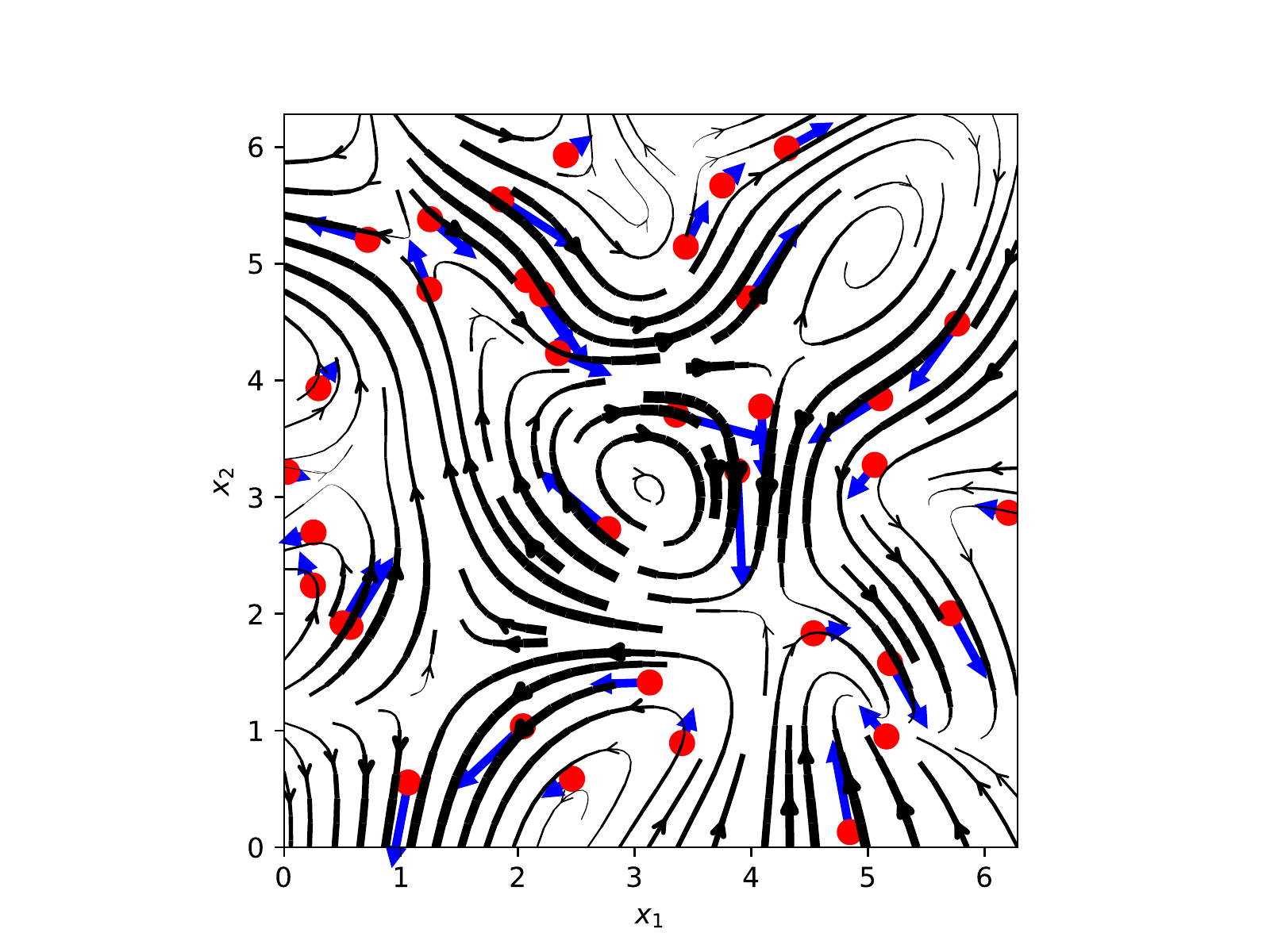}}
  \subfloat[$\#4$]{\label{fg:2.g}\includegraphics[clip,trim={2cm 0 2cm 0},width=0.2\textwidth]{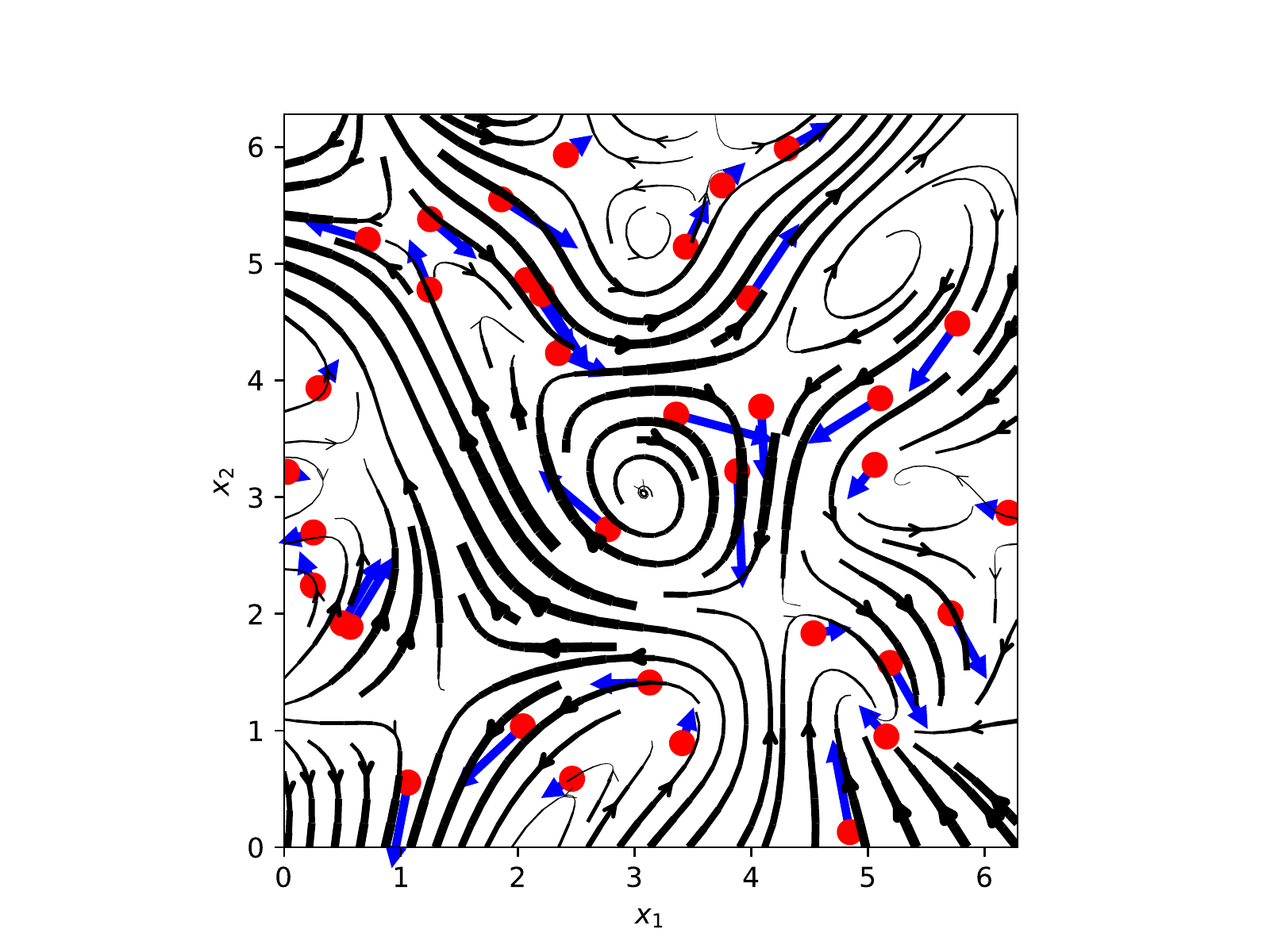}}
  \subfloat[$\#5$]{\label{fg:2.h}\includegraphics[clip,trim={2cm 0 2cm 0},width=0.2\textwidth]{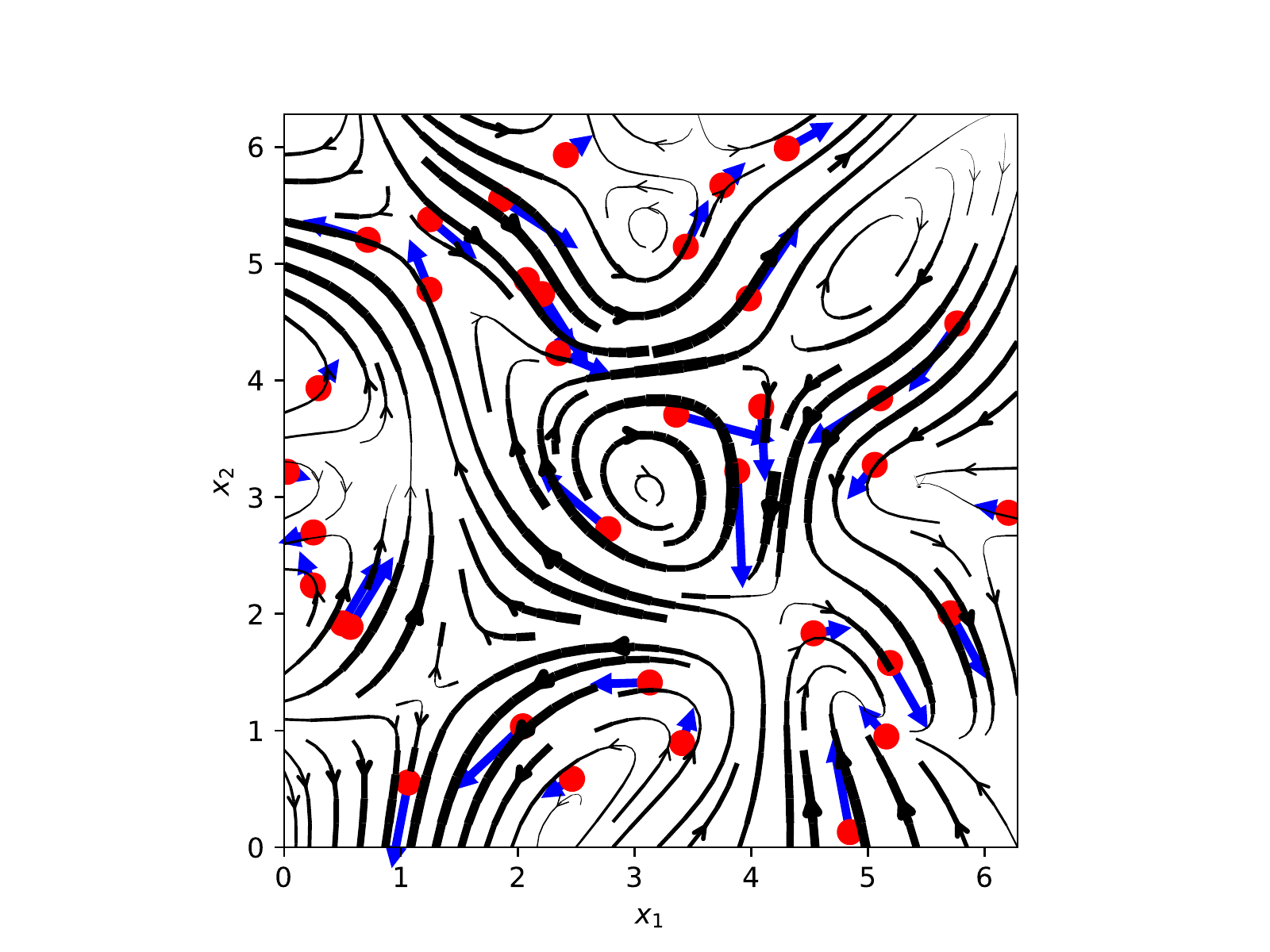}}\\
  \subfloat[$\#6$]{\label{fg:2.i}\includegraphics[clip,trim={2cm 0 2cm 0},width=0.2\textwidth]{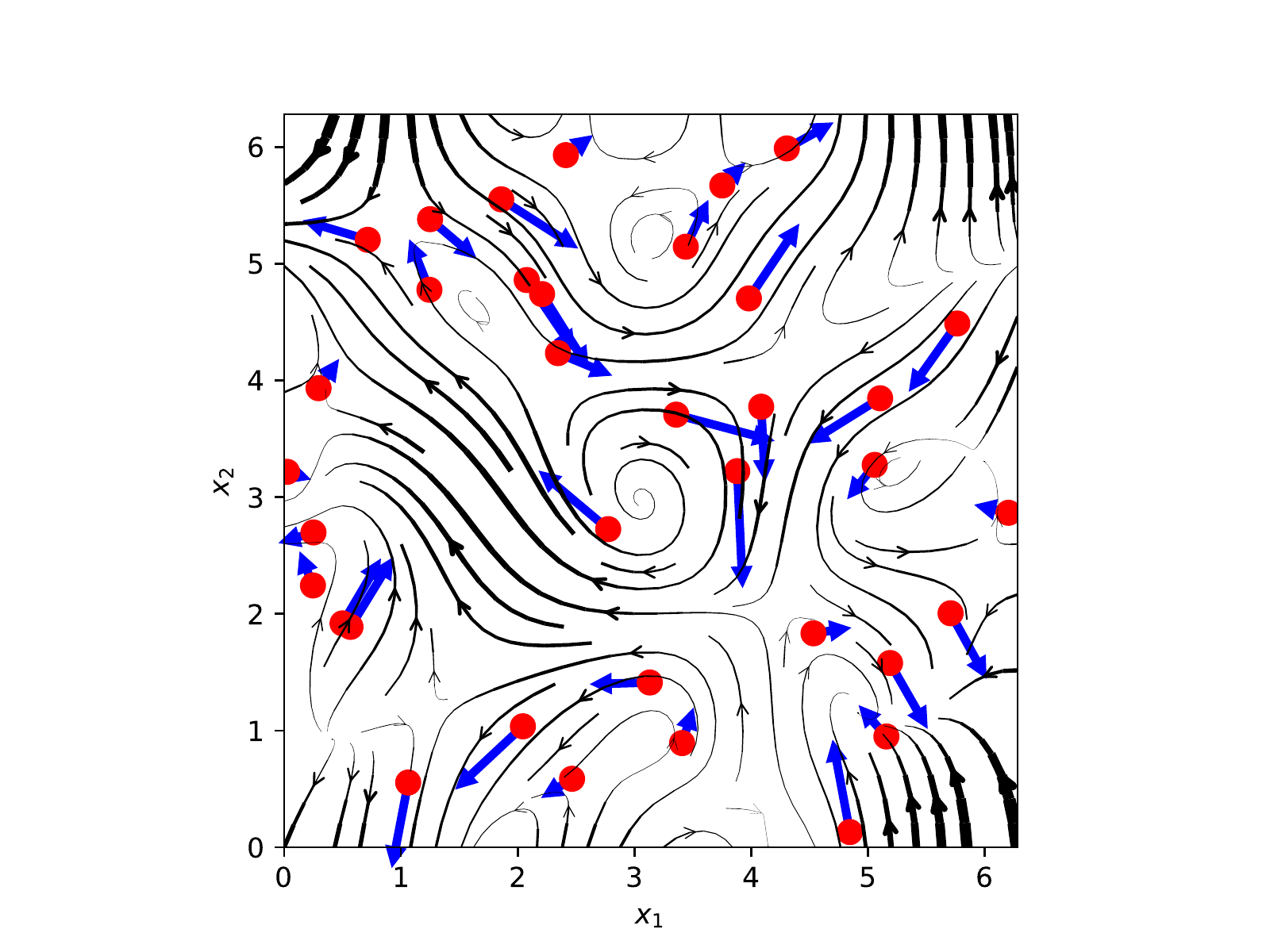}}
  \subfloat[$\#7$]{\label{fg:2.j}\includegraphics[clip,trim={2cm 0 2cm 0},width=0.2\textwidth]{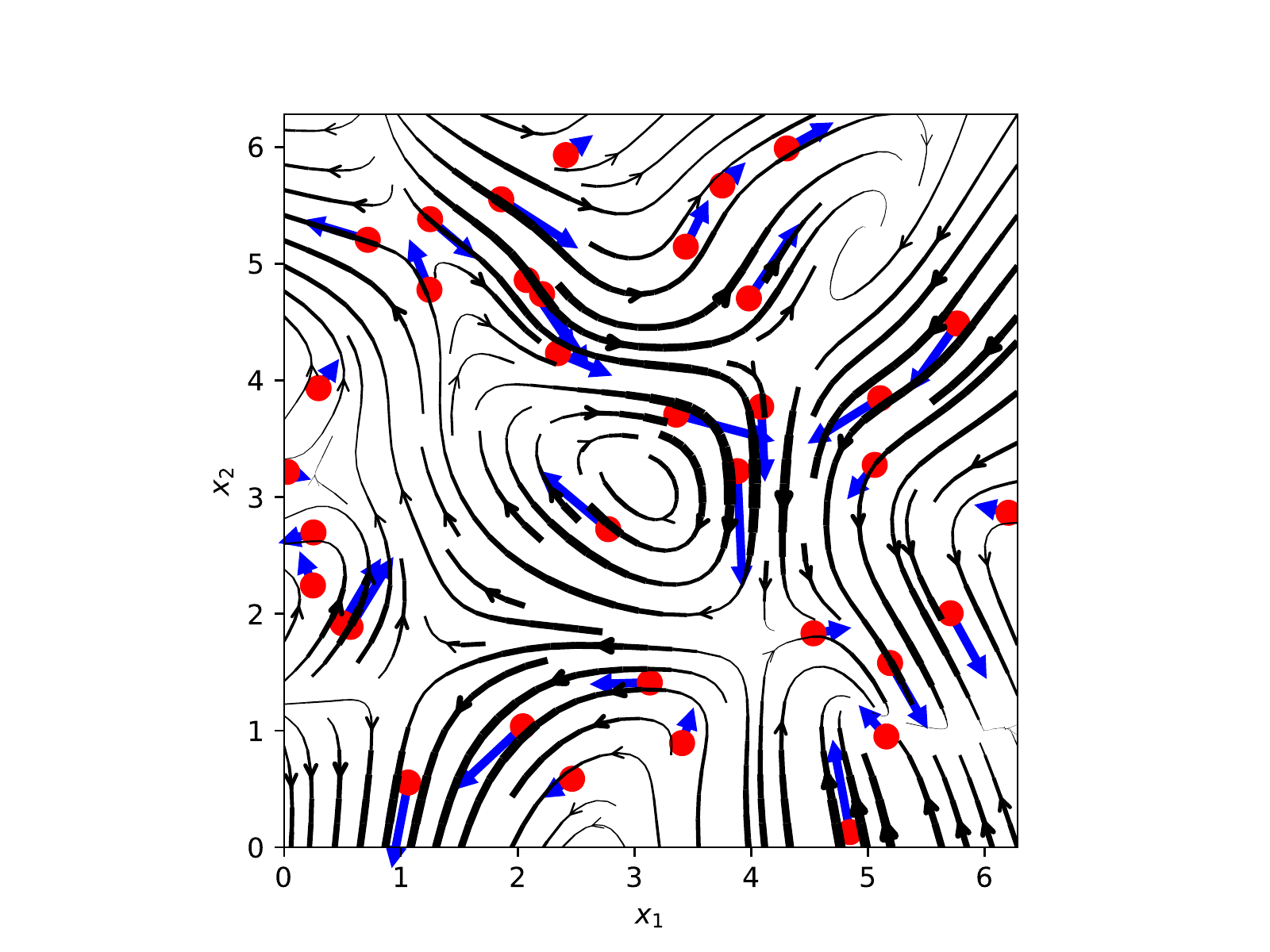}}
  \subfloat[$\#8$]{\label{fg:2.k}\includegraphics[clip,trim={2cm 0 2cm 0},width=0.2\textwidth]{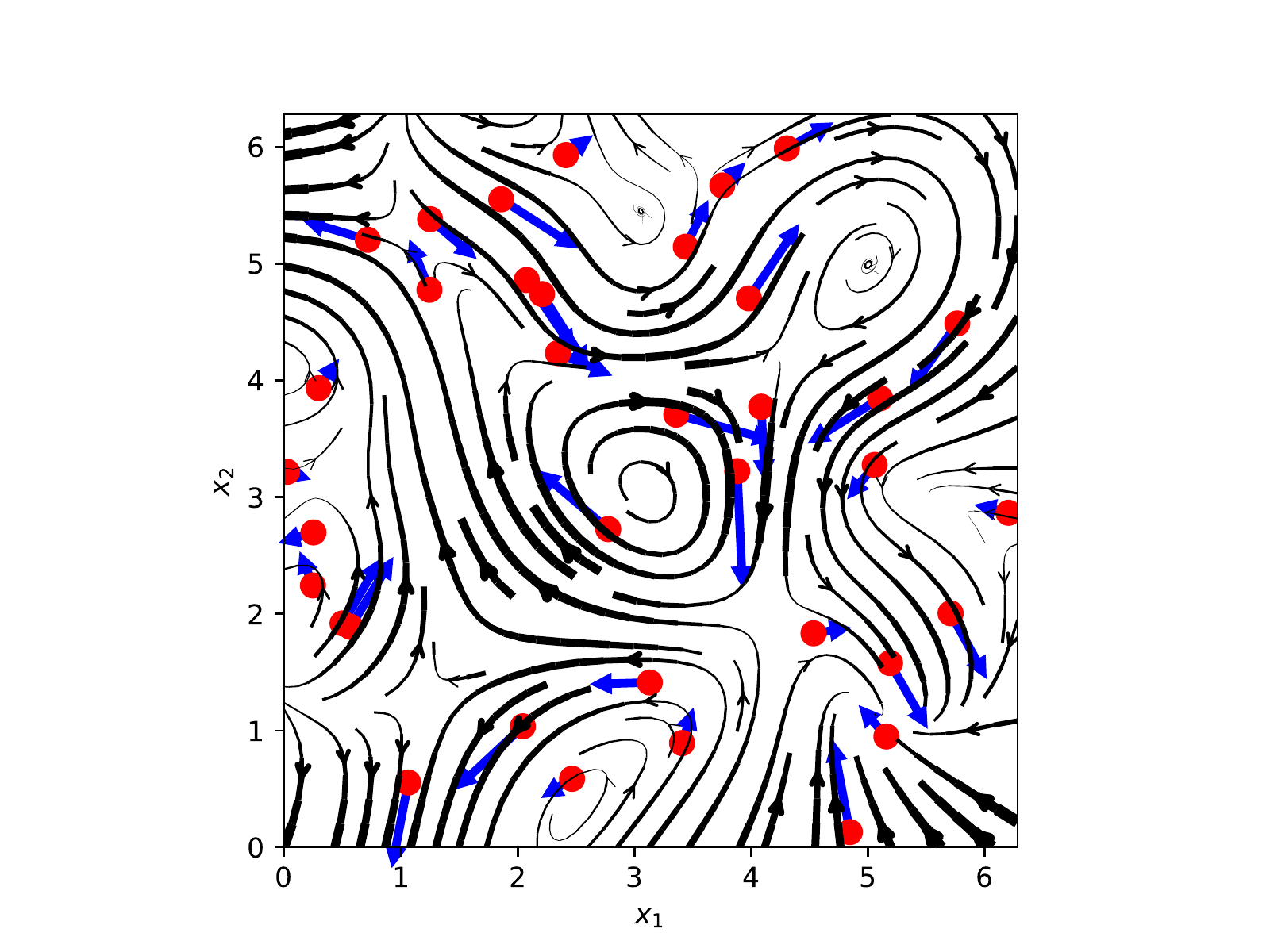}}
  \subfloat[$\#9$]{\label{fg:2.l}\includegraphics[clip,trim={2cm 0 2cm 0},width=0.2\textwidth]{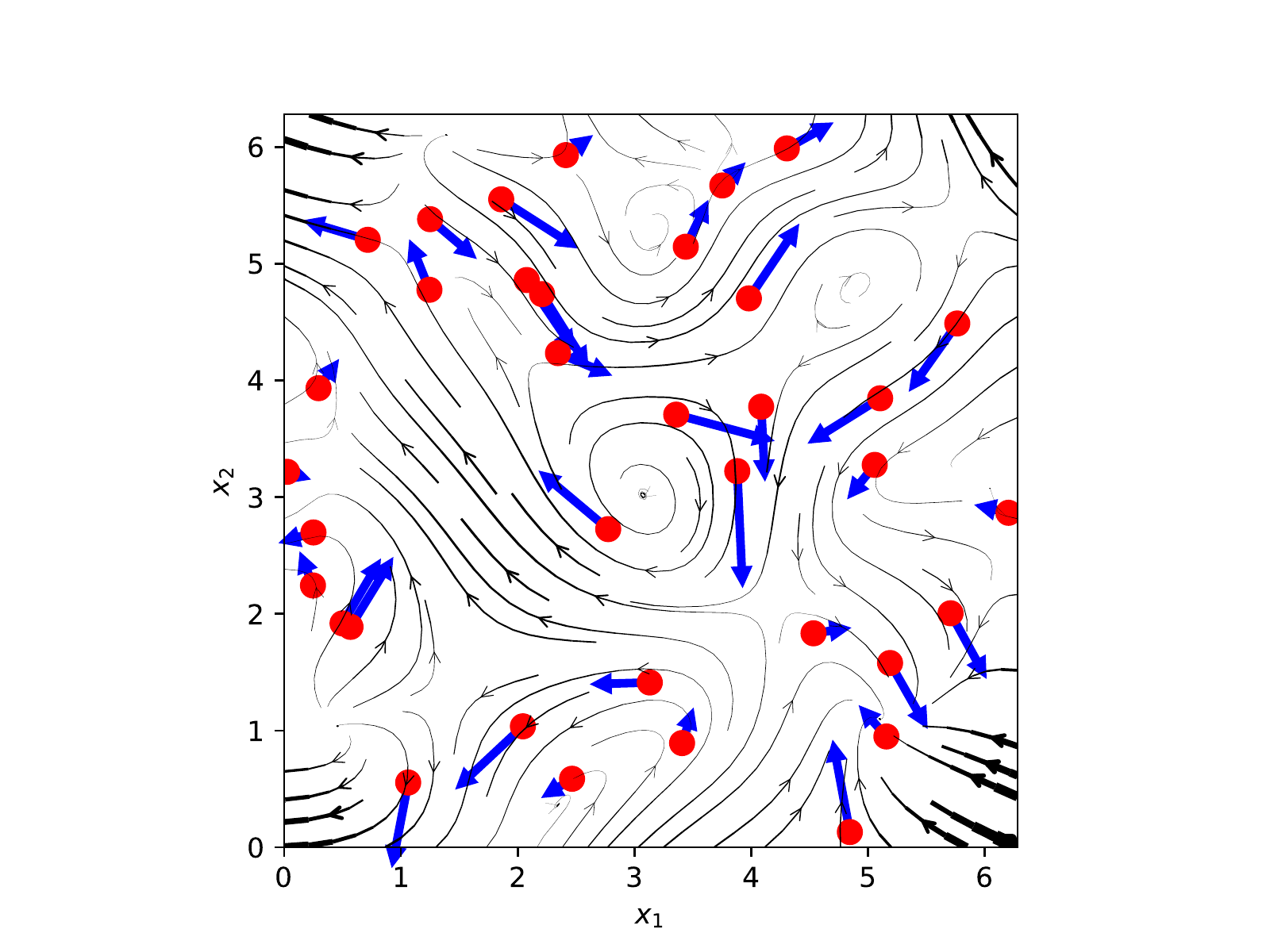}}
  \subfloat[$\#10$]{\label{fg:2.m}\includegraphics[clip,trim={2cm 0 2cm 0},width=0.2\textwidth]{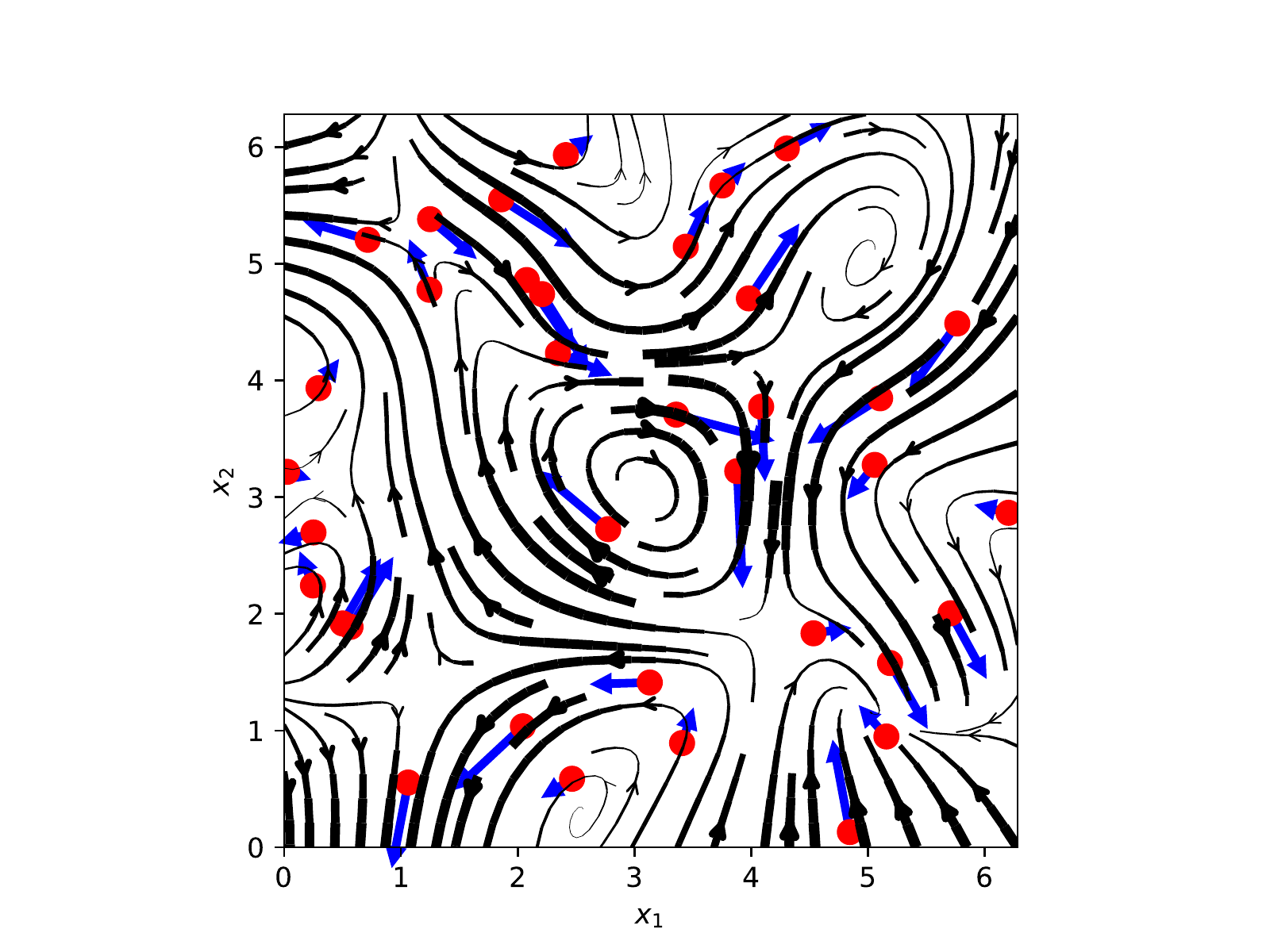}}
  \caption{Radial basis function approximations.}
\label{fg:2.rbf_1}
\end{figure}

\begin{figure}[!htb]
\centering
  \subfloat[$\#1$]{\label{fg:2.n}\includegraphics[clip,trim={2cm 0 2cm 0},width=0.2\textwidth]{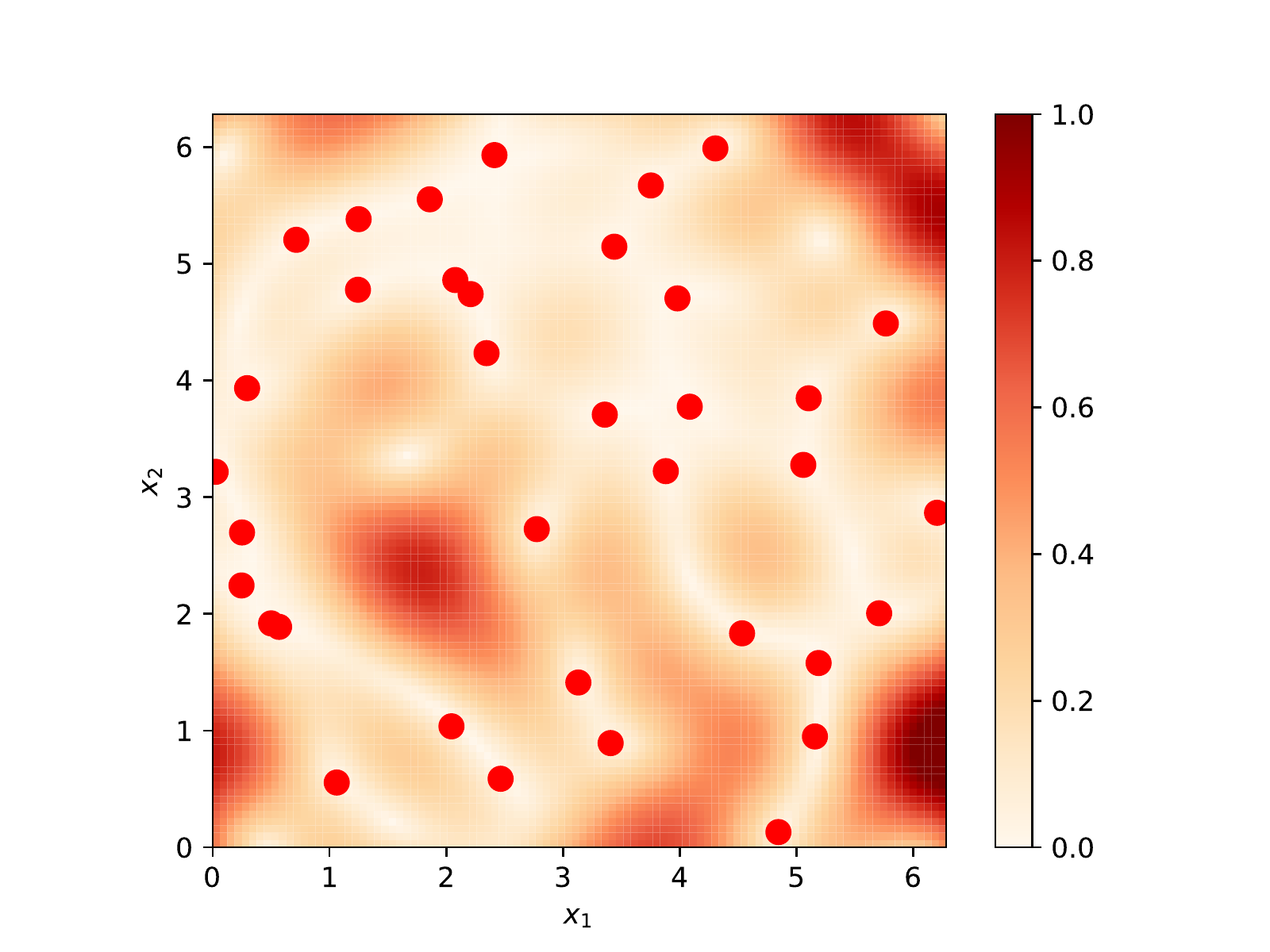}}
  \subfloat[$\#2$]{\label{fg:2.o}\includegraphics[clip,trim={2cm 0 2cm 0},width=0.2\textwidth]{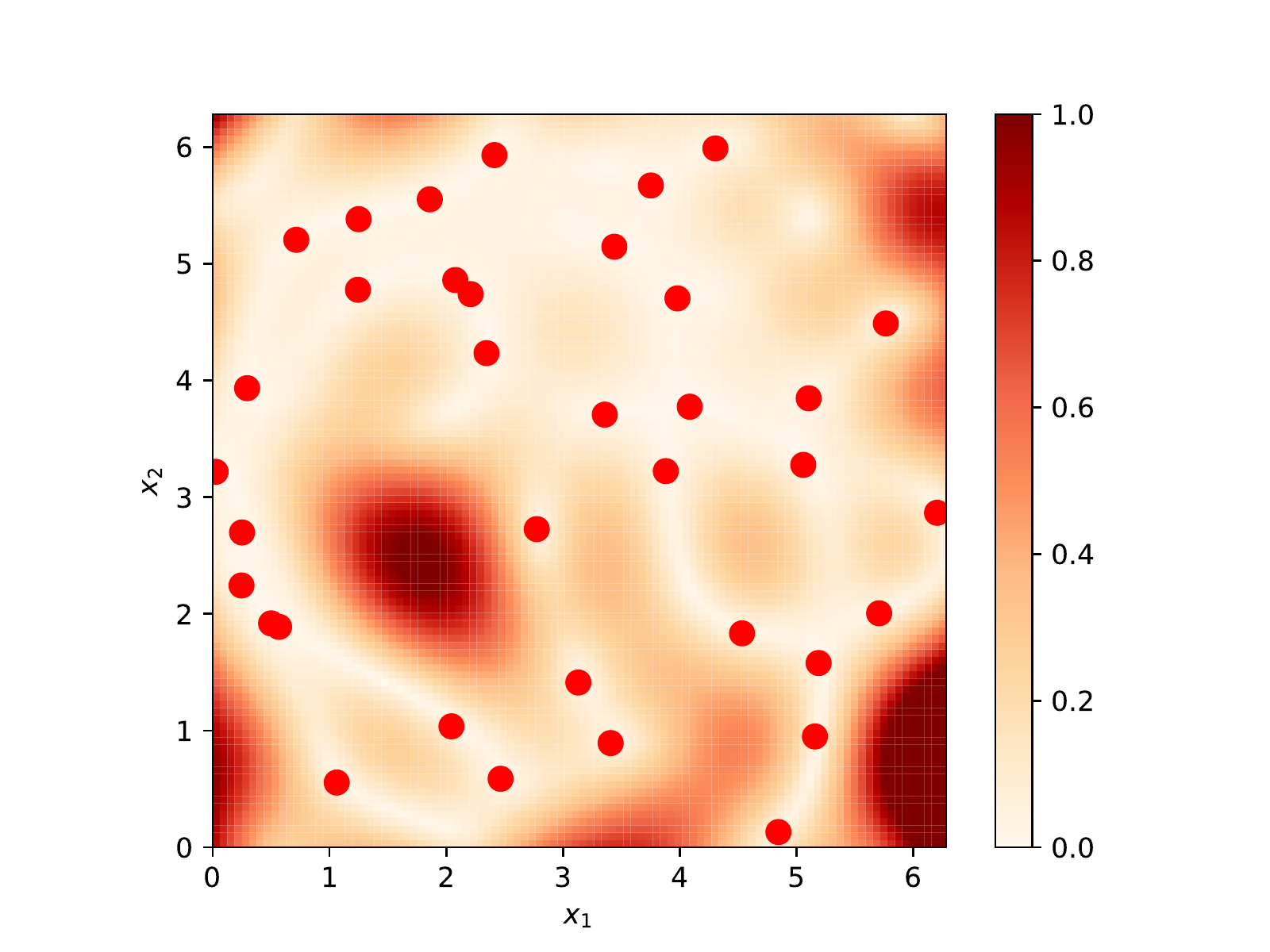}}
  \subfloat[$\#3$]{\label{fg:2.p}\includegraphics[clip,trim={2cm 0 2cm 0},width=0.2\textwidth]{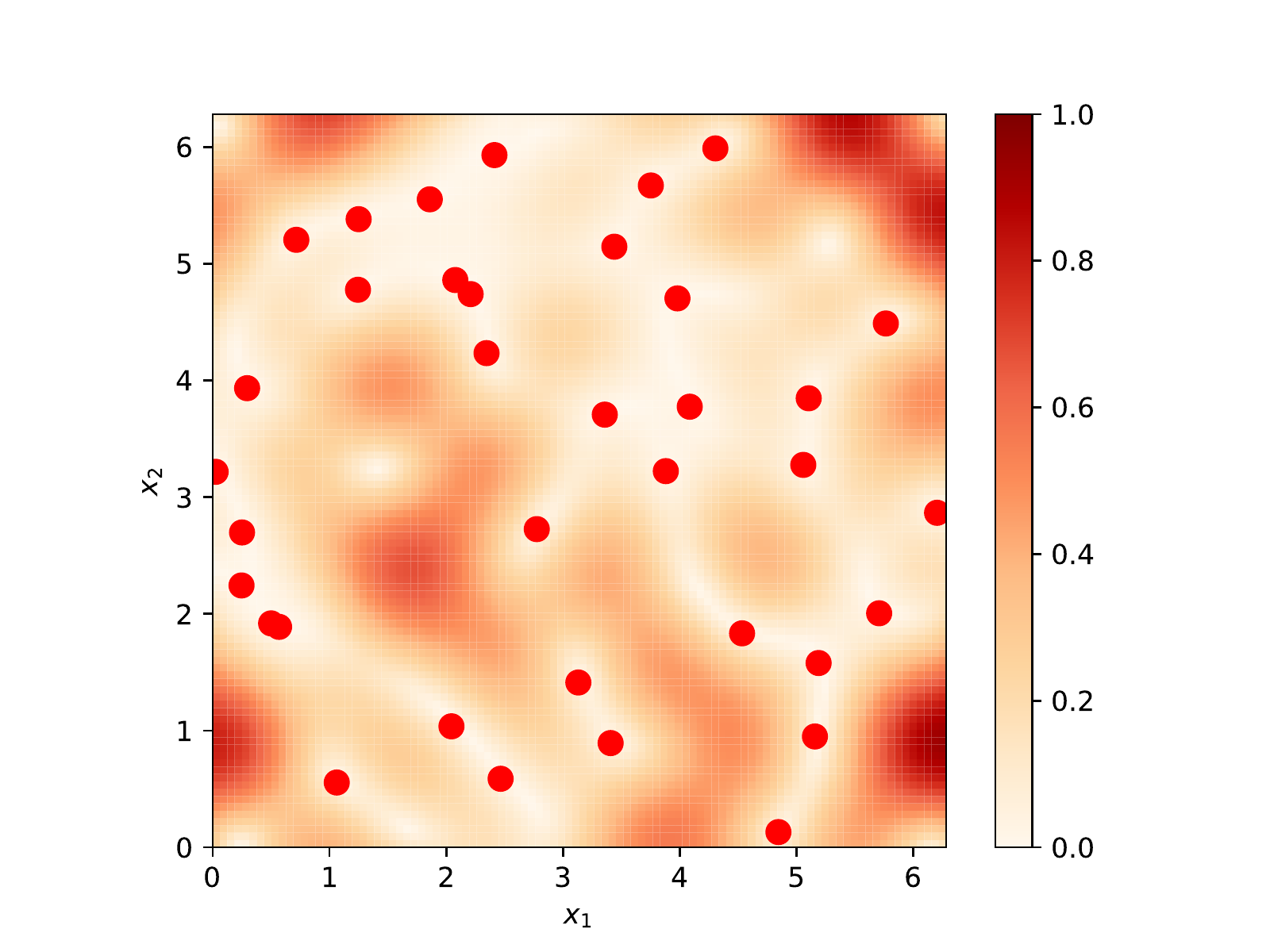}}
  \subfloat[$\#4$]{\label{fg:2.q}\includegraphics[clip,trim={2cm 0 2cm 0},width=0.2\textwidth]{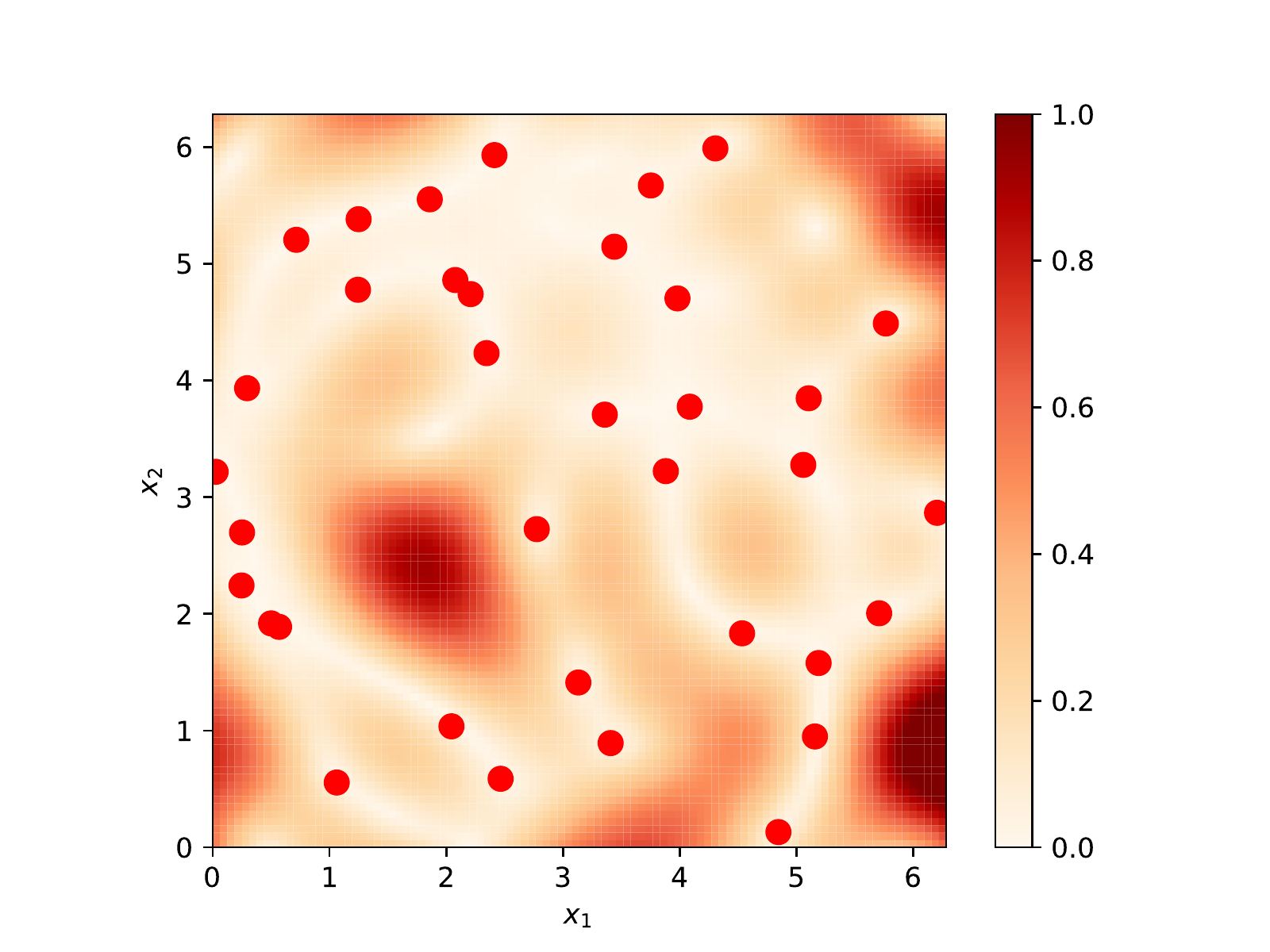}}
  \subfloat[$\#5$]{\label{fg:2.r}\includegraphics[clip,trim={2cm 0 2cm 0},width=0.2\textwidth]{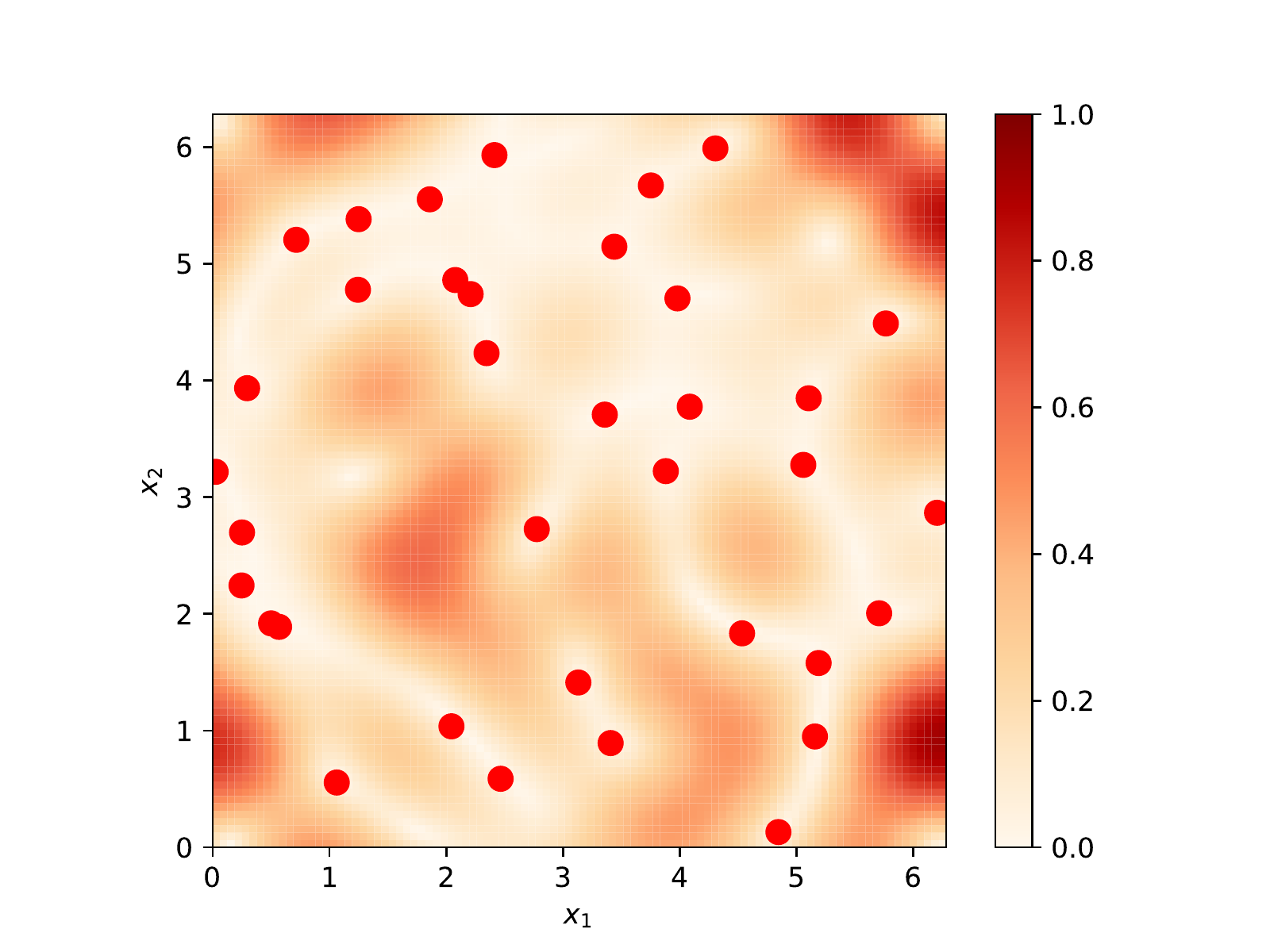}}\\
  \subfloat[$\#6$]{\label{fg:2.s}\includegraphics[clip,trim={2cm 0 2cm 0},width=0.2\textwidth]{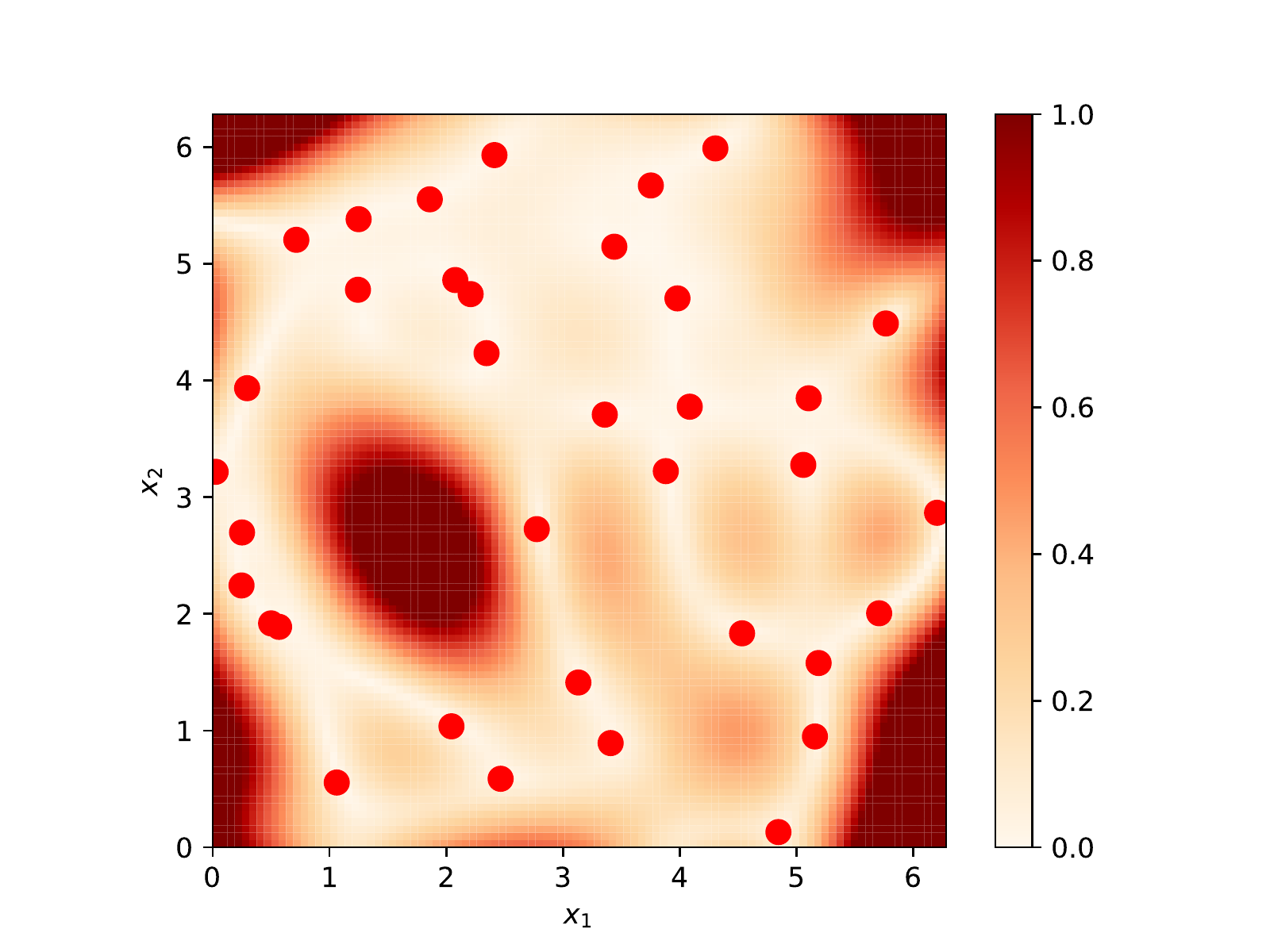}}
  \subfloat[$\#7$]{\label{fg:2.t}\includegraphics[clip,trim={2cm 0 2cm 0},width=0.2\textwidth]{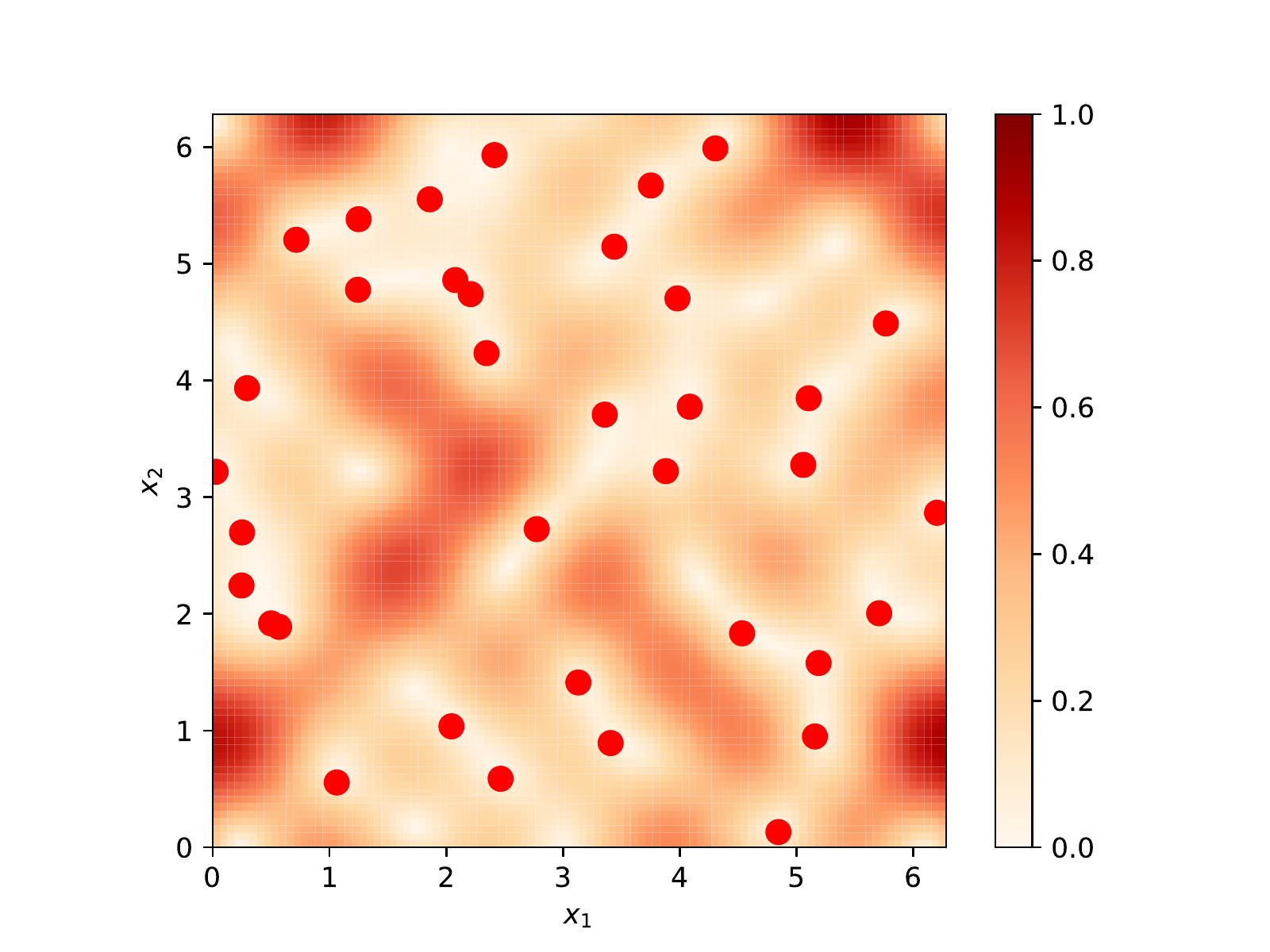}}
  \subfloat[$\#8$]{\label{fg:2.u}\includegraphics[clip,trim={2cm 0 2cm 0},width=0.2\textwidth]{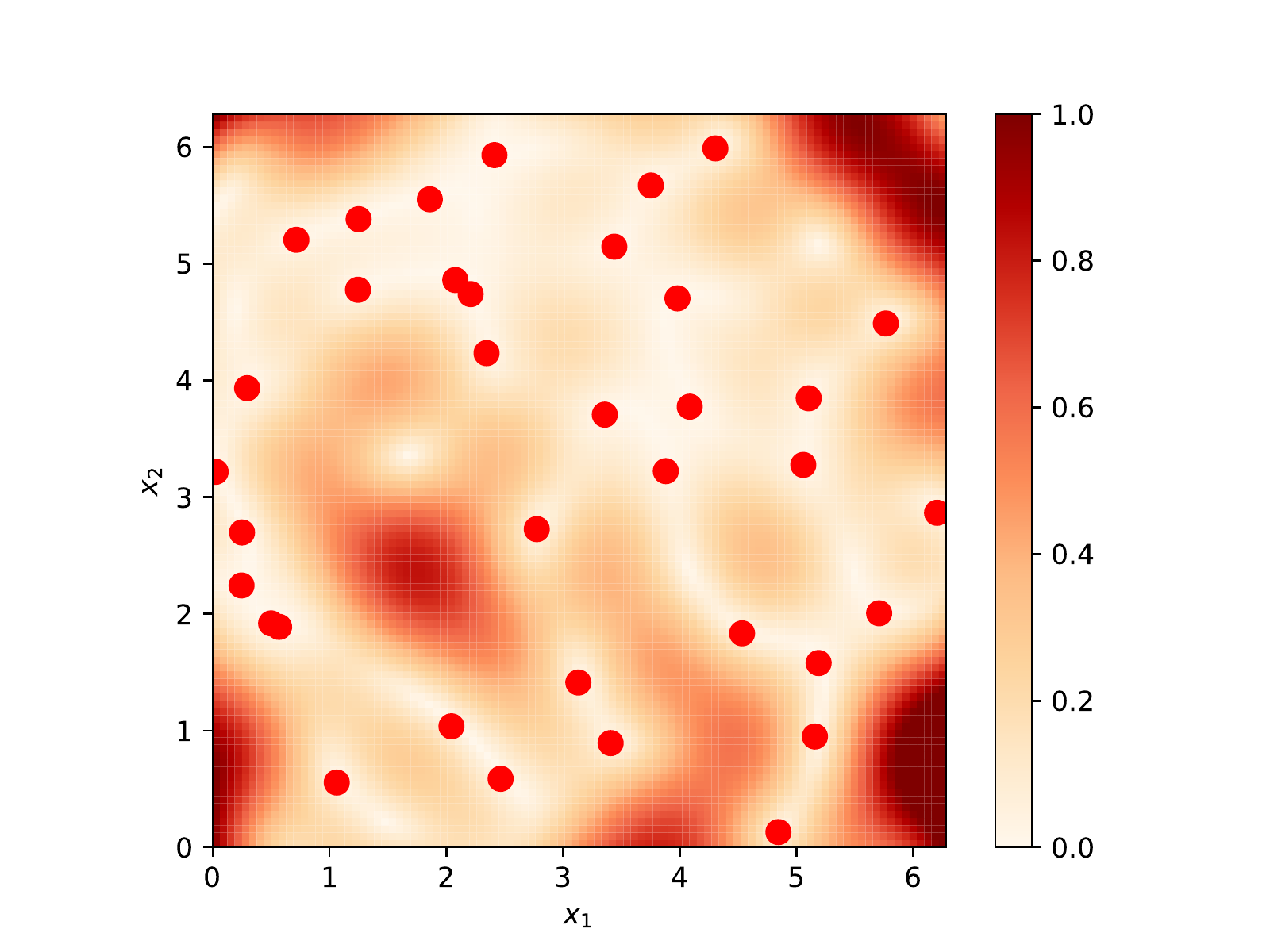}}
  \subfloat[$\#9$]{\label{fg:2.v}\includegraphics[clip,trim={2cm 0 2cm 0},width=0.2\textwidth]{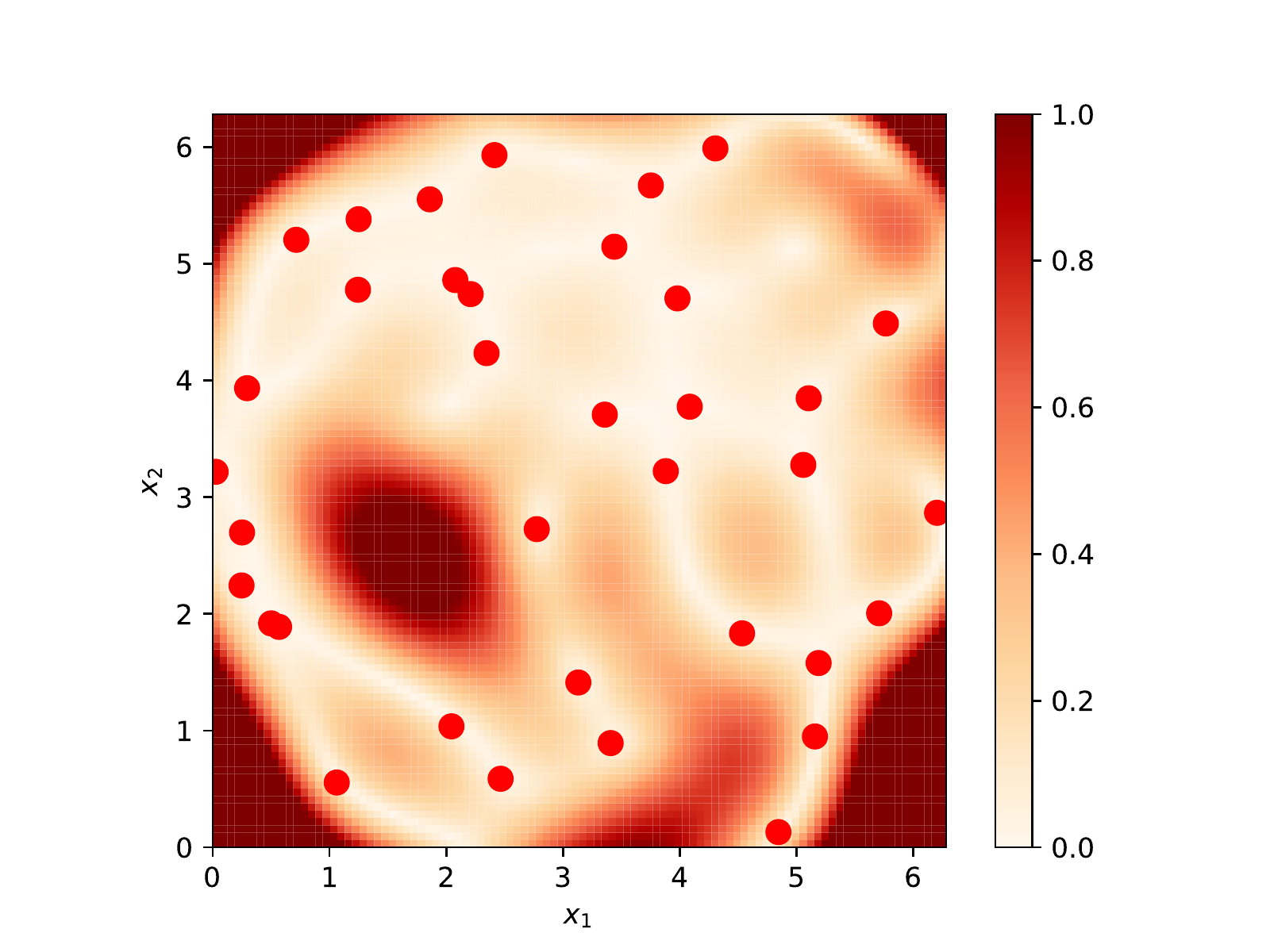}}
  \subfloat[$\#10$]{\label{fg:2.x}\includegraphics[clip,trim={2cm 0 2cm 0},width=0.2\textwidth]{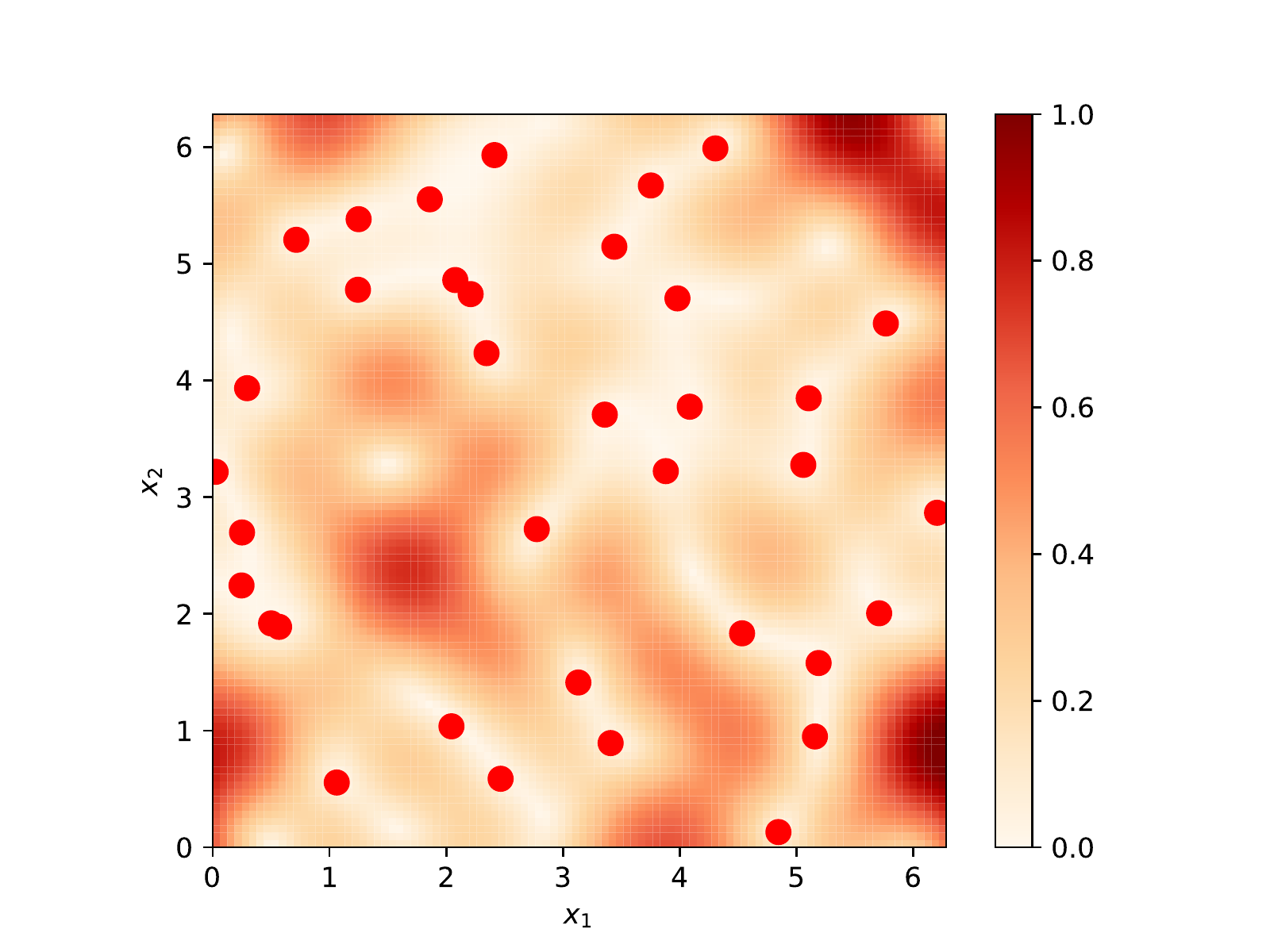}}
  \caption{Error fields of the radial basis function approximations in Figure \ref{fg:2.rbf_1}.}
\label{fg:2.rbf_1_err}
\end{figure}

Figure \ref{fg:2.rbf_1} shows the approximation fields corresponding to the ten radial basis functions. The corresponding error fields of these approximations are given in Figure \ref{fg:2.rbf_1_err}. Clearly, the \SFdf\ approximation is much more accurate than the approximation produce by the classical approximations with radial basis functions. The $L^\infty(e(\bs{x}))$ and $E\coloneqq{L}^2(e(\bs{x}))$ errors of the \SFdf\ approximation are respectively $78$ and $62$ times smaller than the errors obtained with the best approximation using radial basis functions.

\begin{table}
\caption{$L^\infty(e(\bs{x}))$ approximation errors \eqref{eq:intensive.error}}
\begin{tabular}{ c c c c c c c c c c c }
 \SFdf & $\#1$ & $\#2$ & $\#3$ & $\#4$ & $\#5$ & $\#6$ & $\#7$ & $\#8$ & $\#9$ & $\#10$\\
 \hline\\[-10pt]
 0.46 & 39.8 & 47.4 & 38.9 & 36.2 & 36.0 & 76.3 & 42.7 & 49.7 & 144.8 & 42.9
\end{tabular}
\end{table}

\begin{table}
\caption{$E\coloneqq{L}^2(e(\bs{x}))$ approximation errors \eqref{eq:extensive.error}}
\begin{tabular}{ c c c c c c c c c c c }
 \SFdf & $\#1$ & $\#2$ & $\#3$ & $\#4$ & $\#5$ & $\#6$ & $\#7$ & $\#8$ & $\#9$ & $\#10$\\
 \hline\\[-10pt]
 0.024 & 1.7 & 2.0 & 1.7 & 1.7 & 1.5 & 2.9 & 1.9 & 2.0 & 4.7 & 1.8
\end{tabular}
\end{table}

\FloatBarrier

\subsubsection{Second comparison}

The velocity field to be recovered is given by
\begin{equation}\label{eq:nonfourier.several.vortices}
\bs{\upsilon}(\bs{x})=\fr{\log{2}}{5}(-2^{\sin(x_2)}\cos(x_2),2^{1+\sin(2x_1)}\cos(2x_1)).
\end{equation}
Unlike the previous example, in a continuous $L^2$-norm this, field cannot be exactly computed with a finite Fourier series. Here, we assume that only $36$ velocity measurements are given at randomly distributed points in a domain $[0,2\pi]^2$.

\begin{figure}[!htb]
\centering
  \subfloat[Analytical field]{\label{fg:3.a}\includegraphics[clip,trim={1cm 0 1cm 0},width=0.25\textwidth]{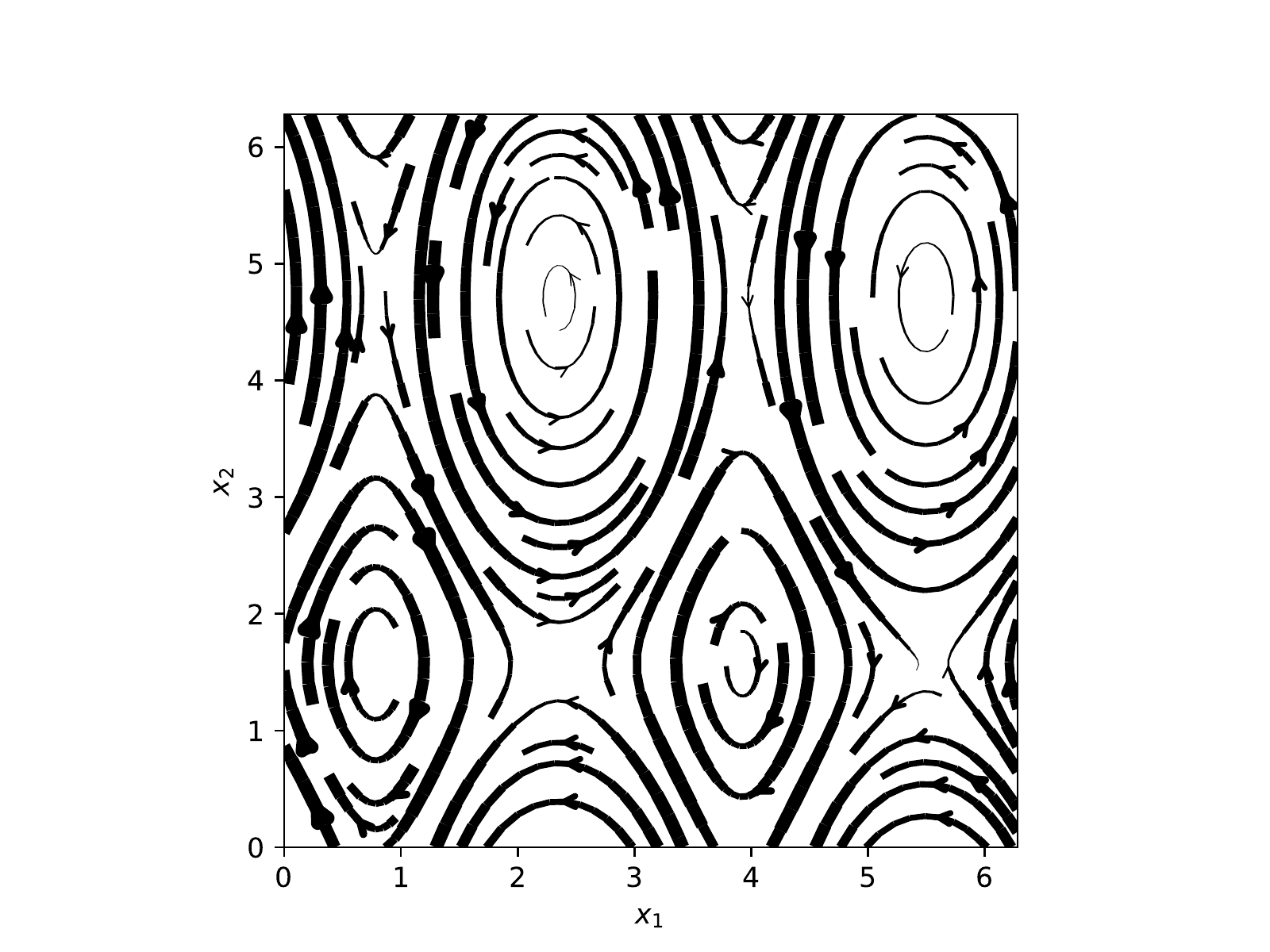}}
  \subfloat[\SFdf\ field]{\label{fg:3.b}\includegraphics[clip,trim={1cm 0 1cm 0},width=0.25\textwidth]{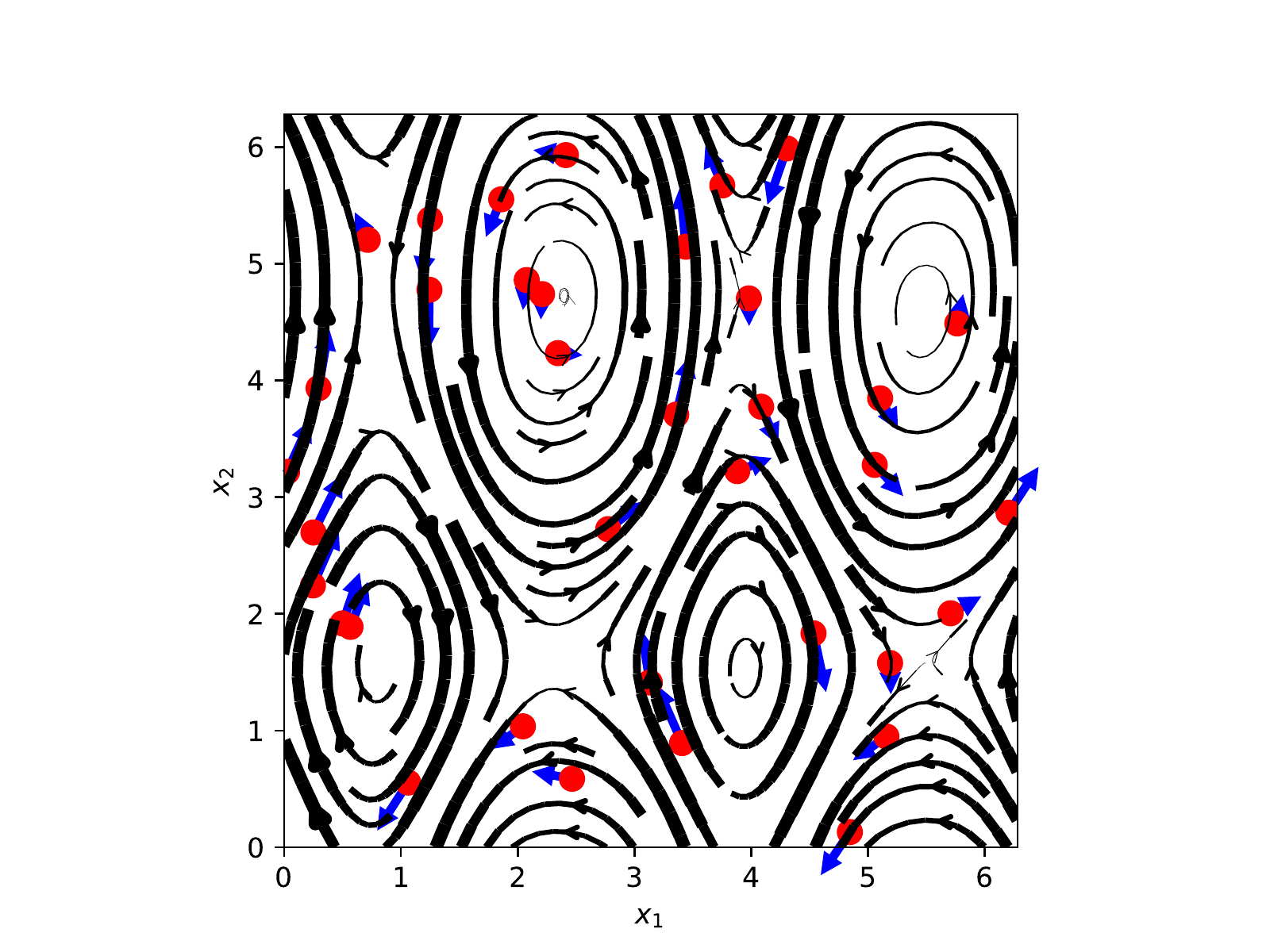}}\\
  \subfloat[Error field $e(\bs{x})$]{\label{fg:3.c}\includegraphics[clip,trim={0cm 0 0cm 0},width=0.25\textwidth]{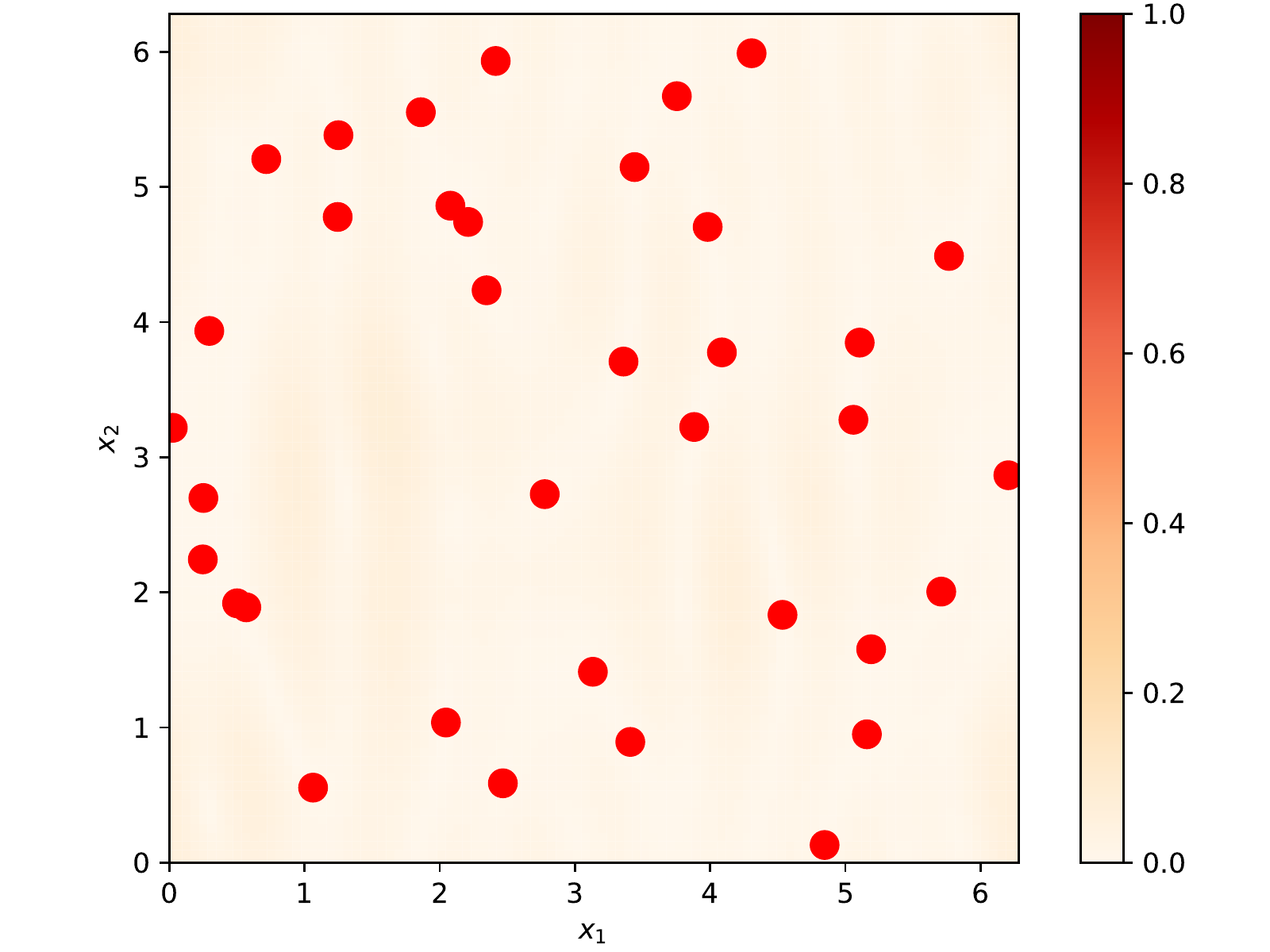}}
  \caption{Reconstruction of the analytical vortices in expression \eqref{eq:nonfourier.several.vortices}}
\label{fg:3.fourier}
\end{figure}

Here, the residual boundary energy was $\varepsilon_{\partial\cl{I}}=50\%$ and the stopping criterion was $\Delta \varepsilon_{\partial\cl{I}}=10^{-3}$. For the fractional Sobolev regularization, we selected $\epsilon=10^{-5}$ and $k=1.5$. After nine outer iterations, we obtained the following index set with $89$ entries
\begin{equation*}
\begin{smallmatrix}
  \rbull & \bbull & \bbull & \bbull & \bbull & \rbull & \rbull & \rbull & \rbull\\
  \rbull & \bbull & \bbull & \bbull & \bbull & \bbull & \bbull & \rbull & \rbull\\
  \rbull & \bbull & \bbull & \bbull & \bbull & \bbull & \bbull & \rbull & \rbull\\
  \rbull & \bbull & \bbull & \bbull & \bbull & \bbull & \bbull & \bbull & \rbull\\
  \rbull & \bbull & \bbull & \bbull & \bbull & \bbull & \bbull & \bbull & \bbull\\
  \bbull & \bbull & \bbull & \bbull & \bbull & \bbull & \bbull & \bbull & \bbull\\
  \bbull & \bbull & \bbull & \bbull & \bbull & \bbull & \bbull & \bbull & \bbull\\
  \bbull & \bbull & \bbull & \bbull & \bbull & \bbull & \bbull & \bbull & \bbull\\
  \bbull & \bbull & \bbull & \bbull & \bbull & \bbull & \bbull & \bbull & \rbull\\
  \rbull & \bbull & \bbull & \bbull & \bbull & \bbull & \bbull & \bbull & \rbull\\
  \rbull & \rbull & \bbull & \bbull & \bbull & \bbull & \bbull & \bbull & \rbull\\
  \rbull & \rbull & \bbull & \bbull & \bbull & \bbull & \bbull & \bbull & \rbull\\
  \rbull & \rbull & \rbull & \rbull & \bbull & \bbull & \bbull & \bbull & \rbull\\
\end{smallmatrix}
\end{equation*}
Panels \subref{fg:3.a}, \subref{fg:3.b}, and \subref{fg:3.c} of Figure \ref{fg:3.fourier} display the analytical field, the \SFdf\ approximation, and the error field of the \SFdf\ approximation, respectively.

\begin{figure}[!htb]
\centering
  \subfloat[$\#1$]{\label{fg:3.d}\includegraphics[clip,trim={2cm 0 2cm 0},width=0.2\textwidth]{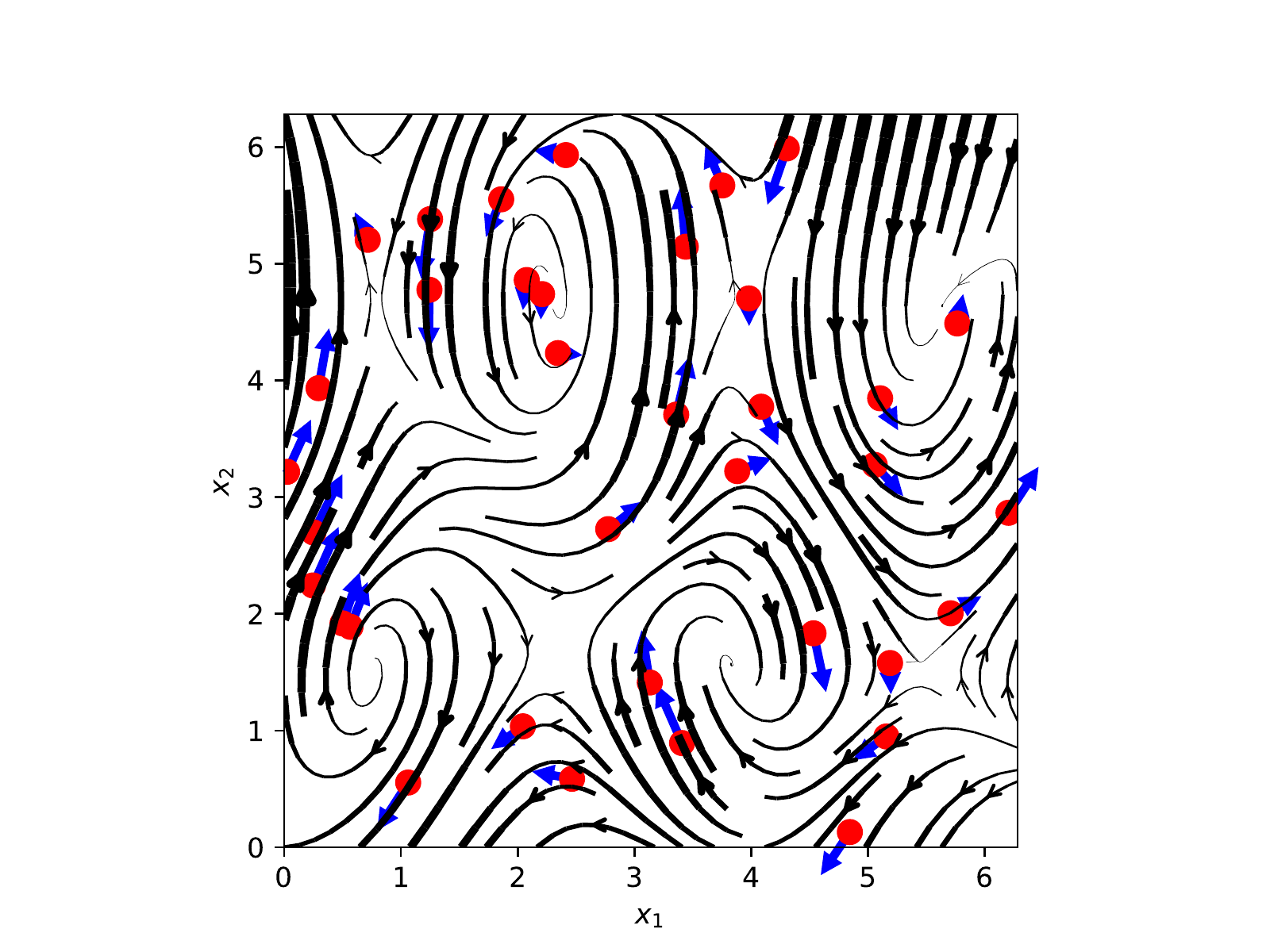}}
  \subfloat[$\#2$]{\label{fg:3.e}\includegraphics[clip,trim={2cm 0 2cm 0},width=0.2\textwidth]{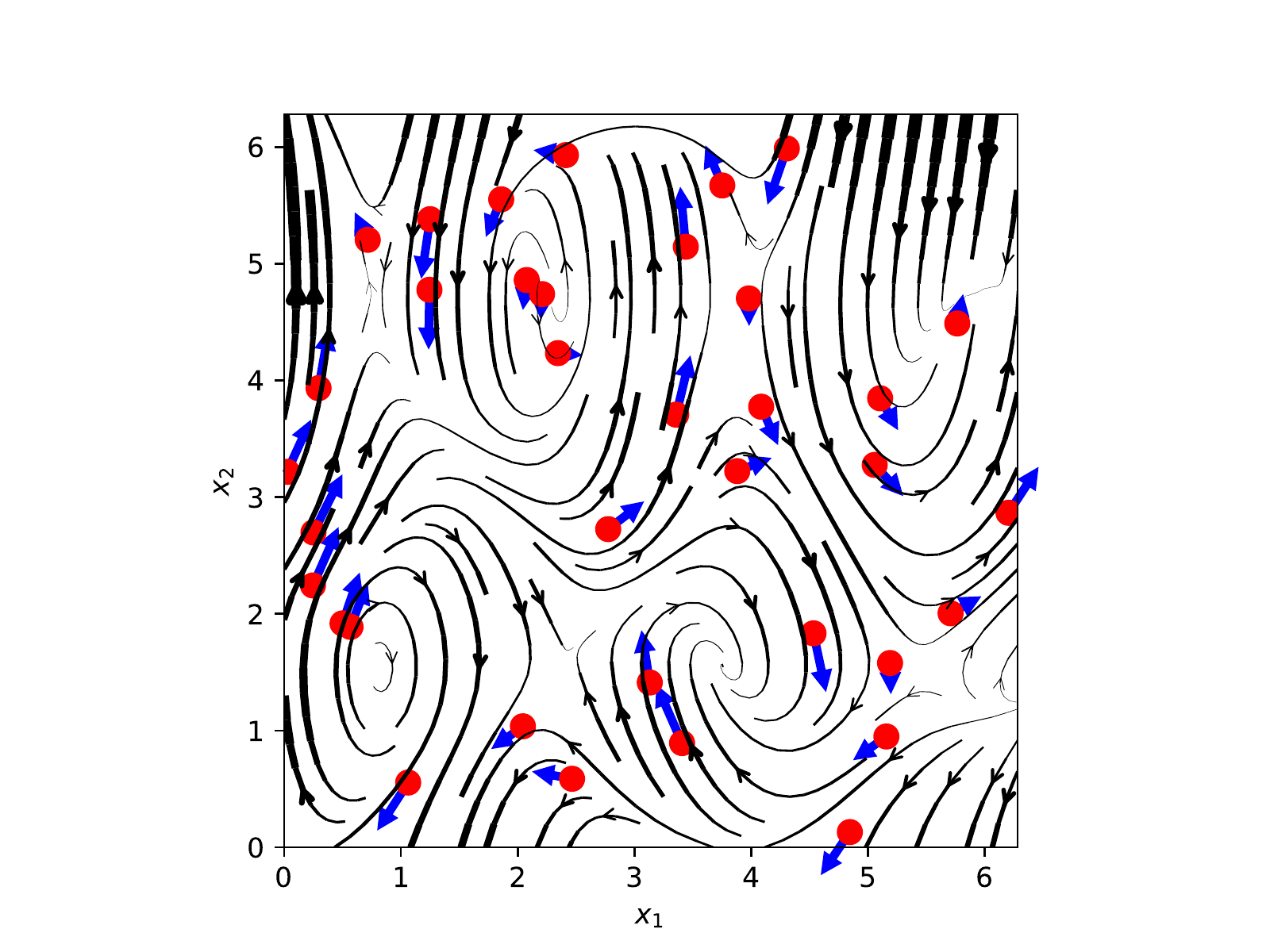}}
  \subfloat[$\#3$]{\label{fg:3.f}\includegraphics[clip,trim={2cm 0 2cm 0},width=0.2\textwidth]{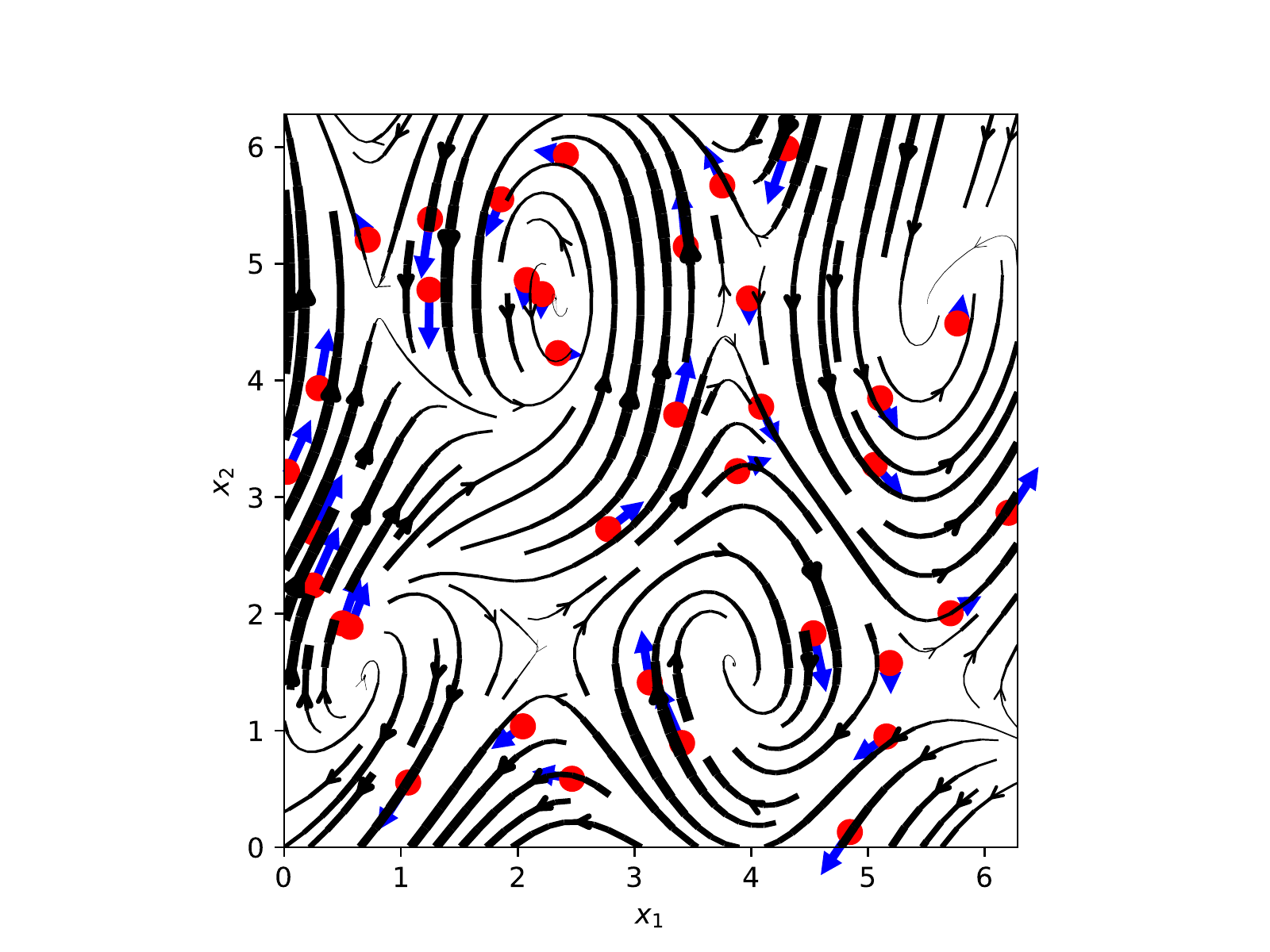}}
  \subfloat[$\#4$]{\label{fg:3.g}\includegraphics[clip,trim={2cm 0 2cm 0},width=0.2\textwidth]{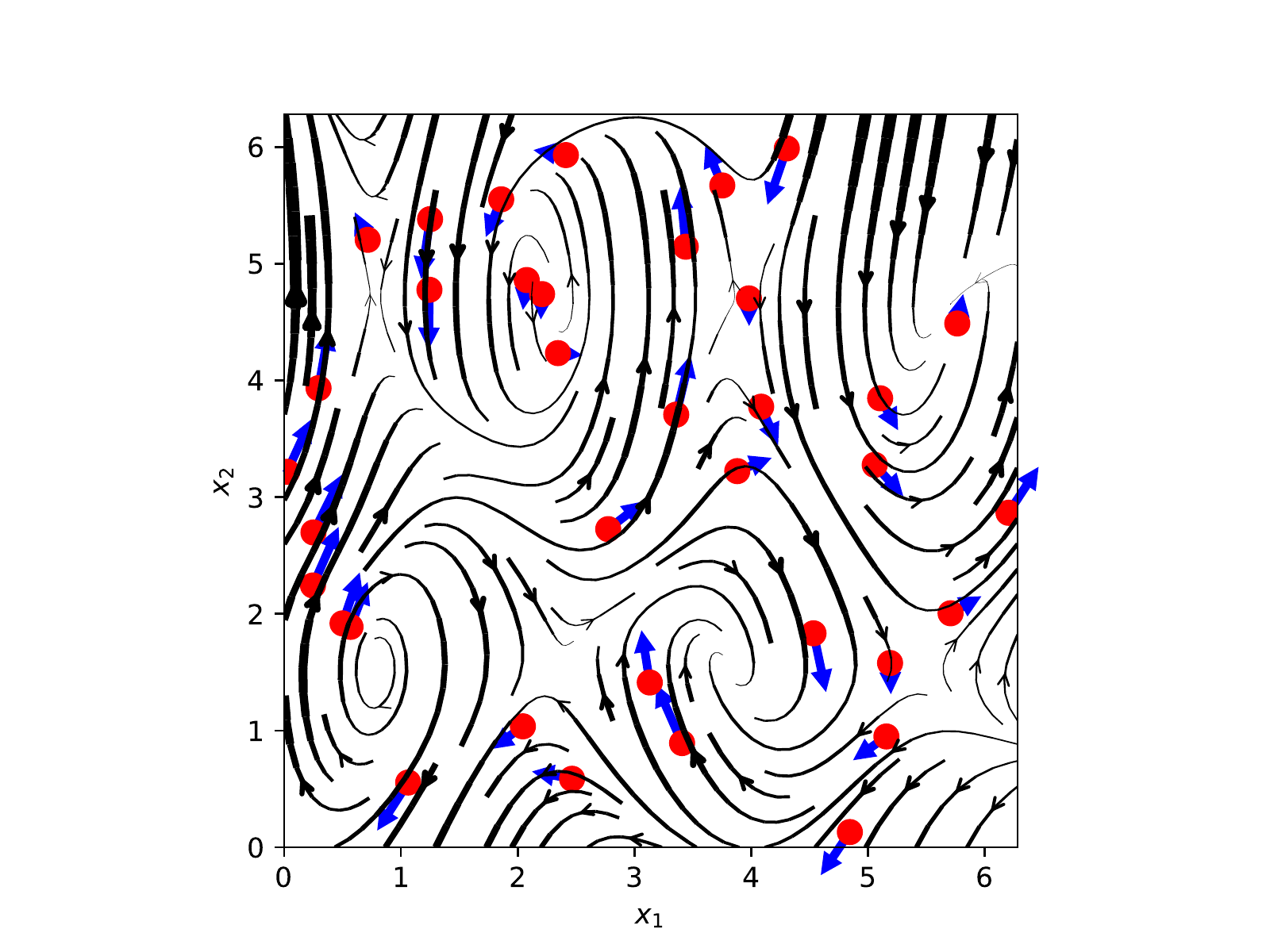}}
  \subfloat[$\#5$]{\label{fg:3.h}\includegraphics[clip,trim={2cm 0 2cm 0},width=0.2\textwidth]{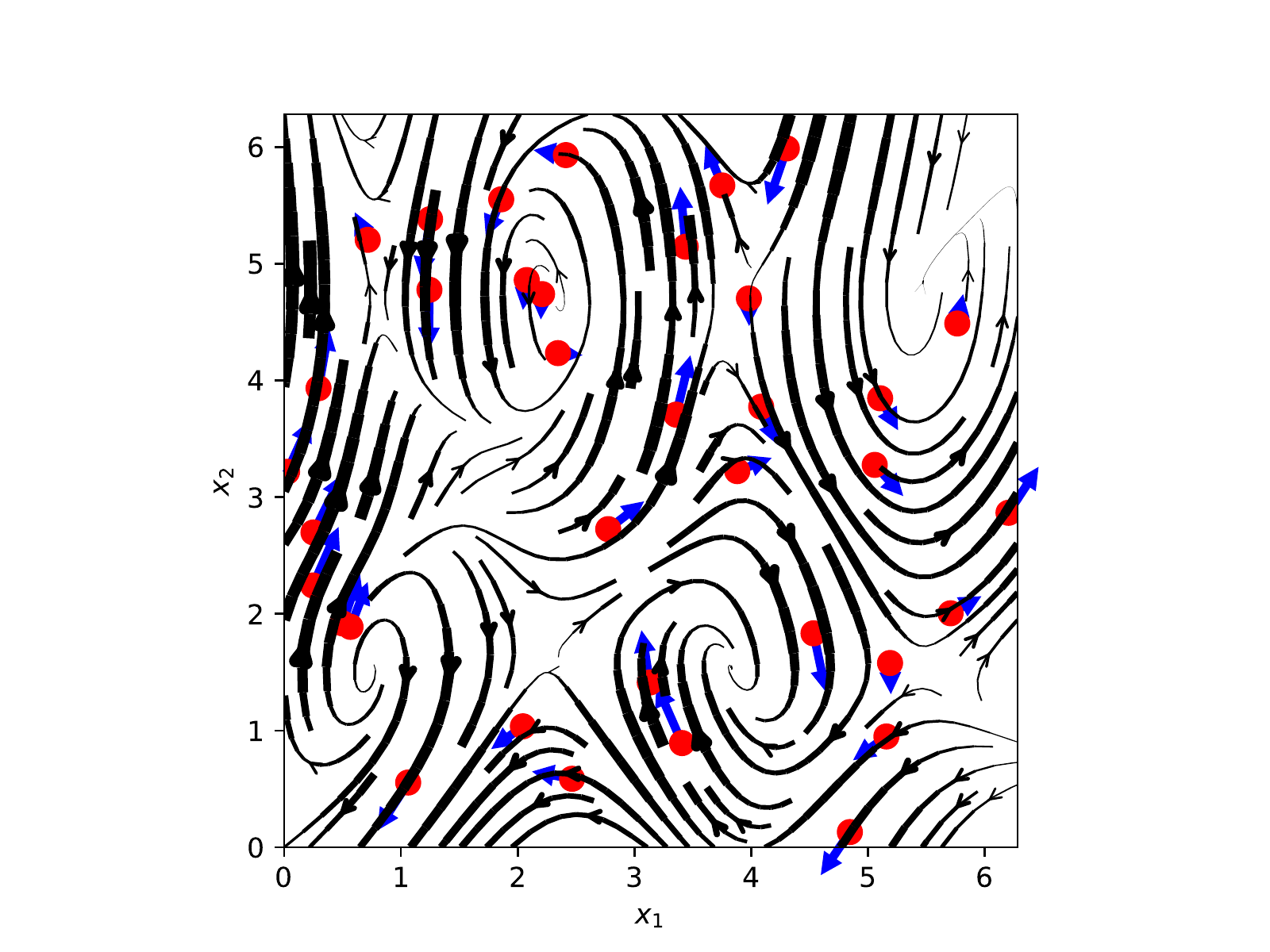}}\\
  \subfloat[$\#6$]{\label{fg:3.i}\includegraphics[clip,trim={2cm 0 2cm 0},width=0.2\textwidth]{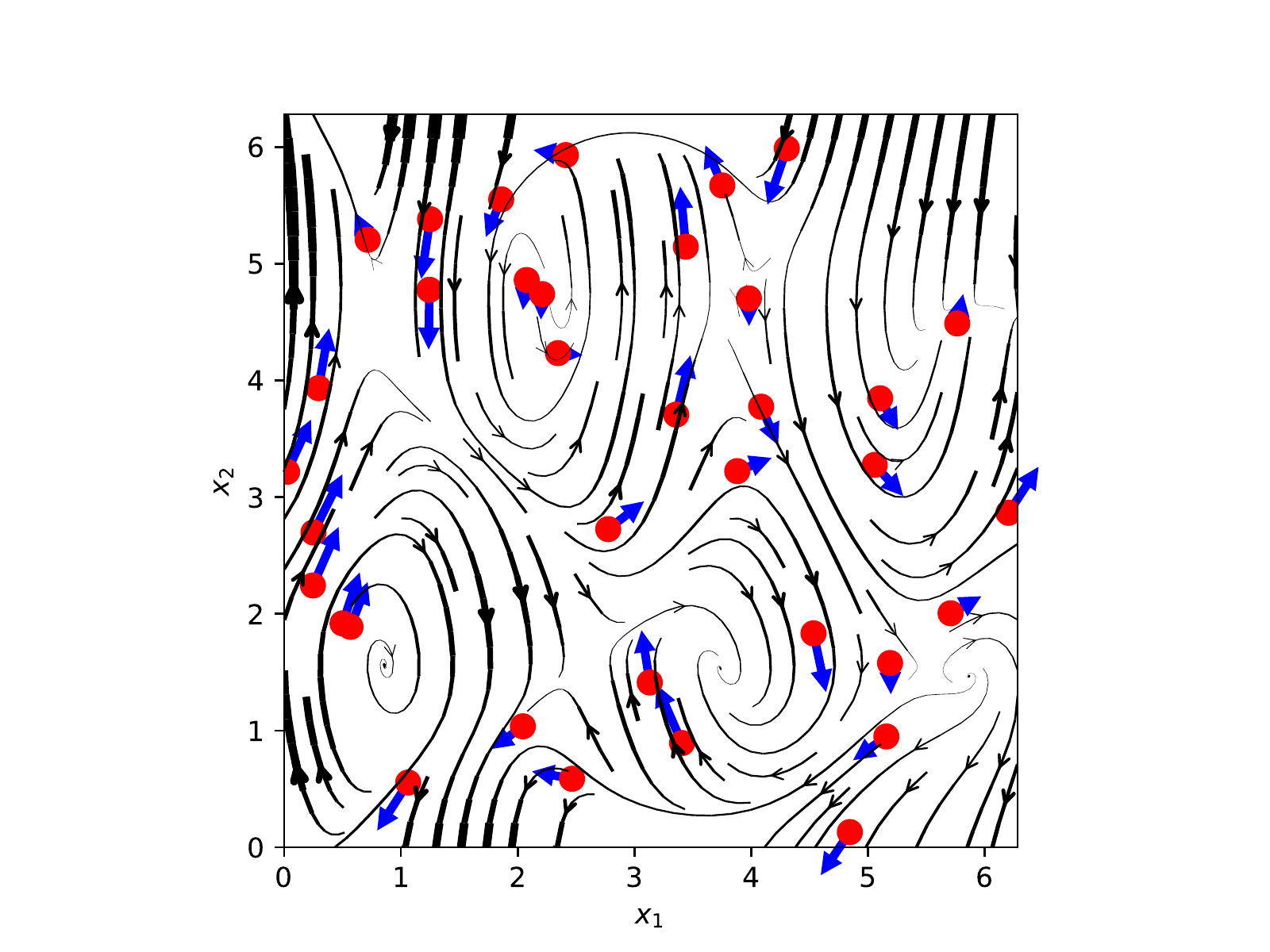}}
  \subfloat[$\#7$]{\label{fg:3.j}\includegraphics[clip,trim={2cm 0 2cm 0},width=0.2\textwidth]{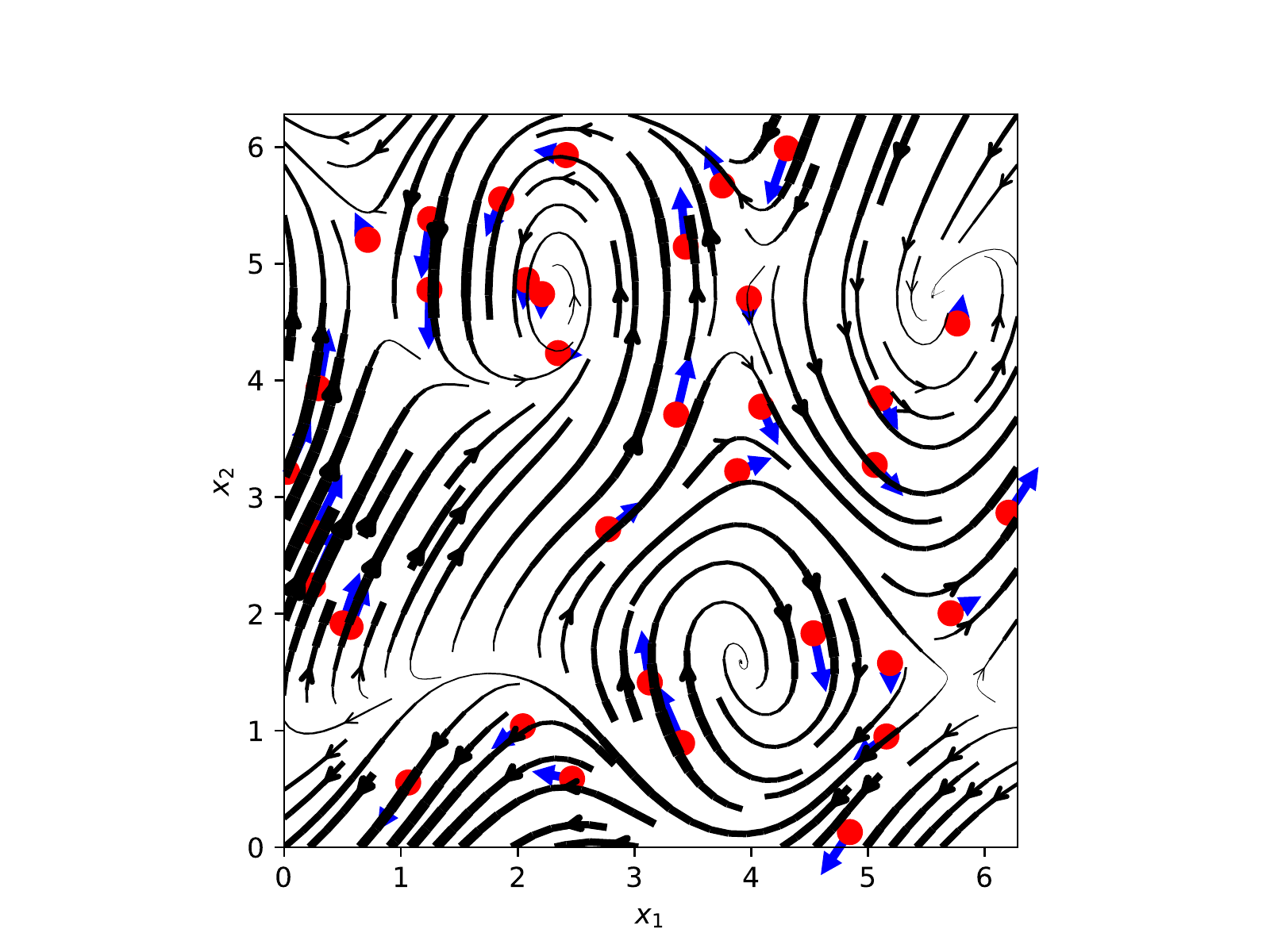}}
  \subfloat[$\#8$]{\label{fg:3.k}\includegraphics[clip,trim={2cm 0 2cm 0},width=0.2\textwidth]{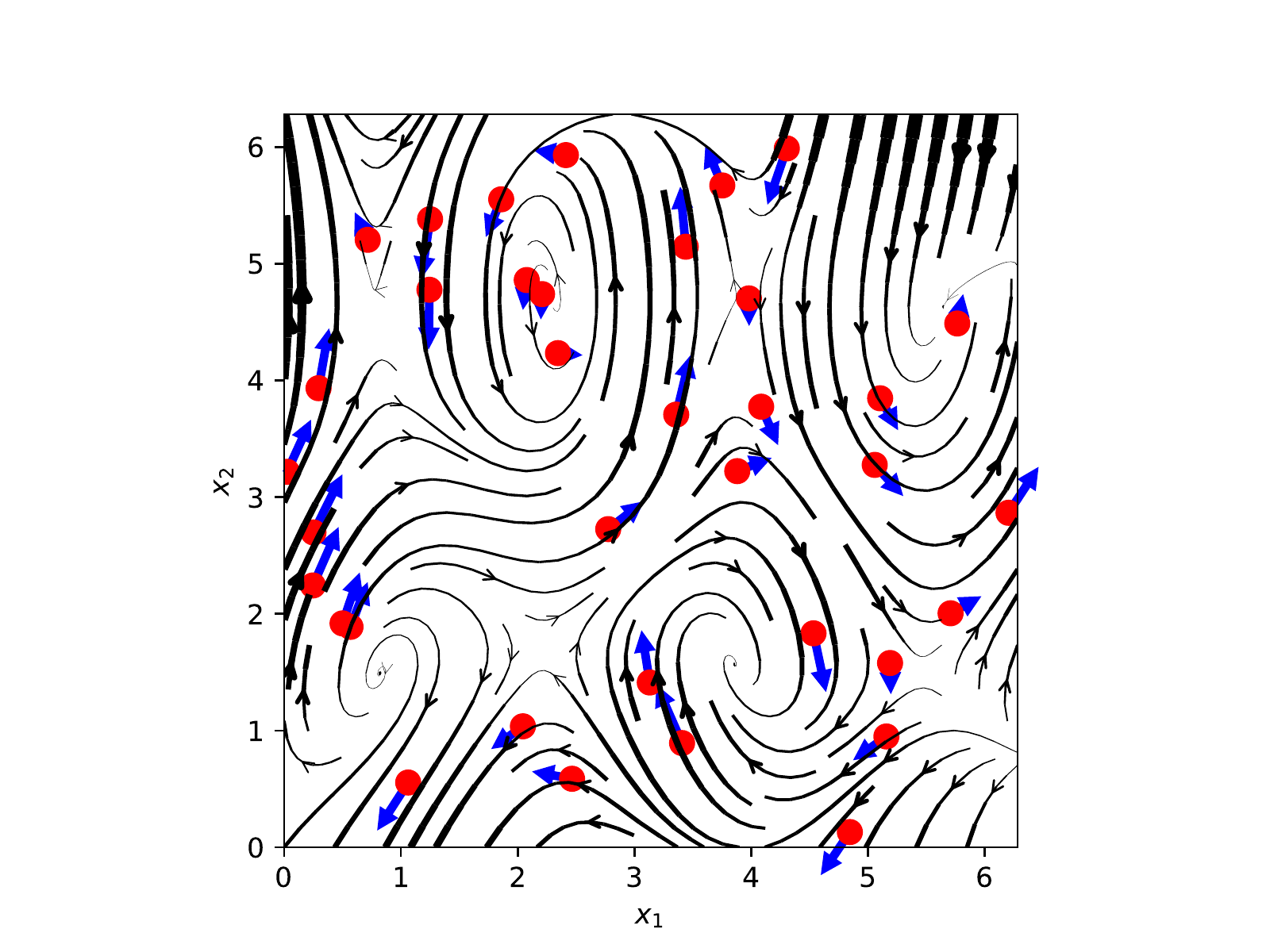}}
  \subfloat[$\#9$]{\label{fg:3.l}\includegraphics[clip,trim={2cm 0 2cm 0},width=0.2\textwidth]{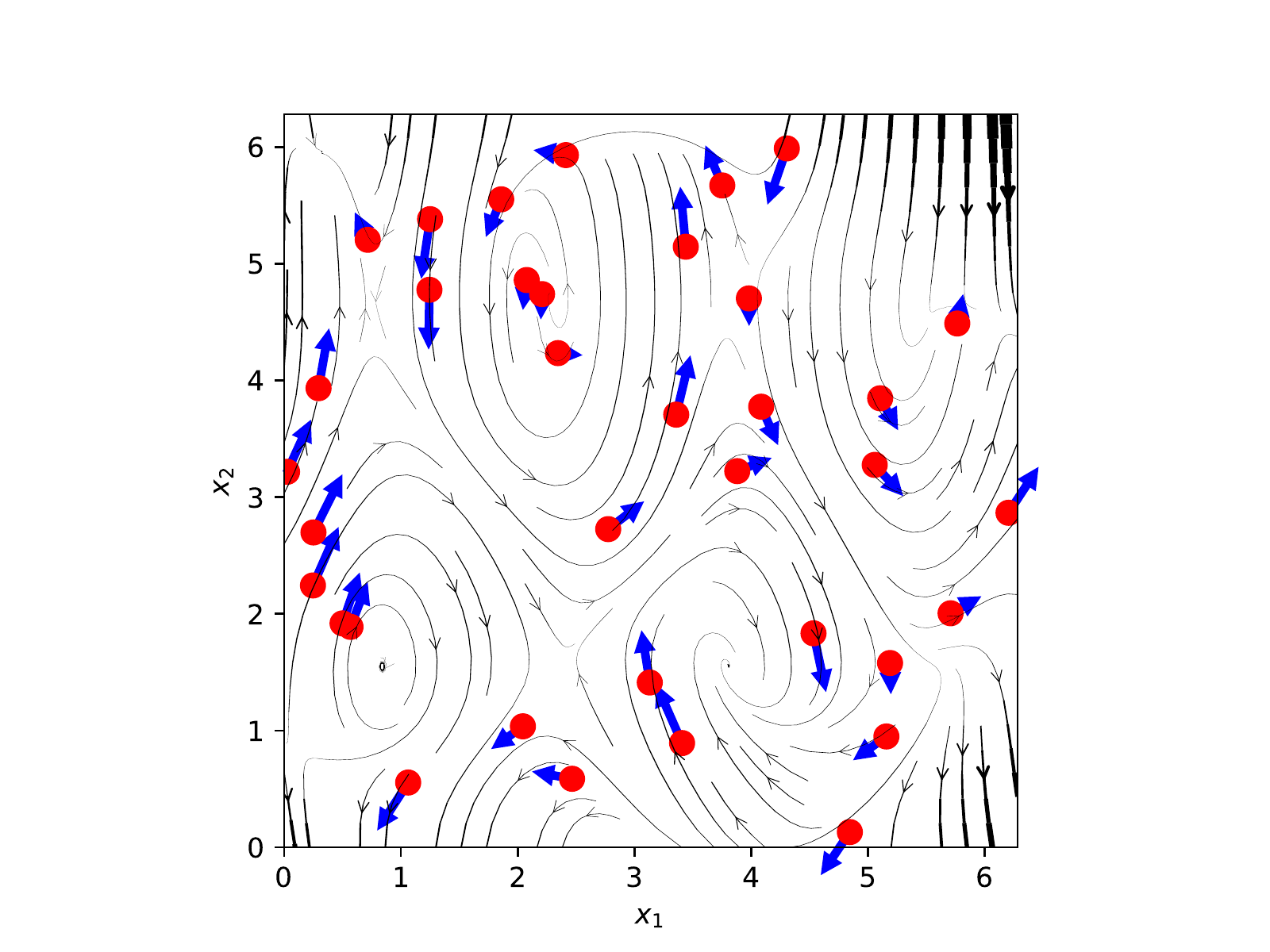}}
  \subfloat[$\#10$]{\label{fg:3.m}\includegraphics[clip,trim={2cm 0 2cm 0},width=0.2\textwidth]{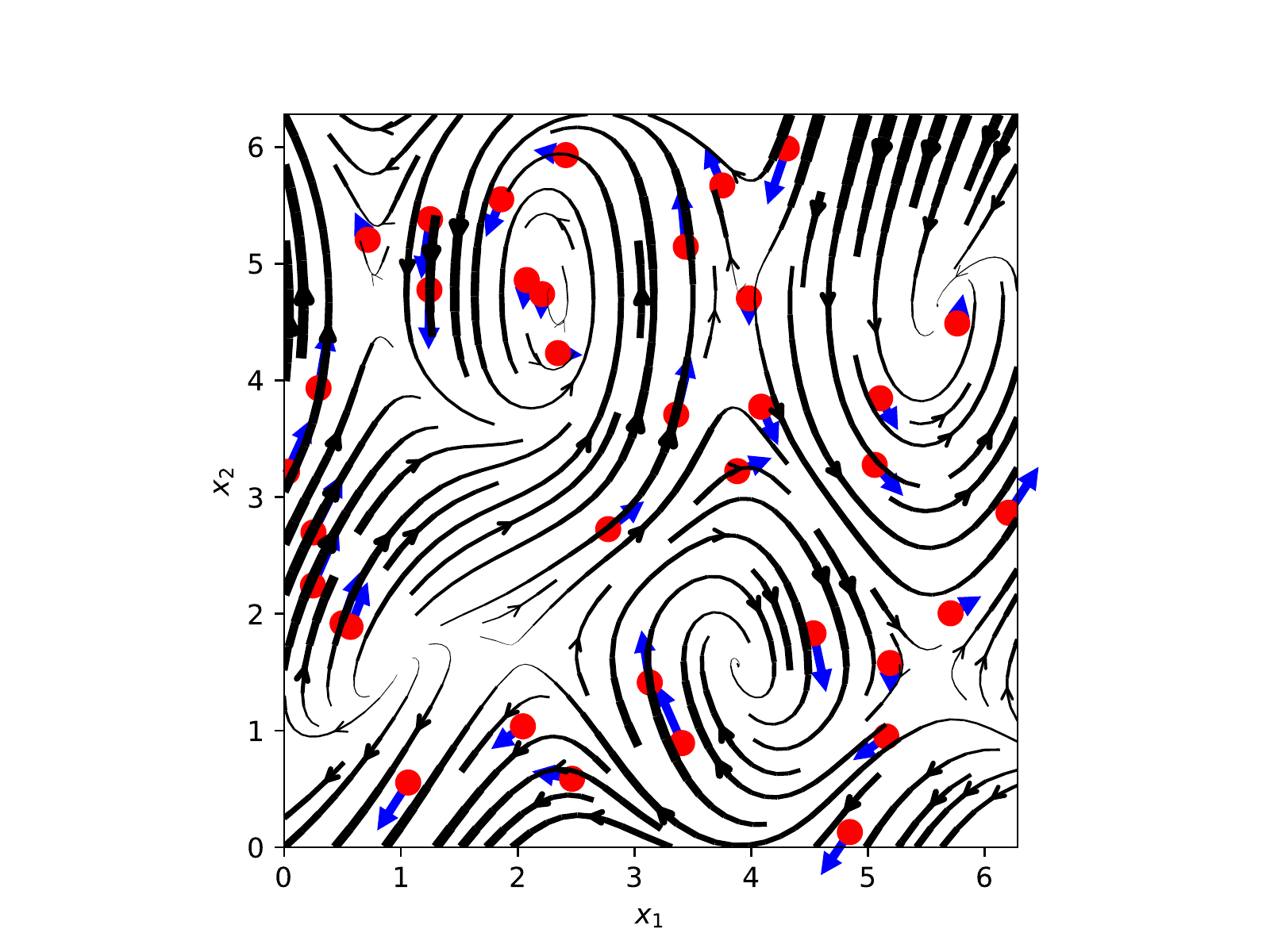}}
  \caption{Radial basis function approximations.}
\label{fg:3.rbf_1}
\end{figure}

\begin{figure}[!htb]
\centering
  \subfloat[$\#1$]{\label{fg:3.n}\includegraphics[clip,trim={2cm 0 2cm 0},width=0.2\textwidth]{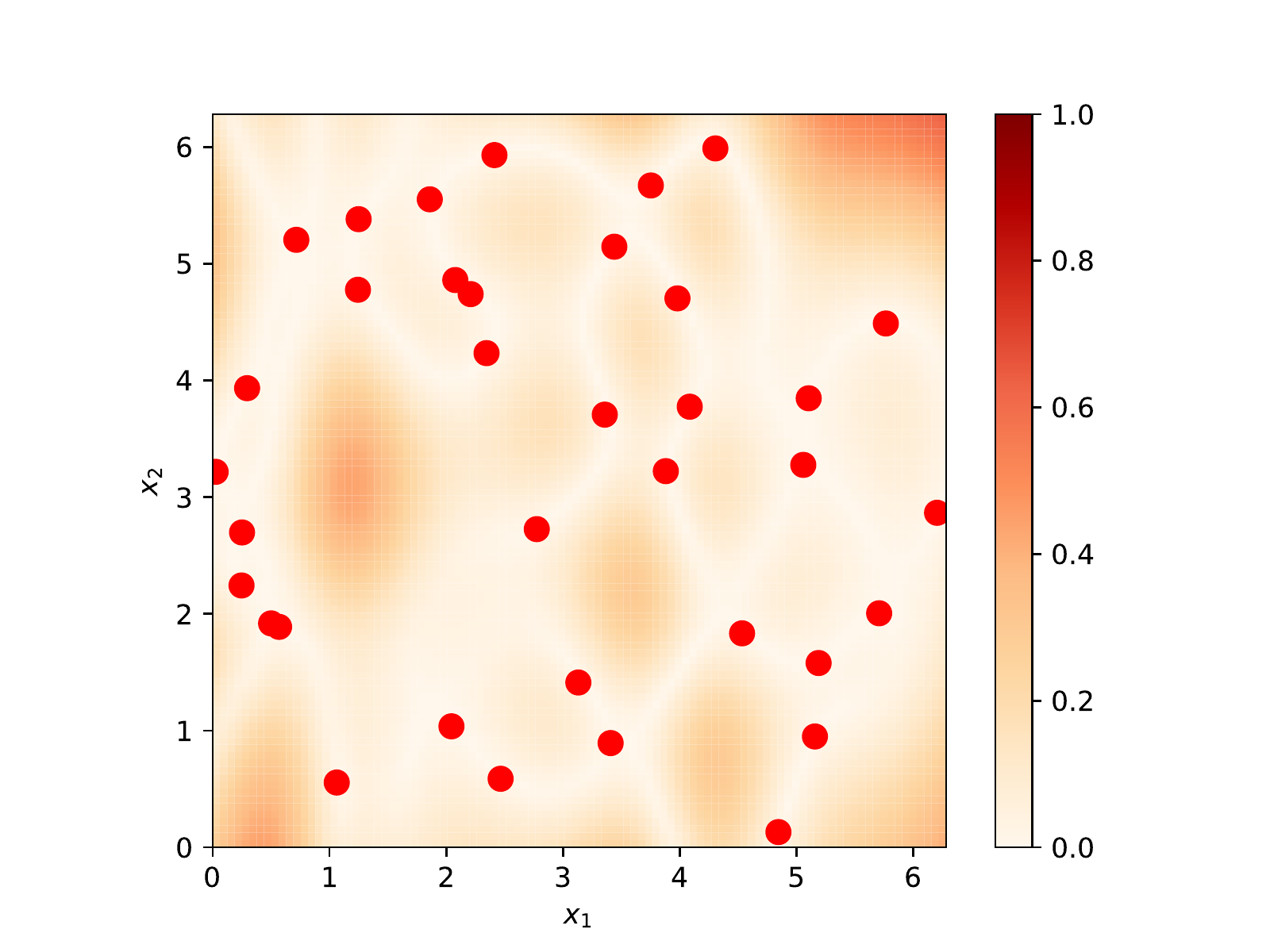}}
  \subfloat[$\#2$]{\label{fg:3.o}\includegraphics[clip,trim={2cm 0 2cm 0},width=0.2\textwidth]{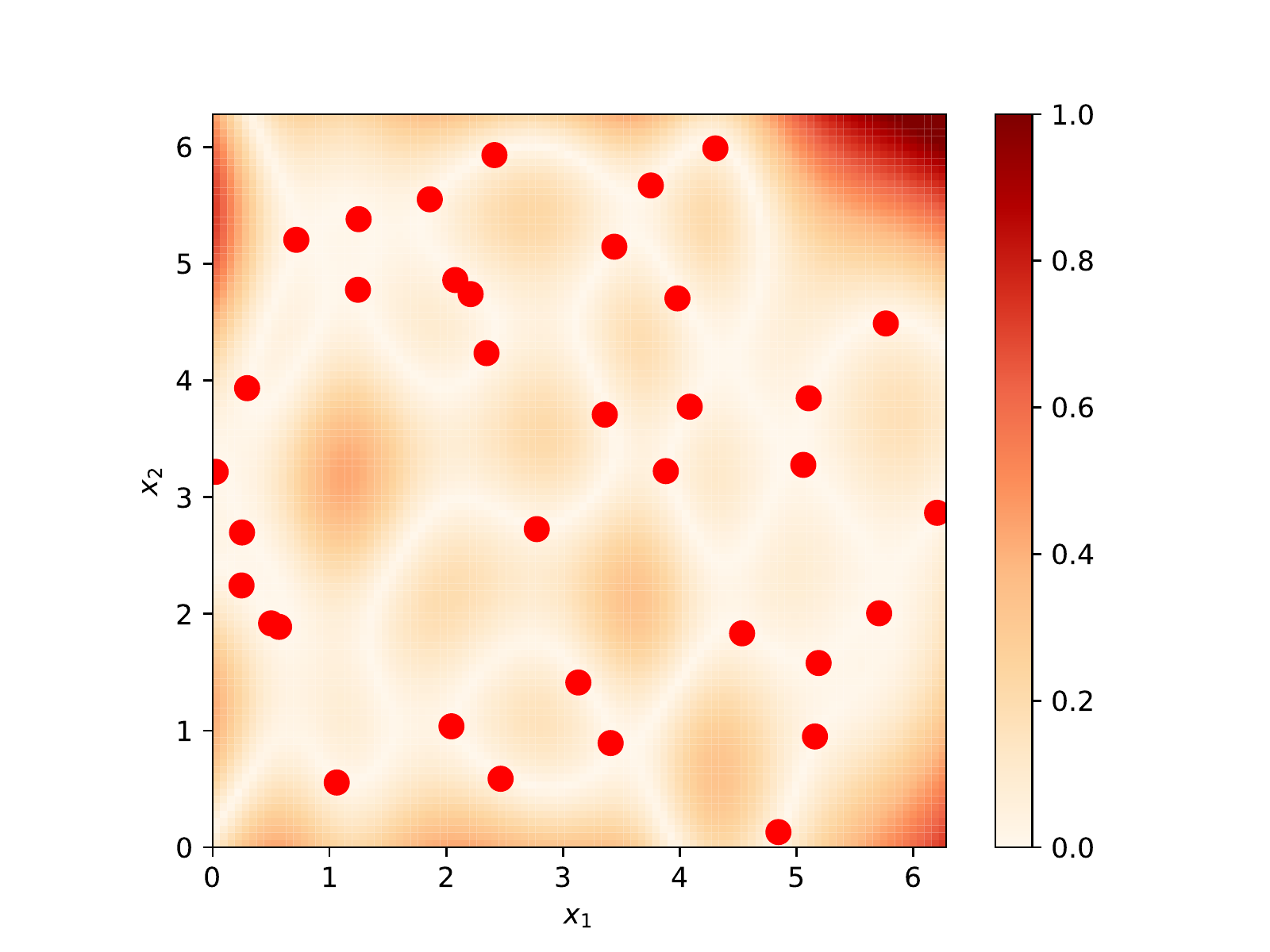}}
  \subfloat[$\#3$]{\label{fg:3.p}\includegraphics[clip,trim={2cm 0 2cm 0},width=0.2\textwidth]{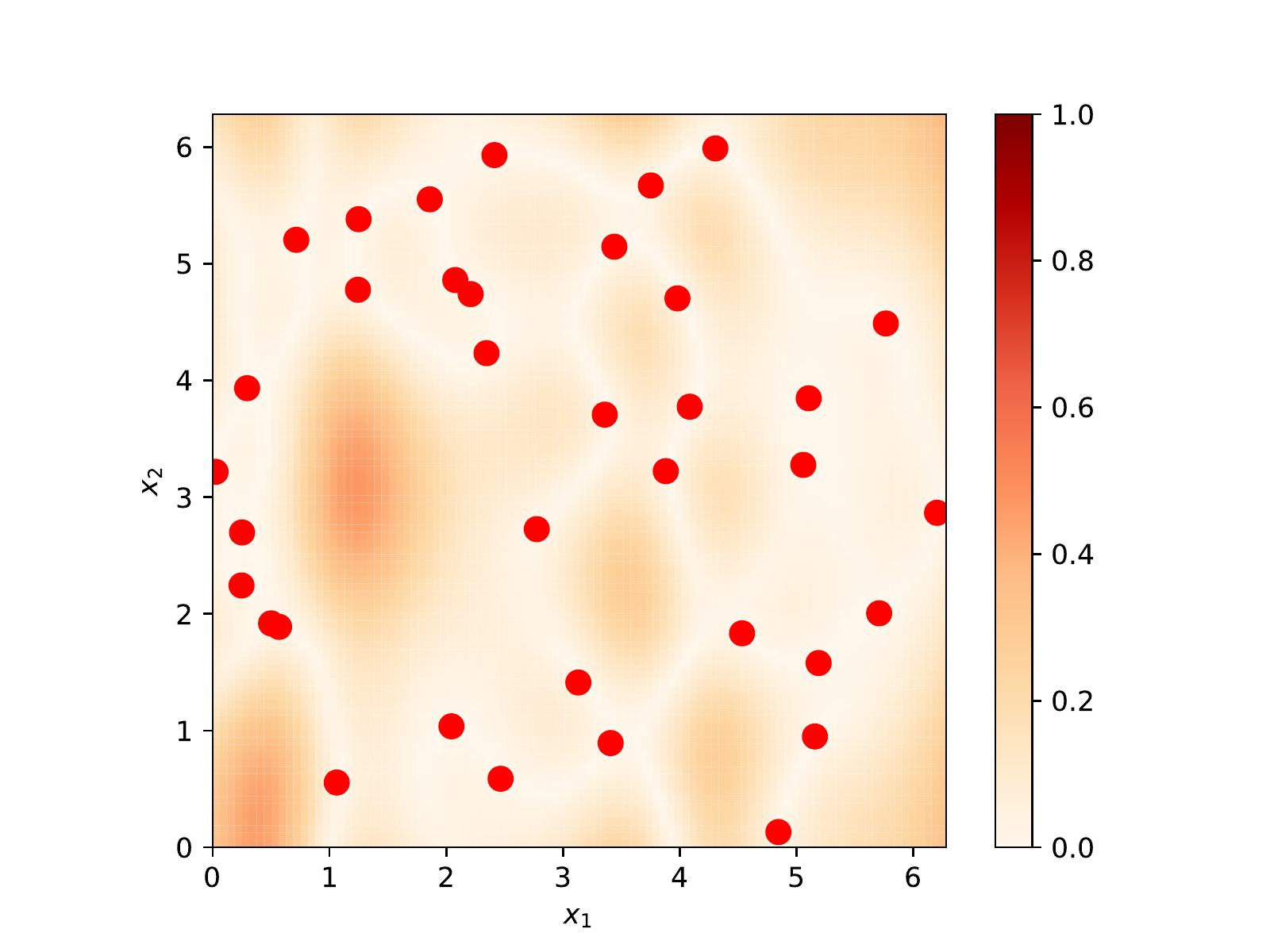}}
  \subfloat[$\#4$]{\label{fg:3.q}\includegraphics[clip,trim={2cm 0 2cm 0},width=0.2\textwidth]{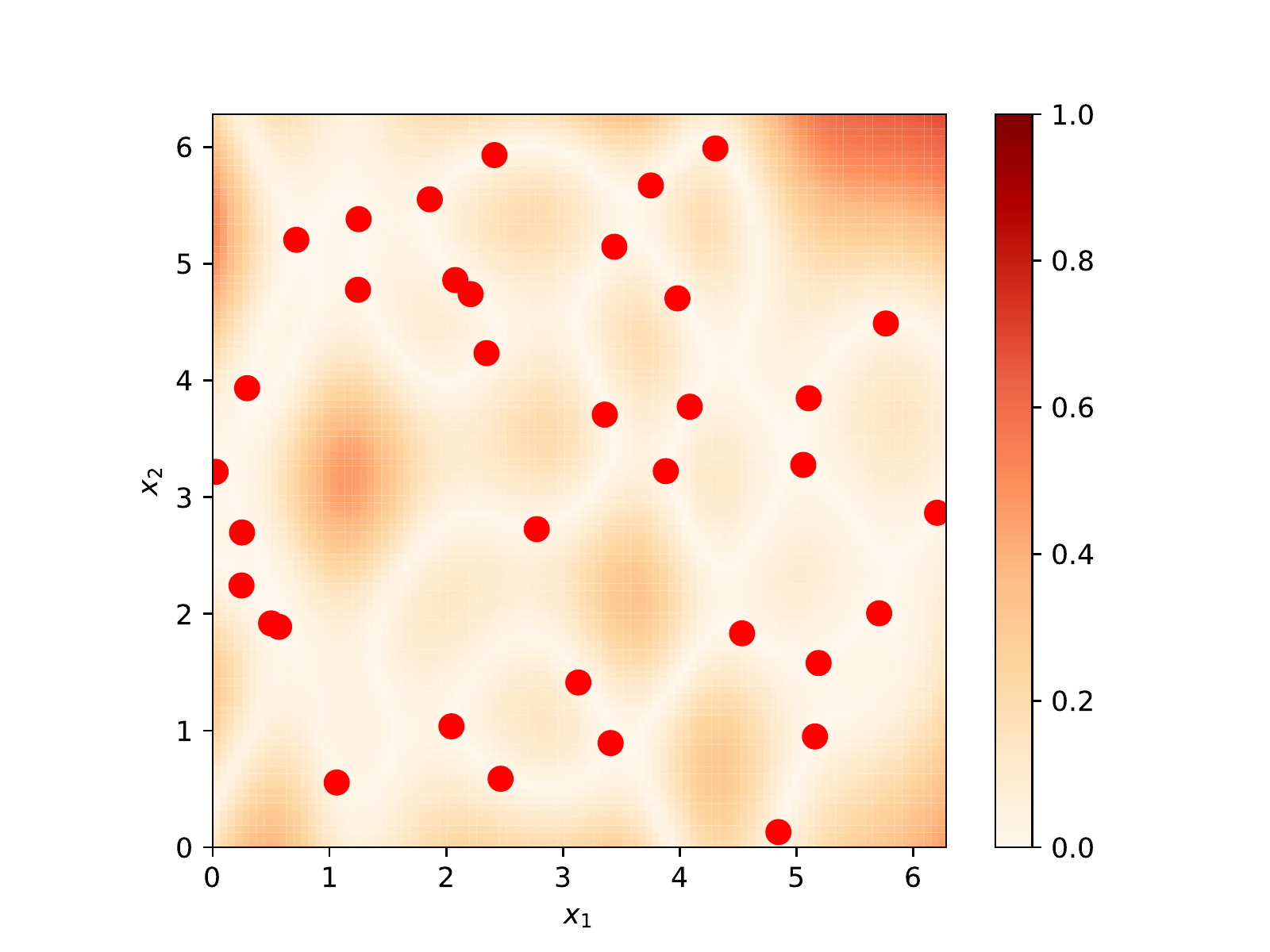}}
  \subfloat[$\#5$]{\label{fg:3.r}\includegraphics[clip,trim={2cm 0 2cm 0},width=0.2\textwidth]{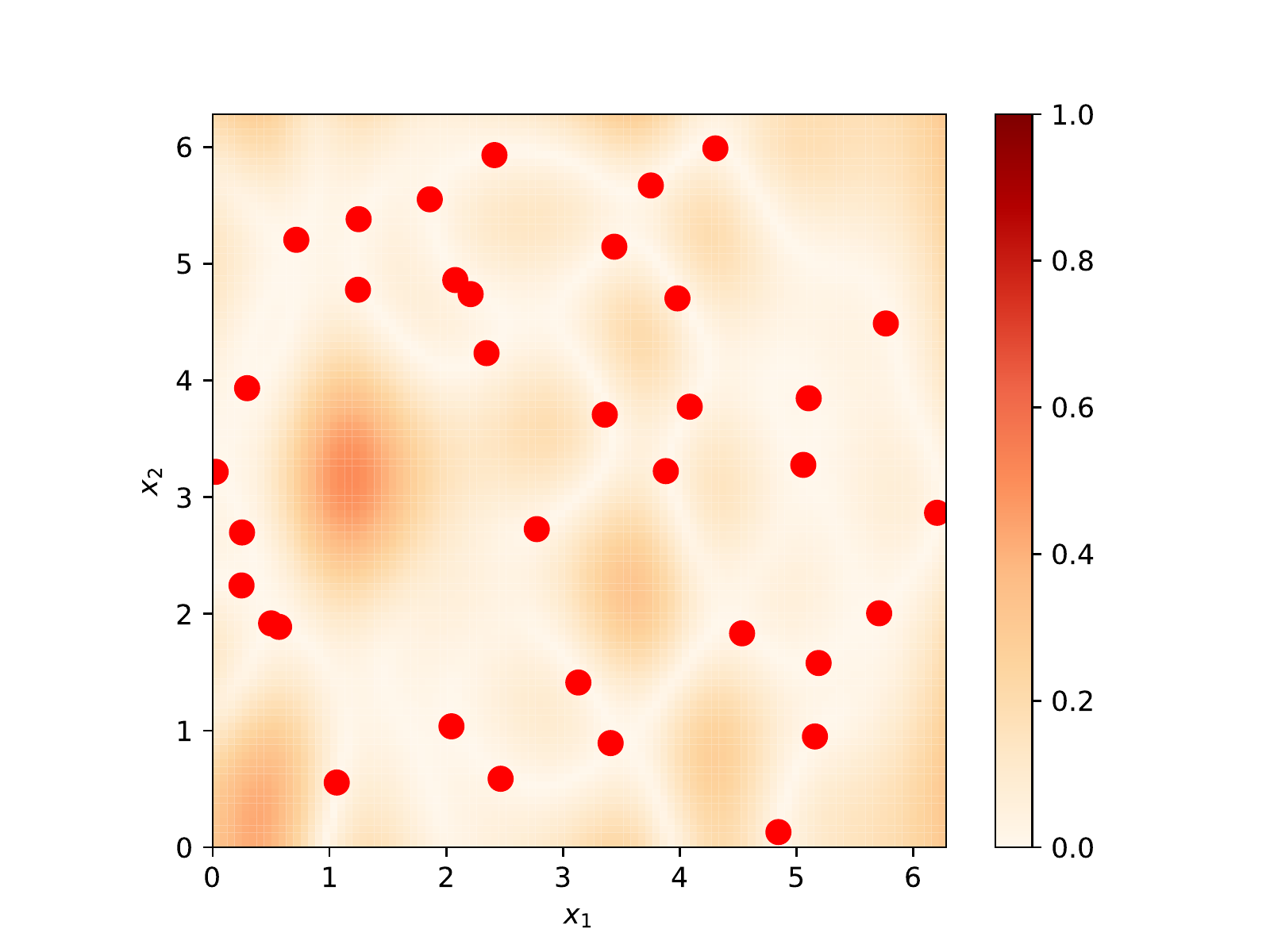}}\\
  \subfloat[$\#6$]{\label{fg:3.s}\includegraphics[clip,trim={2cm 0 2cm 0},width=0.2\textwidth]{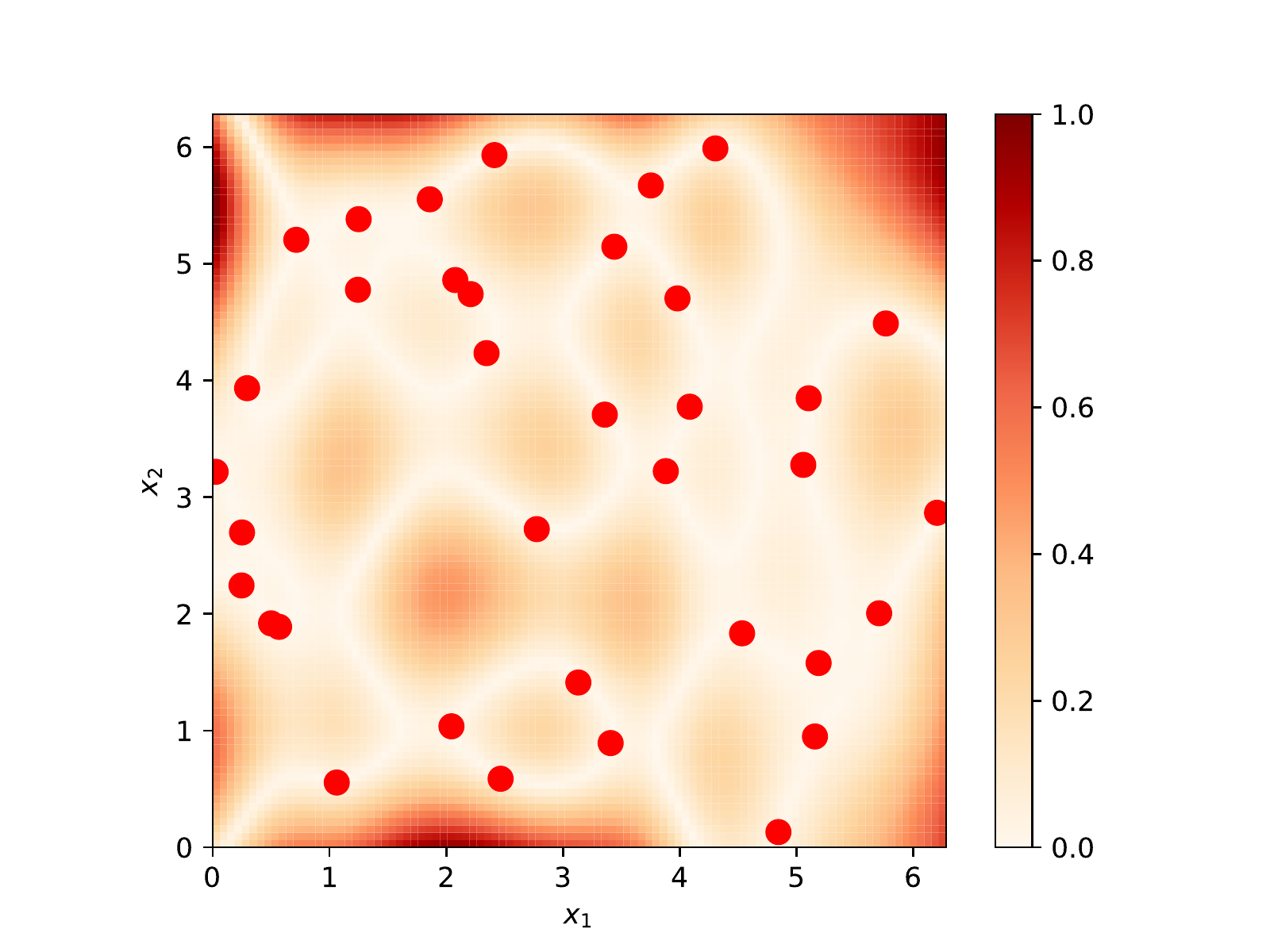}}
  \subfloat[$\#7$]{\label{fg:3.t}\includegraphics[clip,trim={2cm 0 2cm 0},width=0.2\textwidth]{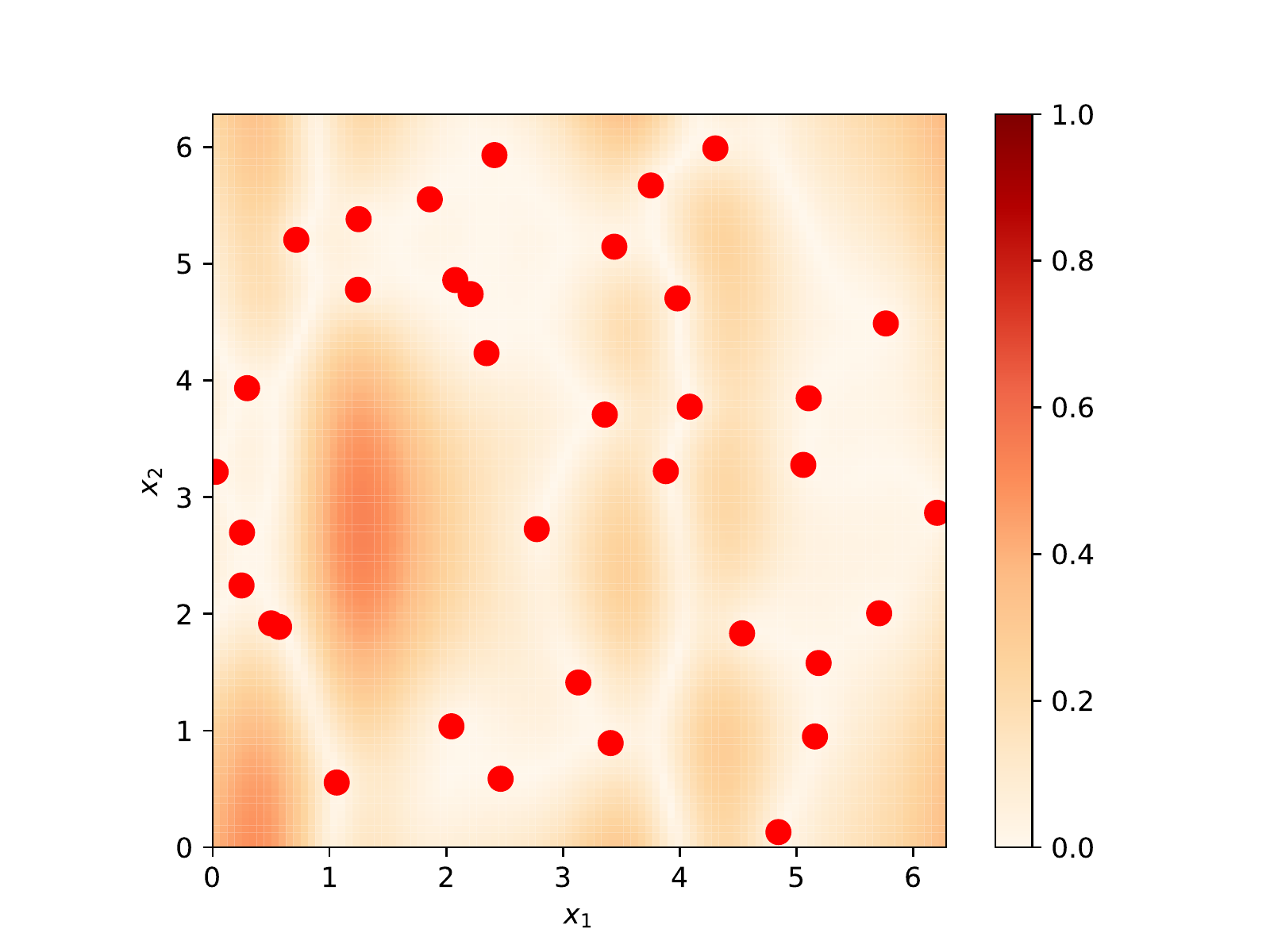}}
  \subfloat[$\#8$]{\label{fg:3.u}\includegraphics[clip,trim={2cm 0 2cm 0},width=0.2\textwidth]{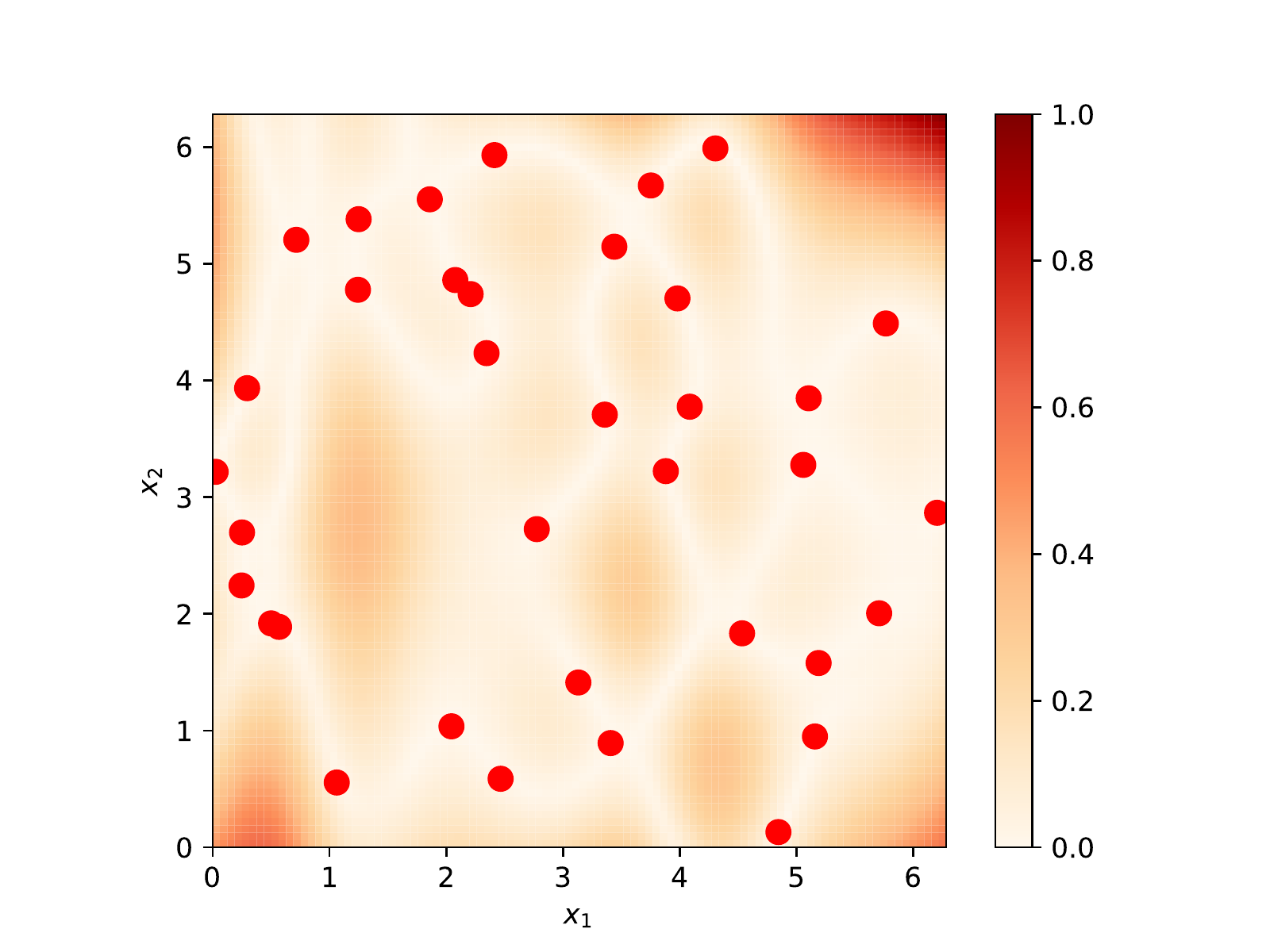}}
  \subfloat[$\#9$]{\label{fg:3.v}\includegraphics[clip,trim={2cm 0 2cm 0},width=0.2\textwidth]{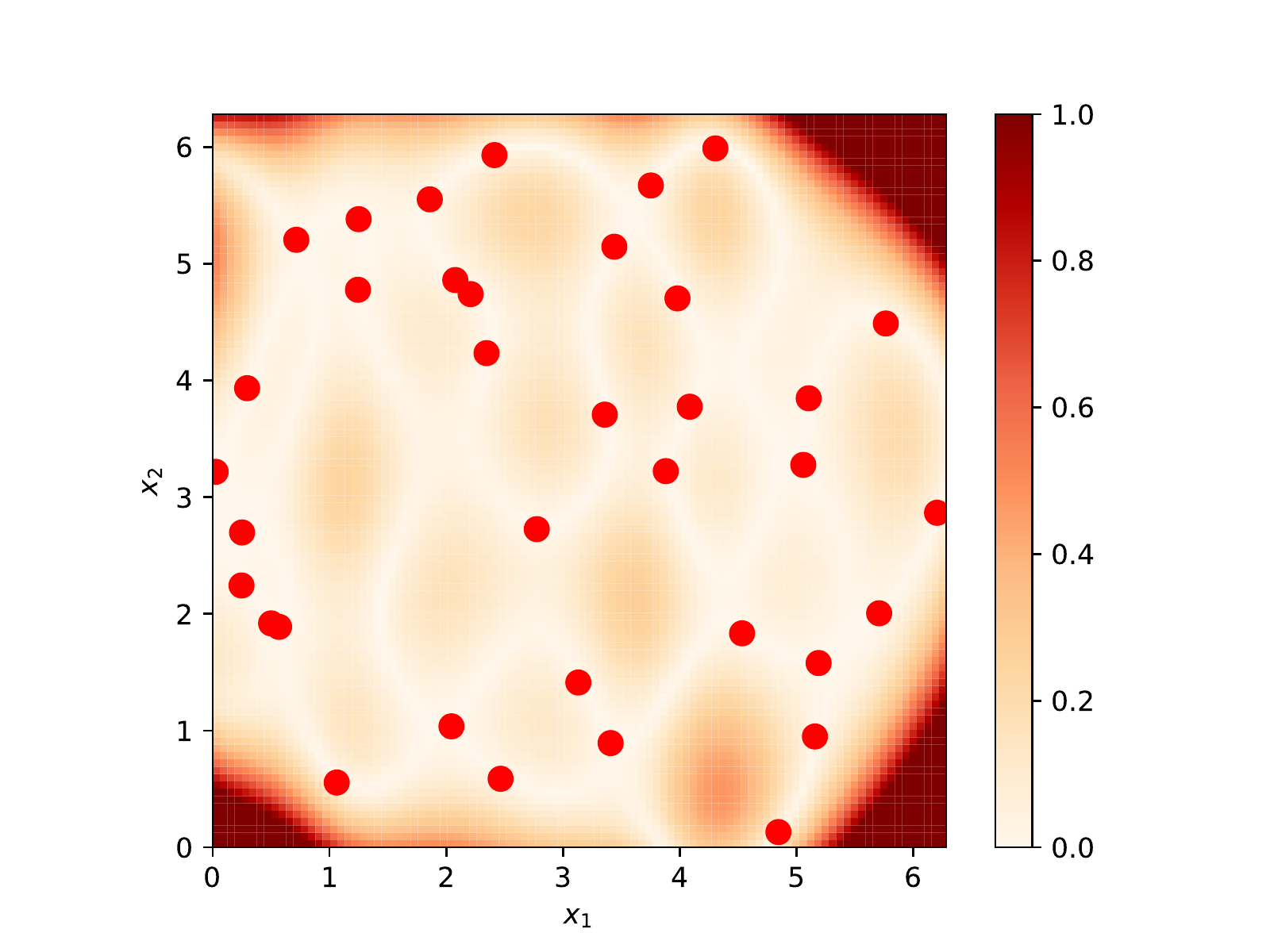}}
  \subfloat[$\#10$]{\label{fg:3.x}\includegraphics[clip,trim={2cm 0 2cm 0},width=0.2\textwidth]{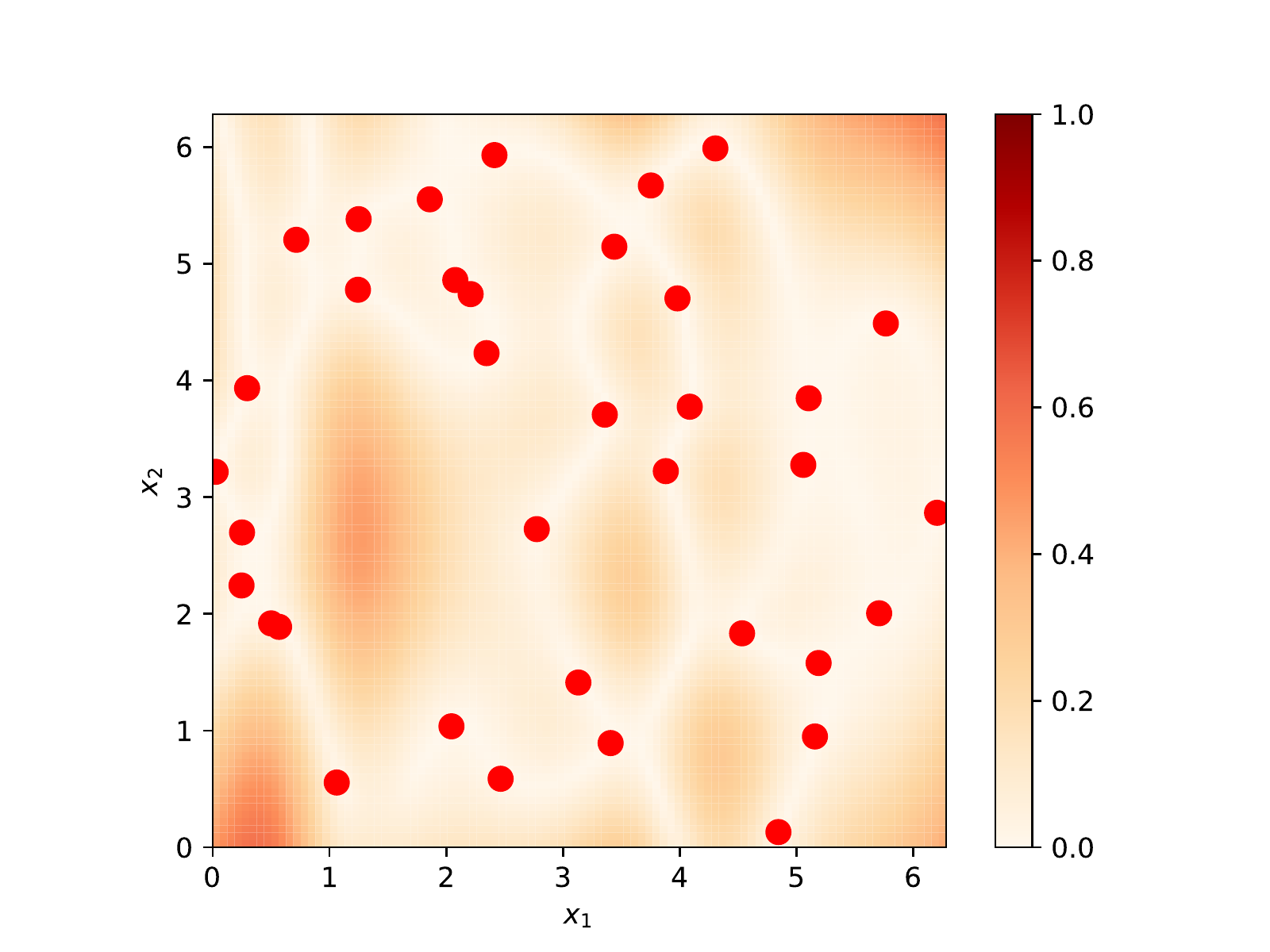}}
  \caption{Error fields ot the radial basis function approximations in Figure \ref{fg:3.rbf_1}.}
\label{fg:3.rbf_1_err}
\end{figure}

Figure \ref{fg:3.rbf_1} presents the approximation fields of the ten radial basis functions and Figure \ref{fg:3.rbf_1_err} depicts their corresponding error fields. The $L^\infty(e(\bs{x}))$ and $E\coloneqq{L}^2(e(\bs{x}))$ errors of the \SFdf\ approximation are respectively $7$ and $6$ times smaller than the errors obtained with the best approximation using radial basis functions. Although the error differences between the \SFdf\ and the radial basis function approximations are much smaller than in the previous examples, the reconstructed fields using radial basis functions present some unphysical features, see Figure \ref{fg:3.rbf_1}. This issue does not occur in the \SFdf\ approximation as may be observed in Figure \ref{fg:3.fourier}.

\begin{table}
\caption{$L^\infty(e(\bs{x}))$ approximation errors \eqref{eq:intensive.error}}
\begin{tabular}{ c c c c c c c c c c c }
 \SFdf & $\#1$ & $\#2$ & $\#3$ & $\#4$ & $\#5$ & $\#6$ & $\#7$ & $\#8$ & $\#9$ & $\#10$\\
 \hline\\[-10pt]
 2.47 & 23.9 & 45.4 & 17.9 & 29.5 & 17.0 & 59.6 & 19.0 & 31.2 & 110.4 & 22.4
\end{tabular}
\end{table}

\begin{table}
\caption{$E\coloneqq{L}^2(e(\bs{x}))$ approximation errors \eqref{eq:extensive.error}}
\begin{tabular}{ c c c c c c c c c c c }
 \SFdf & $\#1$ & $\#2$ & $\#3$ & $\#4$ & $\#5$ & $\#6$ & $\#7$ & $\#8$ & $\#9$ & $\#10$\\
 \hline\\[-10pt]
 0.12 & 0.74 & 1.14 & 0.71 & 0.84 & 0.68 & 1.37 & 0.92 & 0.90 & 3.0 & 0.79
\end{tabular}
\end{table}

\FloatBarrier

\subsection{Immersed boundary: flow around cylinder}

In this example, we aim to reconstruct the velocity field around an immersed cylinder described by $100$ points. Here, we assumed a single velocity measurement; that is,
\begin{equation}
\bs{x}_i = (2\pi,0),\qquad\bs{u}_1(\bs{x}_1) = (1,0),
\end{equation}
in a domain $\cl{D}=[0,2\pi]^2$. The parameters $\varepsilon$ and $k$ were set to $10^{-5}$ and $1.5$, respectively. Additionally, on $x_2=0$ and $x_2=2\pi$ immersed boundaries are used to define free-slip walls with $100$ points. For all immersed boundaries, we set $\bar{\lambda}_B=1$. This simple example aims to depict the capabilities of our method in including immersed boundaries.

\begin{figure}[!htb]
\centering
  \includegraphics[width=0.45\textwidth]{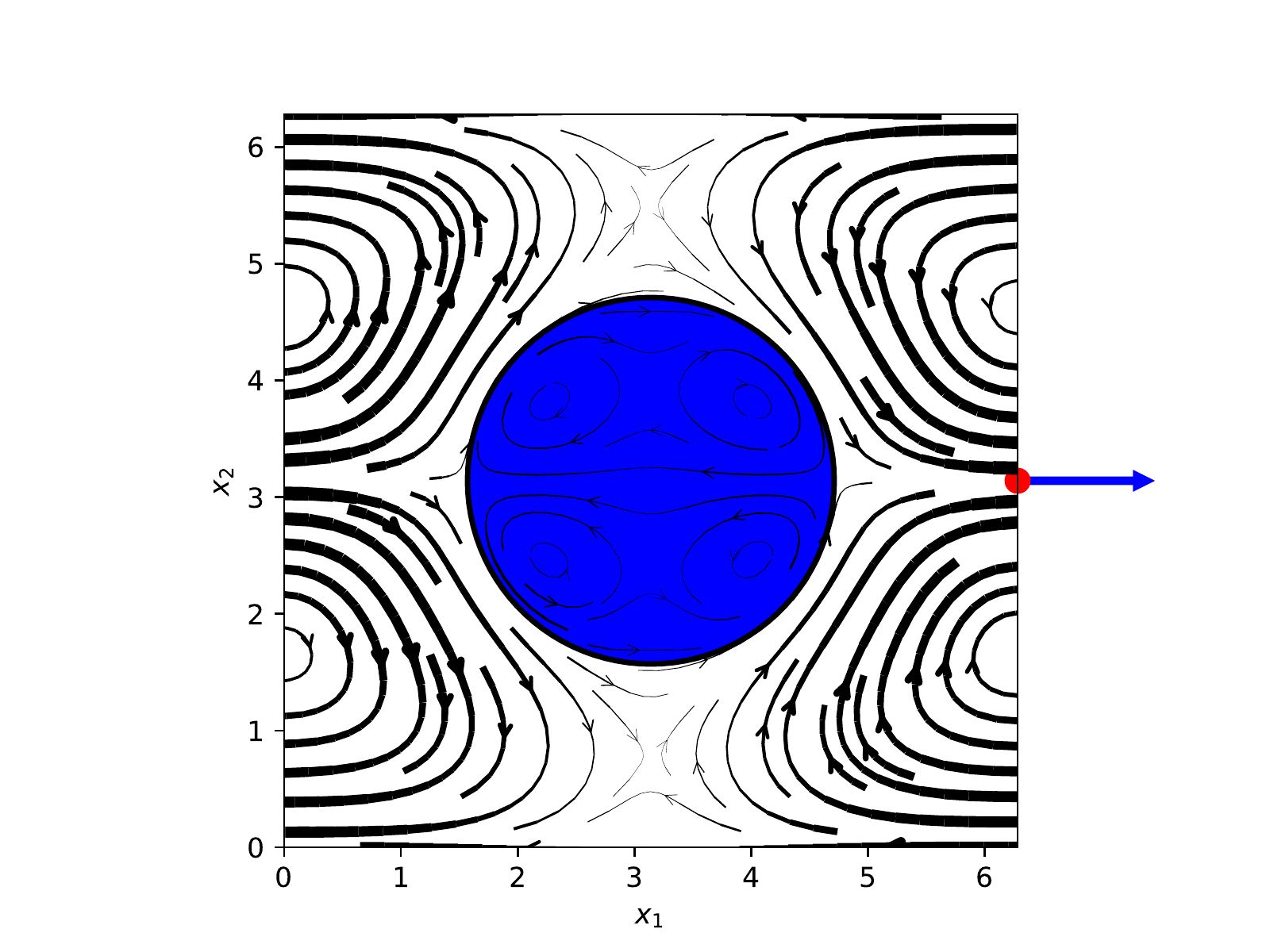}
  \caption{Flow around cylinder with immersed boundary}
\label{fg:4.fourier}
\end{figure}

We set the boundary energy $\varepsilon_{\partial\cl{I}}$ and the stopping criterion $\Delta\varepsilon_{\partial\cl{I}}$ to $80\%$ and $10^{-5}$, respectively. For the fractional Sobolev regularization, we selected $\epsilon=10^{-5}$ and $k=1.5$. After $15$ outer iterations, we obtained the following index set with $73$ entries
\begin{equation*}
\begin{smallmatrix}
  \rbull & \bbull & \bbull & \bbull & \bbull & \bbull & \bbull & \bbull & \rbull\\
  \rbull & \bbull & \bbull & \bbull & \bbull & \bbull & \bbull & \bbull & \rbull\\
  \bbull & \bbull & \bbull & \bbull & \bbull & \bbull & \bbull & \bbull & \bbull\\
  \bbull & \bbull & \bbull & \bbull & \bbull & \bbull & \bbull & \bbull & \bbull\\
  \bbull & \bbull & \bbull & \bbull & \bbull & \bbull & \bbull & \bbull & \bbull\\
  \bbull & \bbull & \bbull & \bbull & \bbull & \bbull & \bbull & \bbull & \bbull\\
  \bbull & \bbull & \bbull & \bbull & \bbull & \bbull & \bbull & \bbull & \bbull\\
  \rbull & \bbull & \bbull & \bbull & \bbull & \bbull & \bbull & \bbull & \rbull\\
  \rbull & \bbull & \bbull & \bbull & \bbull & \bbull & \bbull & \bbull & \rbull\\
\end{smallmatrix}
\end{equation*}
The solution field is given in Figure \ref{fg:4.fourier}.
As expected, the solution field is tangential to the immersed boundaries.

\FloatBarrier

\subsection{Three-dimensional Taylor--Green vortex}

\FloatBarrier

Our computational framework is built for an arbitrary dimension $n$. Thus, to depict this capability, we reconstruct the following three-dimensional Taylor--Green vortex
\begin{equation}
\bs{\upsilon}(\bs{x}) = (\fr{1}{2}\cos(x_1)\sin(x_2)\sin(x_3),\fr{1}{2}\sin(x_1)\cos(x_2)\sin(x_3),-\sin(x_1)\sin(x_2)\cos(x_3)).
\end{equation}
Here, we aim to recover this divergence-free field using $64$ velocity measurements at random points in the domain $\cl{D}=[0,2\pi]^3$.

\begin{figure}[!htb]
\centering
  \includegraphics[width=0.6\textwidth]{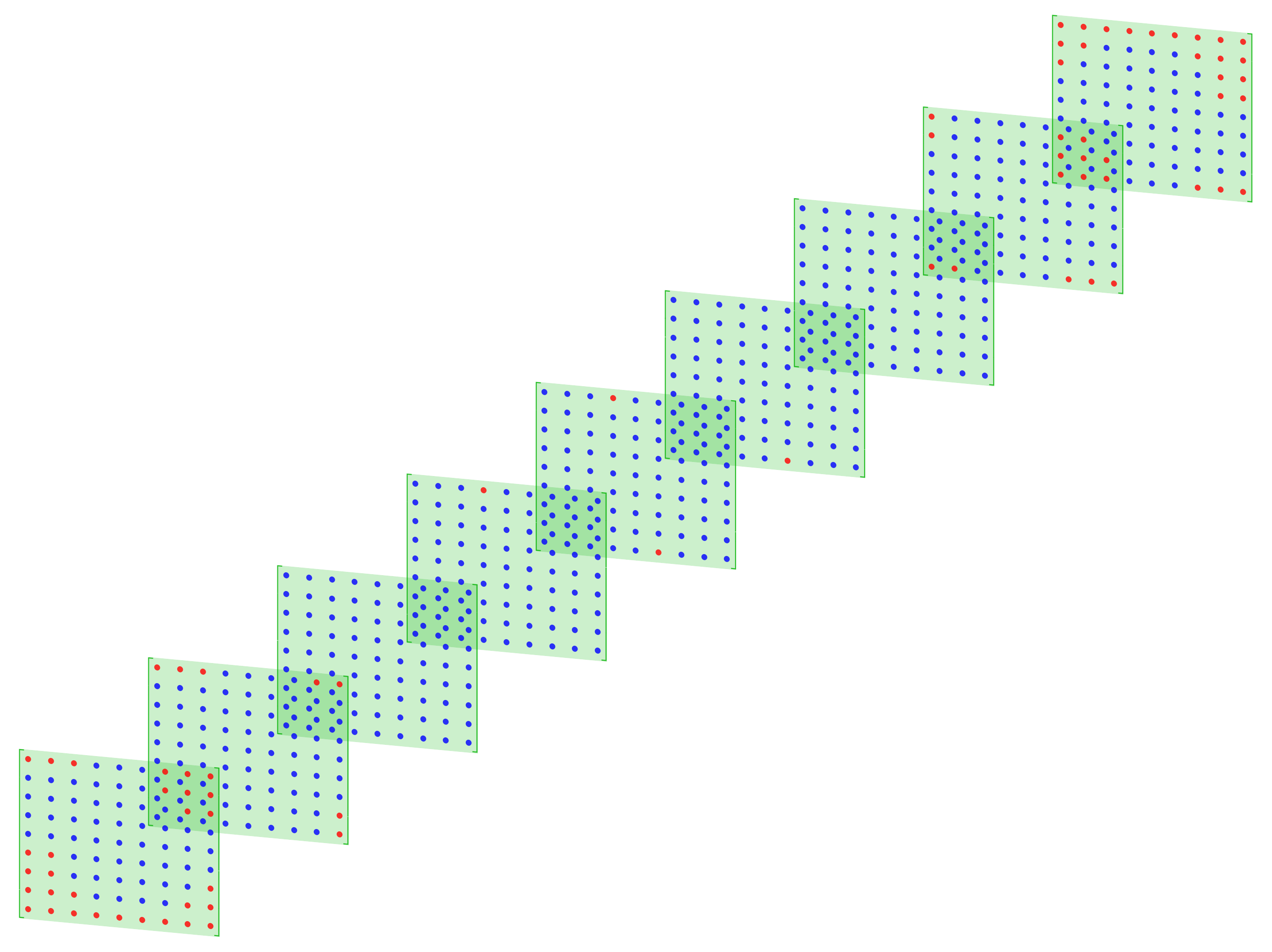}
  \caption{Sparse index set $\cl{I}$ after three iterations.}
  \label{fg:index.set}
\end{figure}

\begin{figure}[!htb]
\centering
  \subfloat[Velocity magnitude isocontours]{\label{fg:6.a}\includegraphics[clip,trim={500 0 500 0},width=0.3\textwidth]{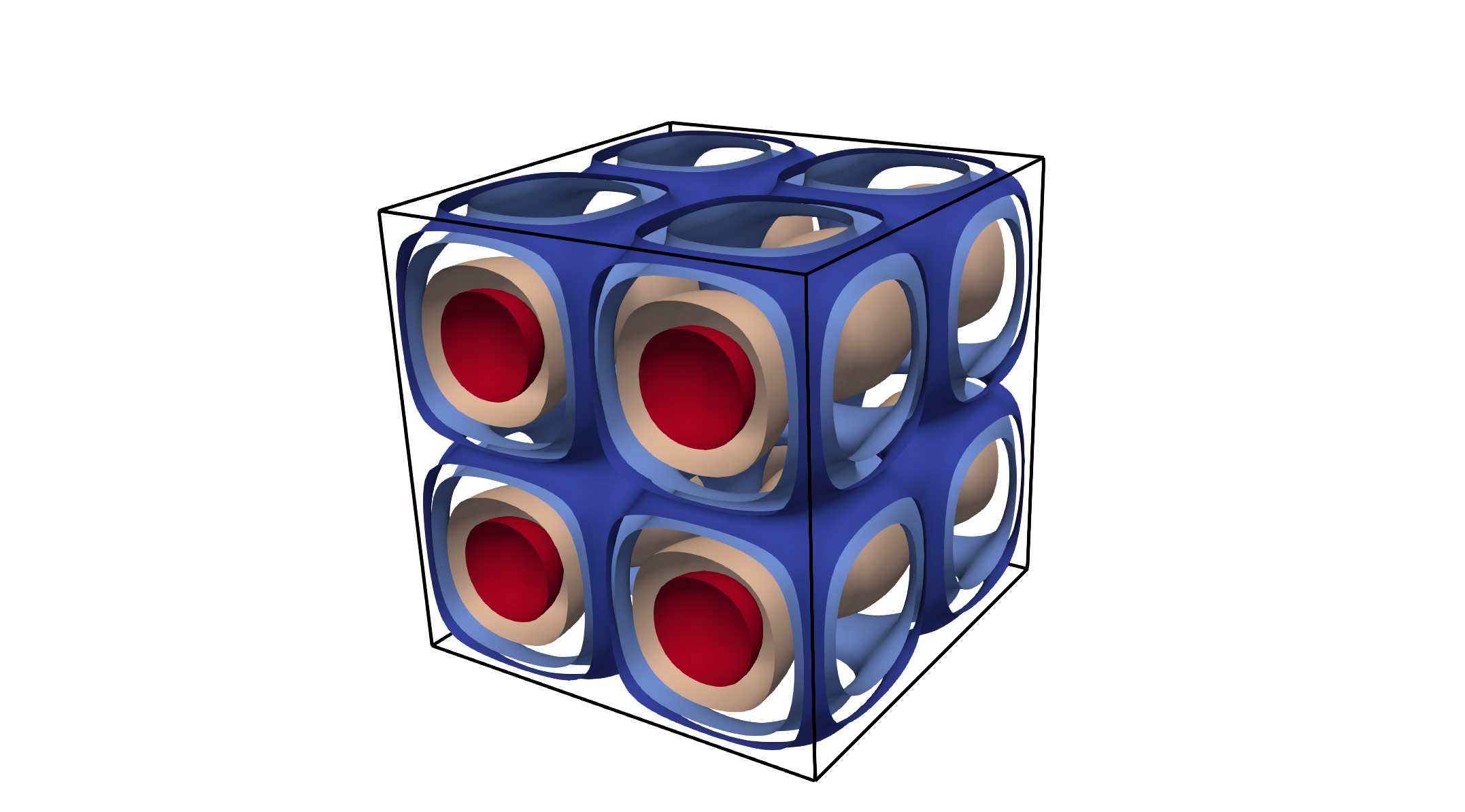}}
  \subfloat[Q-criterion isocontours]{\label{fg:6.b}\includegraphics[clip,trim={500 0 500 0},width=0.3\textwidth]{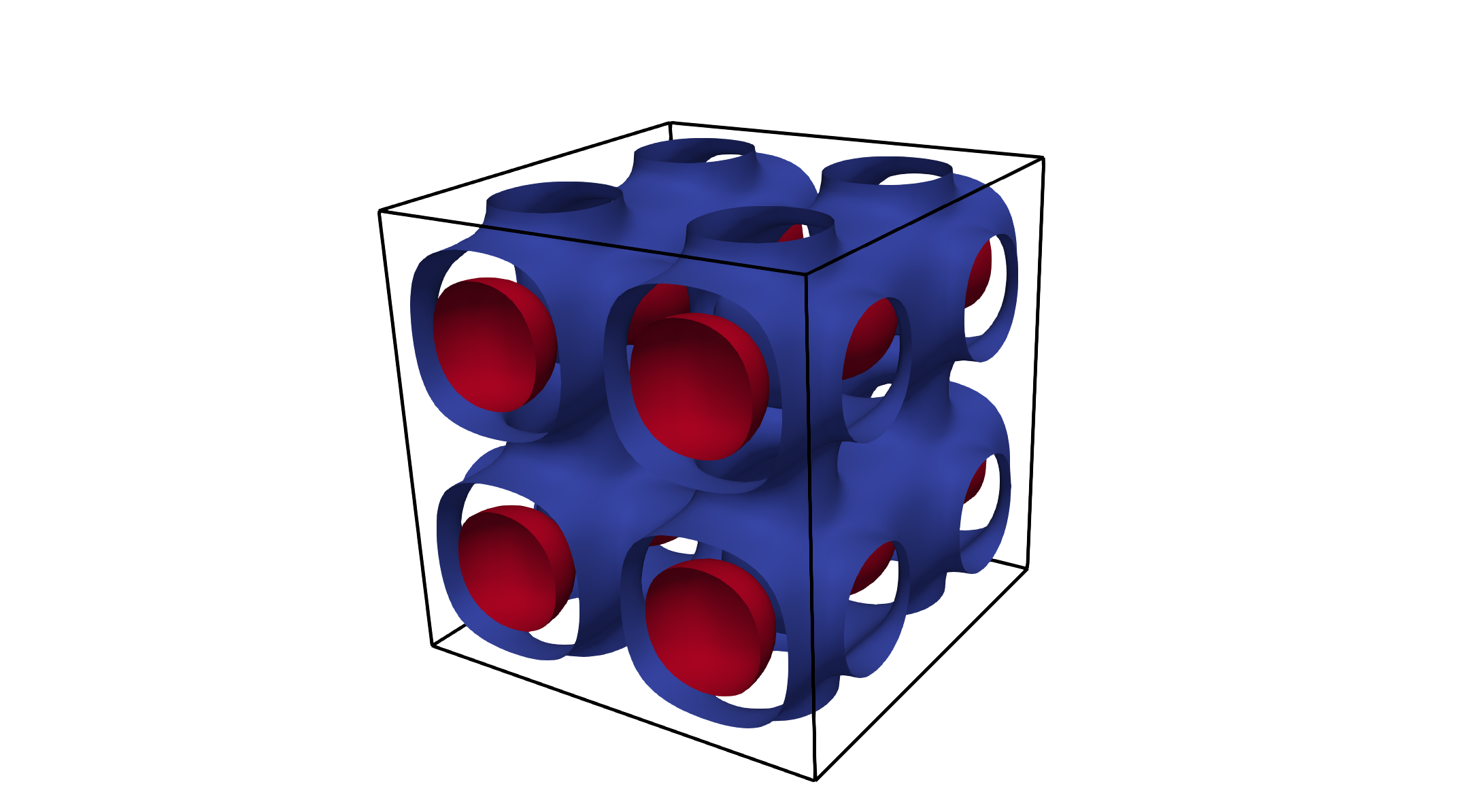}}
  \subfloat[Velocity field]{\label{fg:6.c}\includegraphics[clip,trim={500 0 500 0},width=0.3\textwidth]{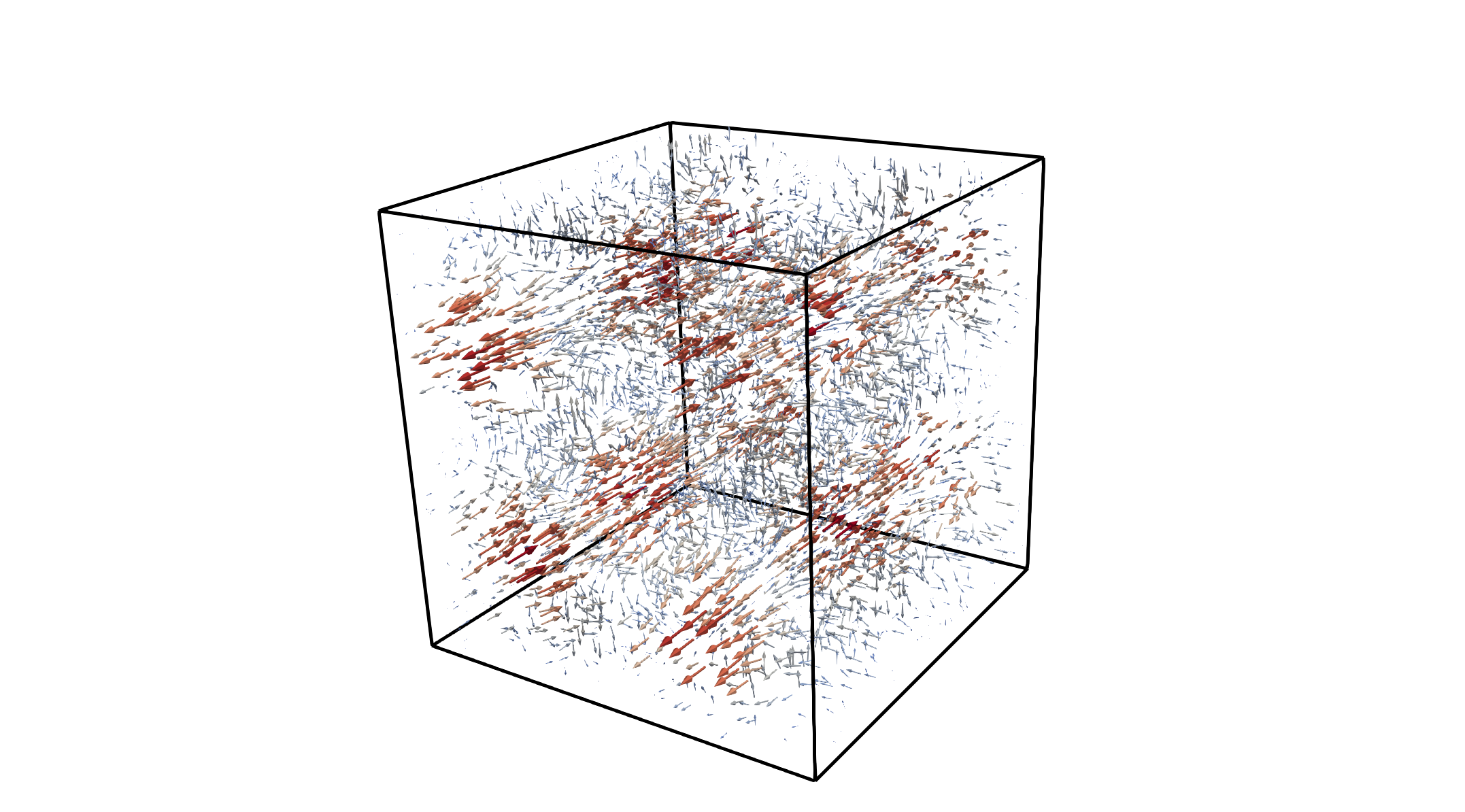}}
  \caption{(Left to right) Velocity magnitude isocontours, $0.125, 0.25, 0.375$, Q-criterion isocontours, $0.33, 0.66$, and velocity vector field}
\label{fg:6}
\end{figure}

\begin{figure}[!htb]
\centering
  \subfloat[$(y,z)$]{\label{fg:7.a}\includegraphics[clip,trim={600 200 600 0},width=0.3\textwidth]{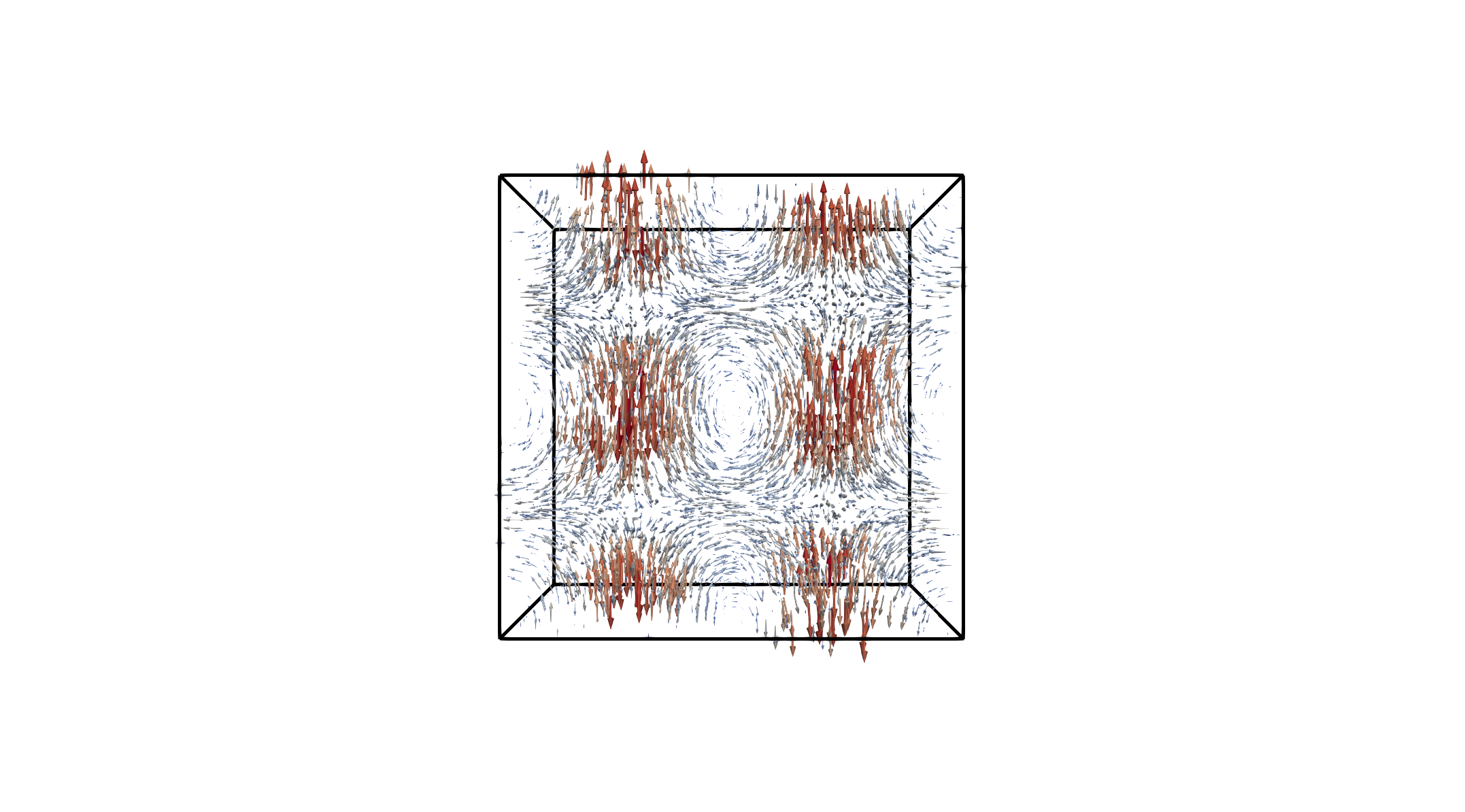}}
  \subfloat[$(x,z)$]{\label{fg:7.b}\includegraphics[clip,trim={600 200 600 0},width=0.3\textwidth]{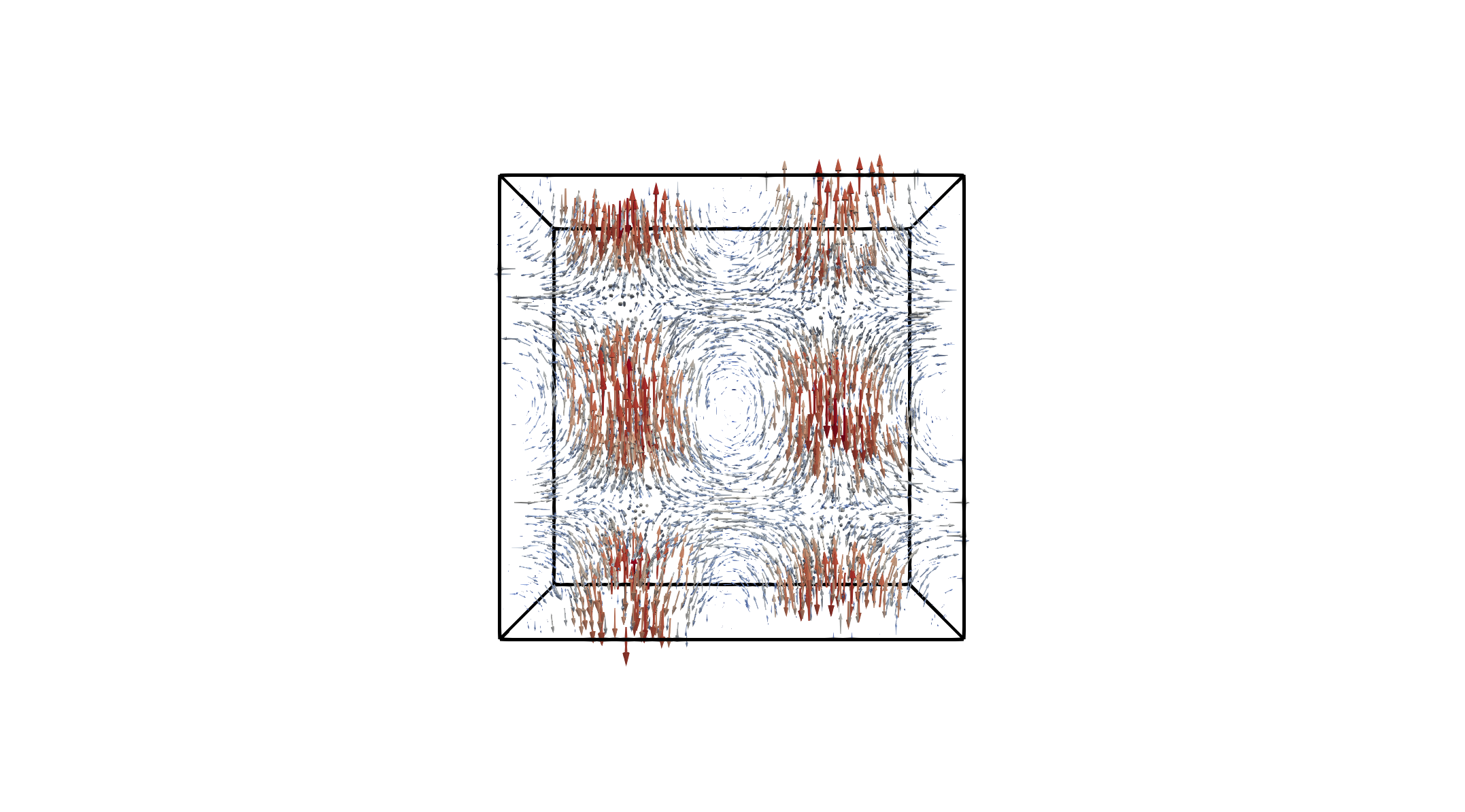}}
  \subfloat[$(x,y)$]{\label{fg:7.c}\includegraphics[clip,trim={600 200 600 0},width=0.3\textwidth]{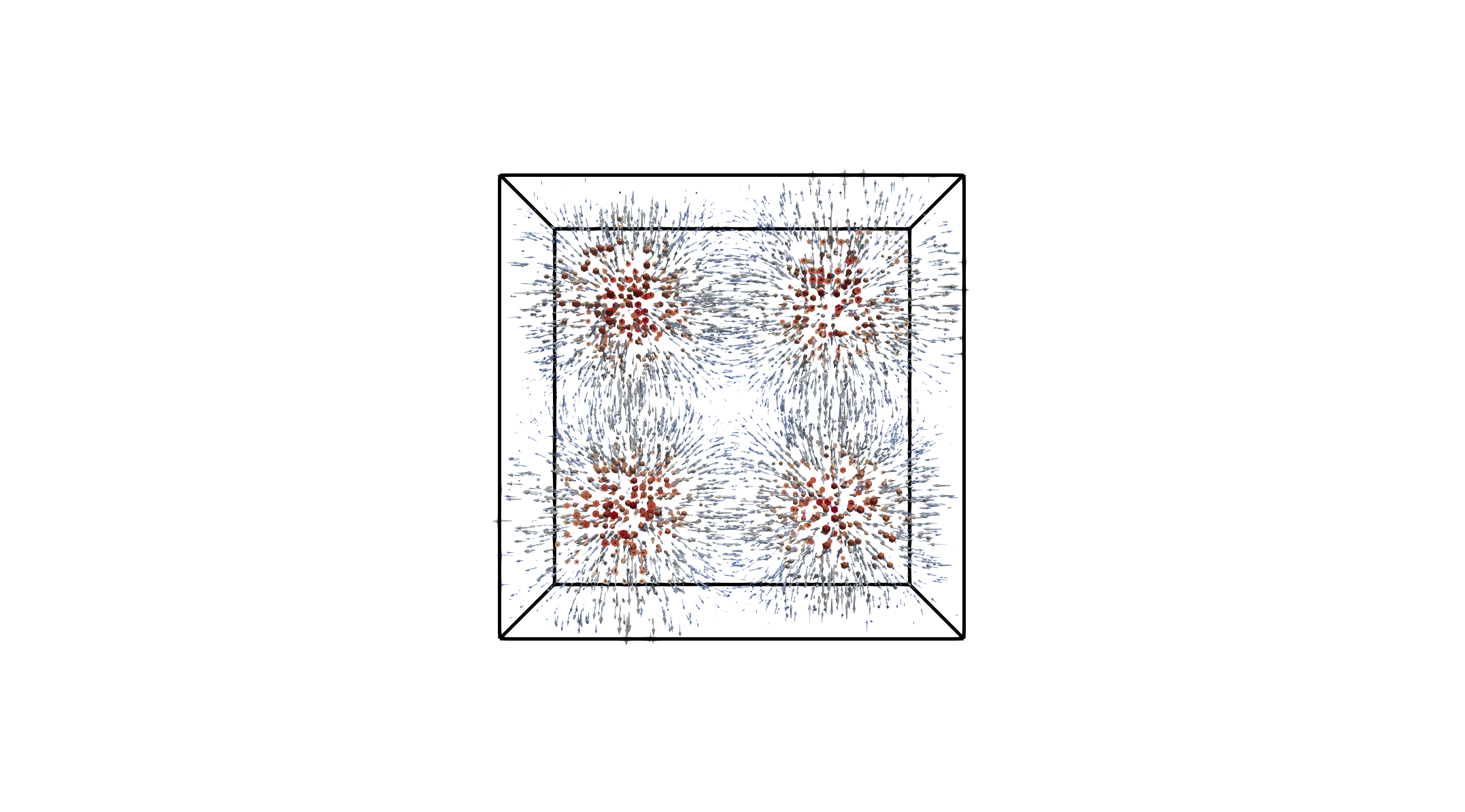}}
  \caption{Three perspective of the reconstructed velocity field.}
\label{fg:7}
\end{figure}

We set the boundary energy $\varepsilon_{\partial\cl{I}}$ and the stopping criterion $\Delta \varepsilon_{\partial\cl{I}}$ to $20\%$ and $10^{-7}$, respectively. For the fractional Sobolev regularization, we selected $\epsilon=10^{-6}$ and $k=1.6$. After three outer iterations, we obtained the index set (Figure \ref{fg:index.set}) with $651$ entries.
Panels \subref{fg:6.a} and \subref{fg:6.b} of Figure \ref{fg:6} display magnitude isocontours ($0.125, 0.25$, and $0.375$) of the solution field and those of the Q-criterion ($0.33$ and $0.66$), respectively. The Q-criterion, which is customary in fluid mechanics, is defined as $Q\coloneqq\fr{1}{2}(\|\sym\Grad\bs{\upsilon}\|^2_F-\|\skw\Grad\bs{\upsilon}\|^2_F)$, where the algebraic operators $\sym$ and $\skw$ are respectively the symmetric and skew-symmetric operators, and $\|\cdot\|^2_F$ denotes the Frobenius norm. Panel \subref{fg:6.c} of Figure \ref{fg:6} is the reconstructed velocity field, and Figure \ref{fg:7} presents this velocity field from different perspectives.

Even though we only had access to $64$ velocity measurements in a three-dimensional space, the actual field and \SFdf\ reconstructed field are indistinguishable.

\FloatBarrier

\subsection{Grid search for the optimal regularization parameter: the Kelvin--Helmholtz instability}

We now aim to reconstruct the velocity field obtained from a Direct Numerical Simulation (DNS) that emulates the Kelvin--Helmholtz instability. We assume accessibility to $50$ velocity measurement points. In all previous simulations, the effect of the regularization was mild. That is, the misfit between the model and the data did not change significantly with $\epsilon$. However, in this example, the underlying physics is more involved, and many features of the flow need to be recovered with a few velocity measurements. Thus, the choice of the regularization parameter $\epsilon$ is crucial to obtain a good approximation. To obtain the optimal regularization parameter $\epsilon^{\mathrm{opt}}$, we solve a sequence of problems to find $\epsilon^{\mathrm{opt}}$ from the L curve, a customary graphical tool for estimating the optimal regularization parameter $\epsilon$. On the L curve, we compare the fractional Sobolev seminorm of the velocity field versus the misfit. The optimal parameter $\epsilon^{\mathrm{opt}}$ is selected as the parameter that minimizes both the fractional Sobolev seminorm of the velocity field and the misfit. Intuitively, $\epsilon^{\mathrm{opt}}$ corresponds to the point obtained by minimizing the distance between the L curve and the origin $(0,0)$.

The DNS was performed with $(1024, 512)$ quadrature points in the domain $\cl{D}=[0,L_{x_1}=2]\times[L_{x_2}/2=-0.5,L_{x_2}/2=0.5]$ over the time window $[0,20]$ with a Reynolds number $Re=10^4$. Along the top and bottom boundaries, we imposed no penetration conditions ($\bs{\upsilon}\cdot\bs{n}=0$) with periodicity in the horizontal direction, $x_1$. The initial condition is given by
\begin{equation}
\left\{
\begin{aligned}
& a = 0.05,\\[4pt]
& \sigma = 0.2,\\[4pt]
& w = -0.5,\\[4pt]
& c = -1.2,\\[4pt]
& \upsilon_1(\bs{x},t=0) = w \tanh\left(\dfrac{x_2}{a}\right),\\[4pt]
& \upsilon_2(\bs{x},t=0) = c \sin\left(\dfrac{2 \pi x_1}{L_{x_1}}\right) \exp\left(-\dfrac{x_2^2}{\sigma^2}\right)+r(\cl{N}(0.1,1),\bs{x}),
\end{aligned}
\right.
\end{equation}
where $r(\cl{N}(0.1,1),\bs{x})$ is a random perturbation with a normal distribution (mean $0.1$ and variance $1$).

\begin{figure}[!htb]
\centering
  \subfloat[Direct Numerical Simulation]{\label{fg:8.a}\includegraphics[width=0.4\textwidth]{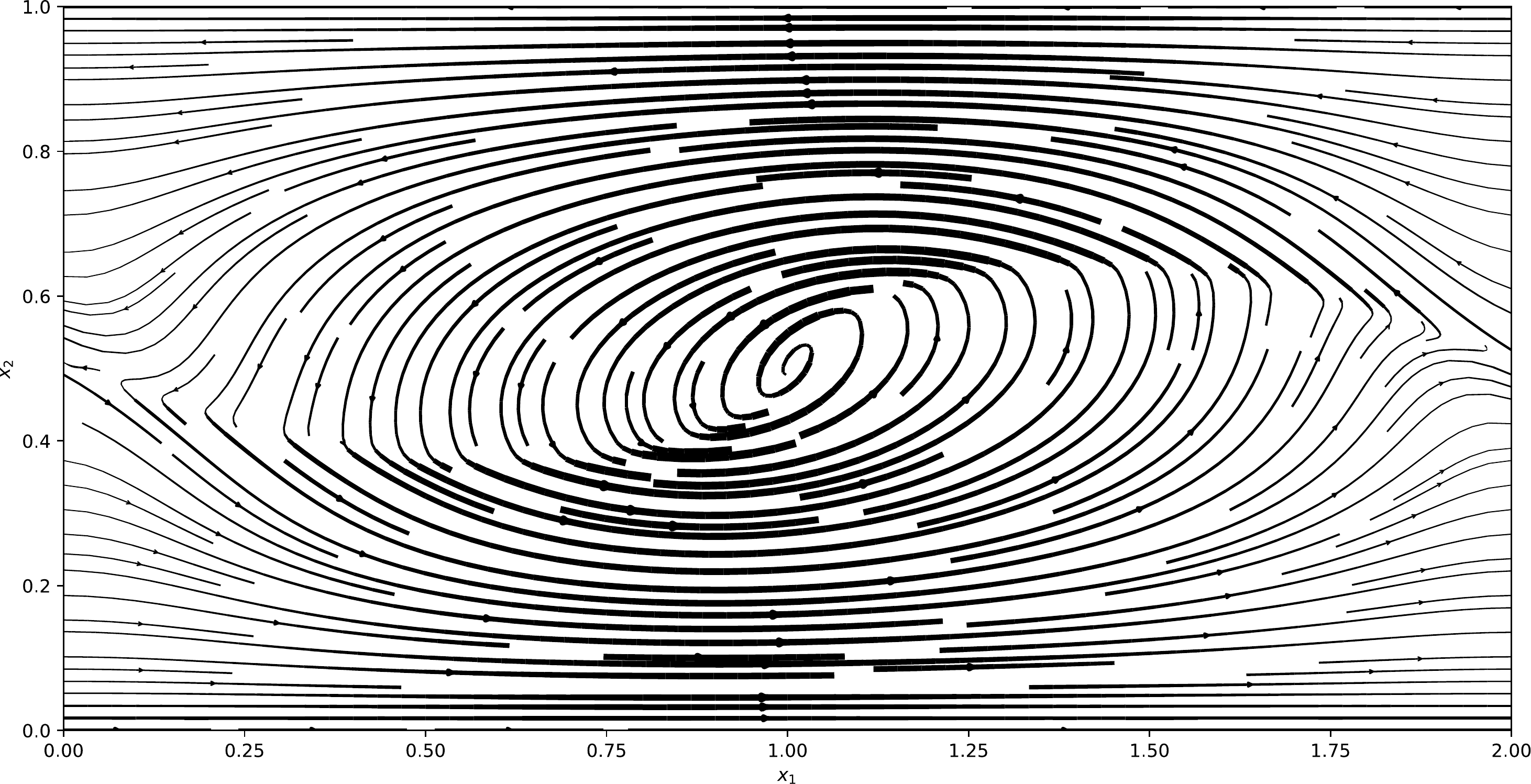}}\hspace{0.5cm}
  \subfloat[\SFdf]{\label{fg:8.b}\includegraphics[width=0.4\textwidth]{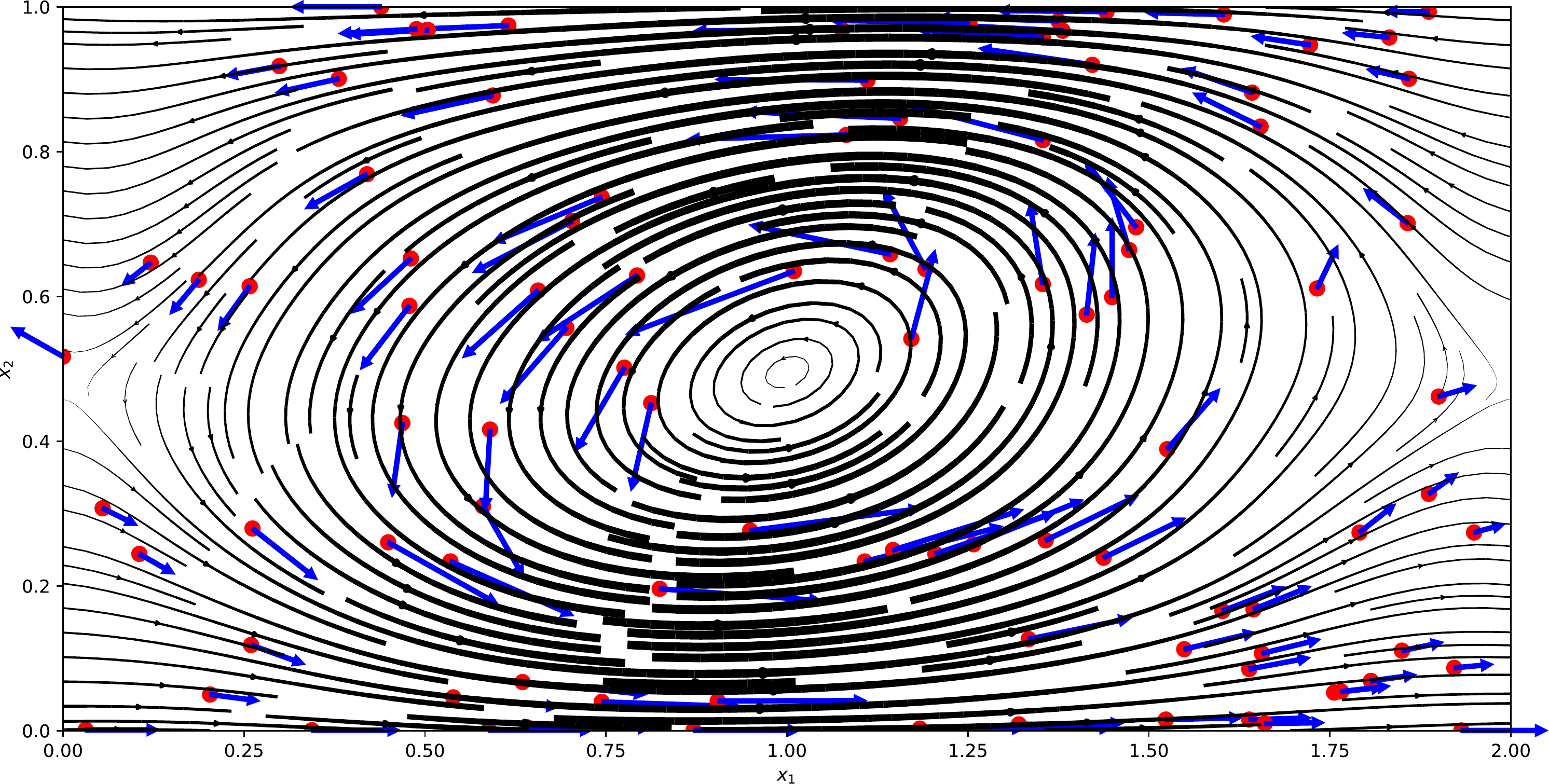}}\\
  \subfloat[Error field $e(\bs{x})$]{\label{fg:8.c}\includegraphics[clip,trim={2cm 0 2cm 0},width=0.55\textwidth]{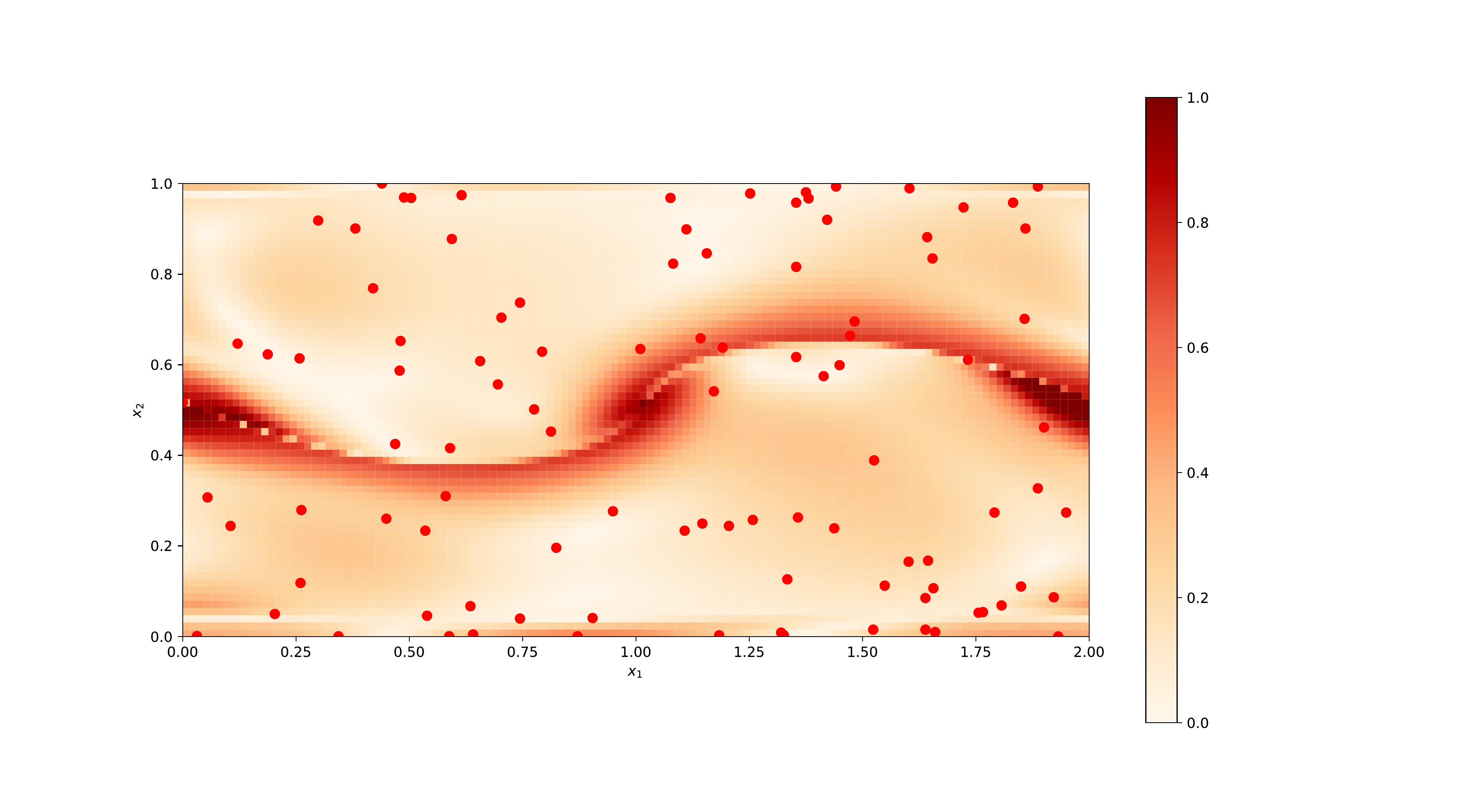}}
  \caption{\protect\subref{fg:8.a} Snapshot of the Direct Numerical Simulation ($t=0.5$ and $Re=10^4$), \protect\subref{fg:8.b} \SFdf\ reconstructed velocity field, and \protect\subref{fg:8.b} error field for $\epsilon=10^{-3}$.}
\label{fg:8.kh_eps}
\end{figure}

To mimic the top and bottow walls, we used immersed boundaries and constructed a ficticious domain of thickness $0.5$ along the top and bottom of the domain, implying that our domain in the vertical direction $x_2$ is actually $L_{x_2}=3$. This example is particularly challenging for the \SFdf\ approximation because a shear layer with a sudden change in the velocity magnitude develops, as can be seen in Figure \ref{fg:8.a} for $t=0.5$.

\begin{figure}[!htb]
\centering
  \includegraphics[width=0.5\textwidth]{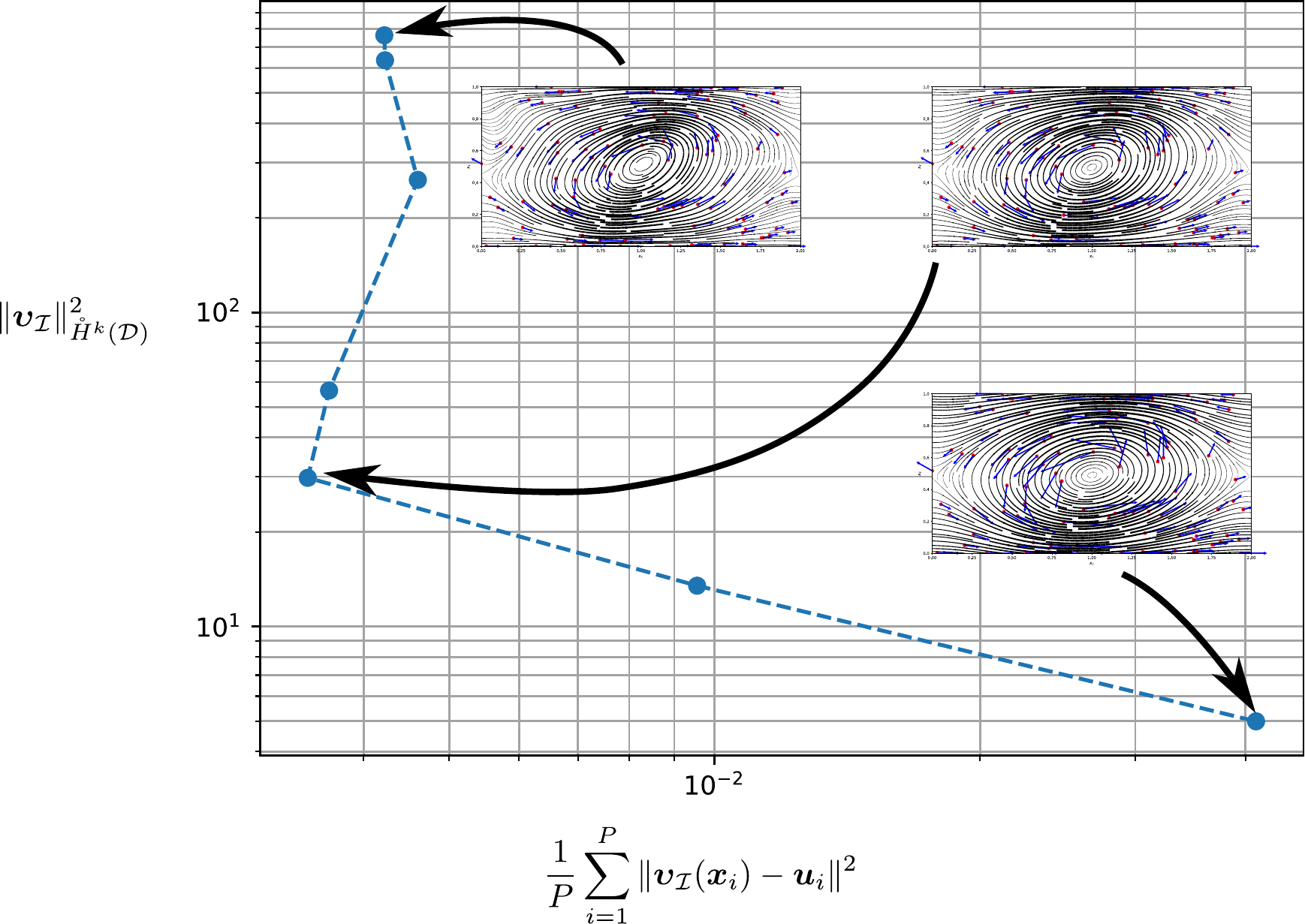}
  \caption{Fractional Sobolev seminorm versus misfit curve in the the Kelvin--Helmholtz simulation.}
  \label{fg:L_curve_kh}
\end{figure}

Figure \ref{fg:L_curve_kh} depicts the L curve, on a log–log scale of the misfit term $\fr{1}{P}\sum_{i=1}^P\norm{\bs{\upsilon}_{\cl{I}}({\bs{x}}_i)-{\bs{u}_i}}{}^2$ versus the regularization (fractional Sobolev seminorm of the velocity field) term $\norm{\bs{\upsilon}}{\mathring{H}^k(\cl{D})}^2$. This curve was computed for $\epsilon$ values in the interval $[10^{-8},10^{-4}]$. Unlike the traditional L curve, the number of degree of freedoms depends on the regularization parameter $\epsilon$ because our \SFdf\ approximation is constructed adaptively. Thus, the L curve departs from the classical L shape.
Within the sequence of index sets $\cl{I}$ related to the curve in Figure \ref{fg:L_curve_kh} lies
\begin{equation*}
\underset{\epsilon=10^{-8}}{
\begin{smallmatrix}
  \rbull & \rbull & \rbull & \rbull & \rbull & \rbull & \bbull & \bbull & \bbull & \rbull & \rbull\\
  \rbull & \rbull & \bbull & \bbull & \bbull & \bbull & \bbull & \bbull & \bbull & \bbull & \rbull\\
  \rbull & \rbull & \bbull & \bbull & \bbull & \bbull & \bbull & \bbull & \bbull & \bbull & \bbull\\
  \bbull & \bbull & \bbull & \bbull & \bbull & \bbull & \bbull & \bbull & \bbull & \bbull & \bbull\\
  \bbull & \bbull & \bbull & \bbull & \bbull & \bbull & \bbull & \bbull & \bbull & \bbull & \bbull\\
  \bbull & \bbull & \bbull & \bbull & \bbull & \bbull & \bbull & \bbull & \bbull & \bbull & \bbull\\
  \bbull & \bbull & \bbull & \bbull & \bbull & \bbull & \bbull & \bbull & \bbull & \rbull & \rbull\\
  \rbull & \bbull & \bbull & \bbull & \bbull & \bbull & \bbull & \bbull & \bbull & \rbull & \rbull\\
  \rbull & \rbull & \bbull & \bbull & \bbull & \rbull & \rbull & \rbull & \rbull & \rbull & \rbull\\
\end{smallmatrix}
}
\Bigg|
\underset{\epsilon=10^{-7}}{
\begin{smallmatrix}
  \rbull & \rbull & \rbull & \rbull & \rbull & \rbull & \bbull & \bbull & \bbull & \rbull & \rbull\\
  \rbull & \rbull & \bbull & \bbull & \bbull & \bbull & \bbull & \bbull & \bbull & \bbull & \rbull\\
  \rbull & \rbull & \bbull & \bbull & \bbull & \bbull & \bbull & \bbull & \bbull & \bbull & \bbull\\
  \bbull & \bbull & \bbull & \bbull & \bbull & \bbull & \bbull & \bbull & \bbull & \bbull & \bbull\\
  \bbull & \bbull & \bbull & \bbull & \bbull & \bbull & \bbull & \bbull & \bbull & \bbull & \bbull\\
  \bbull & \bbull & \bbull & \bbull & \bbull & \bbull & \bbull & \bbull & \bbull & \bbull & \bbull\\
  \bbull & \bbull & \bbull & \bbull & \bbull & \bbull & \bbull & \bbull & \bbull & \rbull & \rbull\\
  \rbull & \bbull & \bbull & \bbull & \bbull & \bbull & \bbull & \bbull & \bbull & \rbull & \rbull\\
  \rbull & \rbull & \bbull & \bbull & \bbull & \rbull & \rbull & \rbull & \rbull & \rbull & \rbull\\
\end{smallmatrix}
}
\Bigg|
\underset{\epsilon=10^{-6}}{
\begin{smallmatrix}
  \rbull & \bbull & \bbull & \bbull & \bbull & \bbull & \bbull & \bbull & \bbull\\
  \bbull & \bbull & \bbull & \bbull & \bbull & \bbull & \bbull & \bbull & \bbull\\
  \bbull & \bbull & \bbull & \bbull & \bbull & \bbull & \bbull & \bbull & \bbull\\
  \bbull & \bbull & \bbull & \bbull & \bbull & \bbull & \bbull & \bbull & \bbull\\
  \bbull & \bbull & \bbull & \bbull & \bbull & \bbull & \bbull & \bbull & \bbull\\
  \bbull & \bbull & \bbull & \bbull & \bbull & \bbull & \bbull & \bbull & \bbull\\
  \bbull & \bbull & \bbull & \bbull & \bbull & \bbull & \bbull & \bbull & \rbull\\
\end{smallmatrix}
}
\Bigg|
\underset{\epsilon=10^{-5}}{
\begin{smallmatrix}
  \rbull & \rbull & \rbull & \rbull & \rbull & \rbull & \bbull & \bbull & \bbull & \bbull & \bbull & \rbull & \rbull\\
  \bbull & \bbull & \bbull & \bbull & \bbull & \bbull & \bbull & \bbull & \bbull & \bbull & \bbull & \rbull & \rbull\\
  \bbull & \bbull & \bbull & \bbull & \bbull & \bbull & \bbull & \bbull & \bbull & \bbull & \bbull & \bbull & \rbull\\
  \bbull & \bbull & \bbull & \bbull & \bbull & \bbull & \bbull & \bbull & \bbull & \bbull & \bbull & \bbull & \rbull\\
  \rbull & \bbull & \bbull & \bbull & \bbull & \bbull & \bbull & \bbull & \bbull & \bbull & \bbull & \bbull & \rbull\\
  \rbull & \bbull & \bbull & \bbull & \bbull & \bbull & \bbull & \bbull & \bbull & \bbull & \bbull & \bbull & \bbull\\
  \rbull & \bbull & \bbull & \bbull & \bbull & \bbull & \bbull & \bbull & \bbull & \bbull & \bbull & \bbull & \bbull\\
  \rbull & \rbull & \bbull & \bbull & \bbull & \bbull & \bbull & \bbull & \bbull & \bbull & \bbull & \bbull & \bbull\\
  \rbull & \rbull & \bbull & \bbull & \bbull & \bbull & \bbull & \rbull & \rbull & \rbull & \rbull & \rbull & \rbull\\
\end{smallmatrix}
}
\end{equation*}

\begin{equation*}
\underset{\epsilon=10^{-4}}{
\begin{smallmatrix}
  \rbull & \rbull & \rbull & \rbull & \rbull & \rbull & \rbull & \bbull & \bbull & \bbull & \bbull & \bbull & \bbull\\
  \rbull & \rbull & \rbull & \bbull & \bbull & \bbull & \bbull & \bbull & \bbull & \bbull & \bbull & \bbull & \bbull\\
  \bbull & \bbull & \bbull & \bbull & \bbull & \bbull & \bbull & \bbull & \bbull & \bbull & \bbull & \bbull & \bbull\\
  \bbull & \bbull & \bbull & \bbull & \bbull & \bbull & \bbull & \bbull & \bbull & \bbull & \bbull & \bbull & \bbull\\
  \bbull & \bbull & \bbull & \bbull & \bbull & \bbull & \bbull & \bbull & \bbull & \bbull & \bbull & \bbull & \bbull\\
  \bbull & \bbull & \bbull & \bbull & \bbull & \bbull & \bbull & \bbull & \bbull & \bbull & \bbull & \bbull & \bbull\\
  \bbull & \bbull & \bbull & \bbull & \bbull & \bbull & \bbull & \bbull & \bbull & \bbull & \bbull & \bbull & \bbull\\
  \bbull & \bbull & \bbull & \bbull & \bbull & \bbull & \bbull & \bbull & \bbull & \bbull & \bbull & \bbull & \bbull\\
  \bbull & \bbull & \bbull & \bbull & \bbull & \bbull & \bbull & \bbull & \bbull & \bbull & \bbull & \bbull & \bbull\\
  \bbull & \bbull & \bbull & \bbull & \bbull & \bbull & \bbull & \bbull & \bbull & \bbull & \rbull & \rbull & \rbull\\
  \bbull & \bbull & \bbull & \bbull & \bbull & \bbull & \rbull & \rbull & \rbull & \rbull & \rbull & \rbull & \rbull\\
\end{smallmatrix}
}
\Bigg|
\underset{\red{\epsilon^{\mathrm{opt}}=10^{-3}}}{
\begin{smallmatrix}
  \rbull & \rbull & \bbull & \bbull & \bbull & \bbull & \bbull & \bbull & \bbull & \bbull & \bbull\\
  \rbull & \rbull & \bbull & \bbull & \bbull & \bbull & \bbull & \bbull & \bbull & \bbull & \bbull\\
  \bbull & \bbull & \bbull & \bbull & \bbull & \bbull & \bbull & \bbull & \bbull & \bbull & \bbull\\
  \bbull & \bbull & \bbull & \bbull & \bbull & \bbull & \bbull & \bbull & \bbull & \bbull & \bbull\\
  \bbull & \bbull & \bbull & \bbull & \bbull & \bbull & \bbull & \bbull & \bbull & \bbull & \bbull\\
  \bbull & \bbull & \bbull & \bbull & \bbull & \bbull & \bbull & \bbull & \bbull & \bbull & \bbull\\
  \bbull & \bbull & \bbull & \bbull & \bbull & \bbull & \bbull & \bbull & \bbull & \bbull & \bbull\\
  \bbull & \bbull & \bbull & \bbull & \bbull & \bbull & \bbull & \bbull & \bbull & \rbull & \rbull\\
  \bbull & \bbull & \bbull & \bbull & \bbull & \bbull & \bbull & \bbull & \bbull & \rbull & \rbull\\
\end{smallmatrix}
}
\Bigg|
\underset{\epsilon=10^{-2}}{
\begin{smallmatrix}
  \rbull & \rbull & \bbull & \bbull & \bbull & \bbull & \bbull\\
  \bbull & \bbull & \bbull & \bbull & \bbull & \bbull & \bbull\\
  \bbull & \bbull & \bbull & \bbull & \bbull & \bbull & \bbull\\
  \bbull & \bbull & \bbull & \bbull & \bbull & \bbull & \bbull\\
  \bbull & \bbull & \bbull & \bbull & \bbull & \bbull & \bbull\\
  \bbull & \bbull & \bbull & \bbull & \bbull & \bbull & \bbull\\
  \bbull & \bbull & \bbull & \bbull & \bbull & \rbull & \rbull\\
\end{smallmatrix}
}
\end{equation*}
where the `optimal' regularization parameter is $\epsilon^{\mathrm{opt}}=10^{-3}$.

In Figure \ref{fg:8.kh_eps}, we present the DNS velocity field \ref{fg:8.a}, the \SFdf\ field \ref{fg:8.b}, and the pointwise error field \ref{fg:8.c}.

%% file: section-4.tex
%auto-ignore
\section{Temporal approximation}
\label{sc:compress.sensing}

We now approximate the time series of velocity fields measured at various locations, where each location is fixed in time. Given a set of $P$ spatial coordinates $\{\bs{x}_p\}_{p = 1}^P$ and times $t_1, t_2, \ldots, t_T$, we let $\bs{u}_{m,p} \in \R^n$ denote a measurement at point $\bs{x}_p \in \R^n$ at time $t_m$. We denote all measurements at time $t_m$ as $\bs{u}_m$ $\in \R^{nP}$. We seek to recreate a flow with divergence-free velocity $\bs{\upsilon} = \bs{\upsilon}(\bs{x},t)$ from the measurements $\bs{u}_m$ at $t = t_1, \ldots, t_T$. Note that $\bs{u}_m$ can be expressed as
\begin{equation}
    \bs{u}_m = \sum_{k=1}^{nP} \lambda_k(t_m) \bs{e}_k,
  \label{eq:snapshot-expansion}
\end{equation}
where $\{\bs{e}_k\}_{k=1}^{nP}$ is any basis of $\R^{nP}$, and $\lambda_k(t_m) \in \R$ are the corresponding time-evolution coefficients, which are uniquely determined.

Next, let
$\Pi_{\cl{V}}\colon\bb{R}^{nP}\rightarrow\cl{V}\subset{H}^{k}$
denote the discrete spatial projection operator that maps the given measurements
$ \bs{u}_m \in \R^{nP}$ to the corresponding Sparse Fourier divergence-free \SFdf\ approximation at a fixed time. To this end, we solve problem
\eqref{eq:optimization}; that is,
\begin{equation}
\bs{\upsilon}_{\cl{I}}^{\mathrm{opt}}(\bs{x},t_m) \coloneqq\Pi_{\cl{V}}(\bs{u}_m)(\bs{x}).
\end{equation}
As problem \eqref{eq:optimization} is a quadratic optimization problem (see Remark \ref{rk:perturbation}) with
linear constraints, the optimal solution is also linear with
respect to the data $\bs{u}_m$. By \eqref{eq:snapshot-expansion}, we may express the reconstructed velocity field as a time-varying linear
combination of $nP$ base velocity fields; that is,
\begin{equation}
\bs{\upsilon}_{\cl{I}}^{\mathrm{opt}}(\bs{x},t_m) =
\sum_{k=1}^{nP}\lambda_{k}(t_m) \, \Pi_{\cl{V}}\left(\bs e_{k}\right).
\label{eq:time_compression_exact}
\end{equation}
Applying $\Pi_{\cl{V}}$, we find $nP$ base velocity fields
independently of the number of time points $T$. As $nP$ is smaller than $T$, the procedure saves considerable
computational time.

To further increase the efficiency, we exploit the low-rank approximation of the data. Consider the $T \times nP$ data matrix
\begin{equation}
\bs{u} =\left(
  \begin{array}{c}
\bs{u}_{1}^\T\\
\dots\\
\bs{u}_{T}^\T
  \end{array}\right)
=\begin{pmatrix}
    u_{1,1,1} & \dots & u_{1,P,1} & \ldots &  u_{1,1,n} & \dots & u_{1,P,n} \\
    \vdots & \ddots & \vdots & \ddots & \vdots & \ddots & \vdots \\
    u_{T,1,1} & \dots & u_{T,P,1} & \ldots &  u_{T,1,n} & \dots & u_{T,P,n} \\ \\
  \end{pmatrix}
\in\R^{T \times nP},
\label{eq:u_matrix}
\end{equation}
where $u_{m,p,i}$ denotes the \emph{i}-th coordinate of $\bs{u}_{m, p} \in \R^n$.
The best $K$-rank approximation ($K < nP$) of this matrix is given by the truncated SVD decomposition (see, for instance \cite{EckartYoung1936}) and reads
\begin{equation}
\bs{u} \approx \bs{U}_{K}\bs{\Sigma}_{K}\bs{V}_{K}^{\trans},
\label{eq:t-approx:svd}
\end{equation}
where $\bs{U}_{K}\in\R^{T\times K}$ and $\bs{V}_{K}^{\trans}\in\R^{K\times nP}$
such that $\bs{U}_{K}^{\trans}\bs{U}_{K}=\bs{V}_{K}^{\trans}\bs{V}_{K}=\id_{K}$, $\id_K$ is the $K \times K$
identity matrix,
and $\bs{\Sigma}_{K}=\diag(\sigma_{1},\dots,\sigma_{K})$ with
$\sigma_1 \geq \dots \geq \sigma_K$.

The columns of $\bs{V}_{K}$ form an orthonormal basis $\bs{e}_1, \ldots, \bs{e}_K$ of $\R^K \subset \R^{nP}$. Analogous to \eqref{eq:snapshot-expansion}, $\bs{u}_m$ can be approximated as
\begin{equation}
    \bs{u}_m \approx \sum_{k=1}^{K} \lambda_k(t_m) \bs{e}_k.
  \label{eq:snapshot-expansion-approximate}
\end{equation}
Finally, in view of \eqref{eq:time_compression_exact}, \eqref{eq:t-approx:svd} and \eqref{eq:snapshot-expansion-approximate}, we arrive at an approximation of the reconstructed field, defined at time points $t_1, \ldots, t_m$, that is,
\begin{equation}
\begin{aligned}
\bs{\upsilon}_{\cl{I}}^{\mathrm{opt}}(\bs{x},t_m) &\approx
\sum_{k=1}^{K}\lambda_{k}(t_m) \, \Pi_{\cl{V}}\left(\bs e_{k}\right), \\
\lambda_k(t_m) & = \bs{u}_m \cdot \bs{e}_k.
\end{aligned}
\label{eq:time_compression_approx}
\end{equation}
\begin{rmk}[Efficiency]
The reconstructed flow in expression \eqref{eq:time_compression_approx} is only defined for discrete set of time points $t_1, \ldots, t_T$. In practice this becomes a limitation if the measurements are sparse in time. Note however that the computational cost only depends on $K$ and is \emph{independent} of $T$.
\end{rmk}

%% file: section-5.tex
%auto-ignore
\section{Numerical experiments on the spatiotemporal approximation}
\label{sc:numerics.spatio.temporal}

\begin{figure}[ht]
  \centering
  \subfloat[Standard deviation of $\upsilon_1$]{%
    \includegraphics[width=0.6\textwidth]{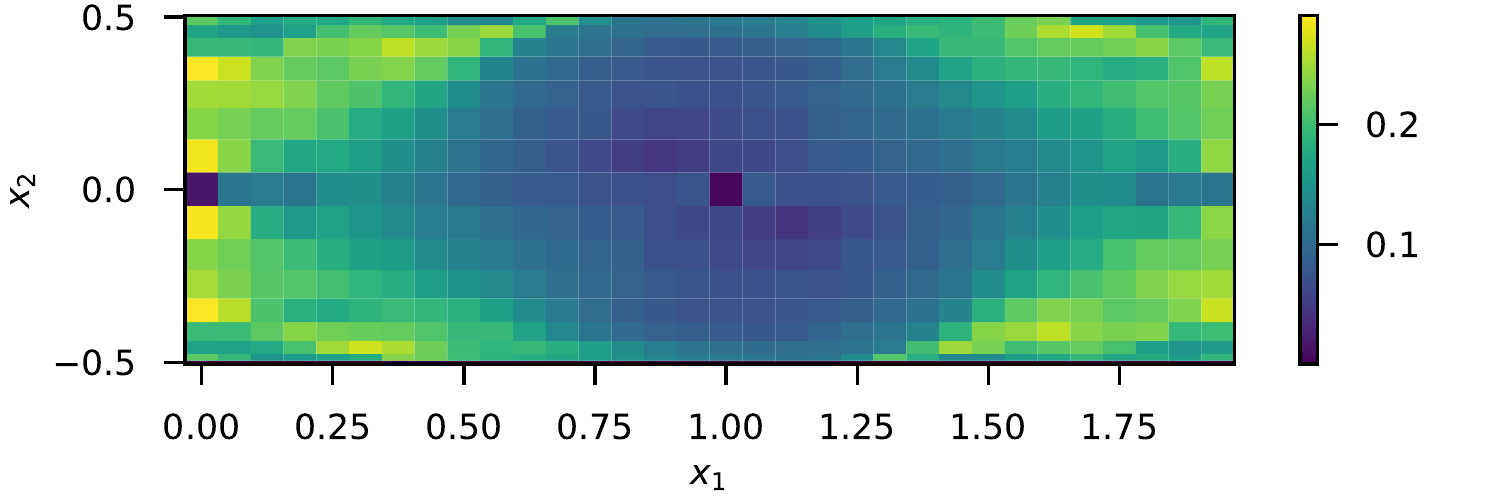}%
    \label{fig:sta:vel-stddev-u}
  }\\
  \subfloat[Standard deviation of $\upsilon_2$]{%
    \includegraphics[width=0.6\textwidth]{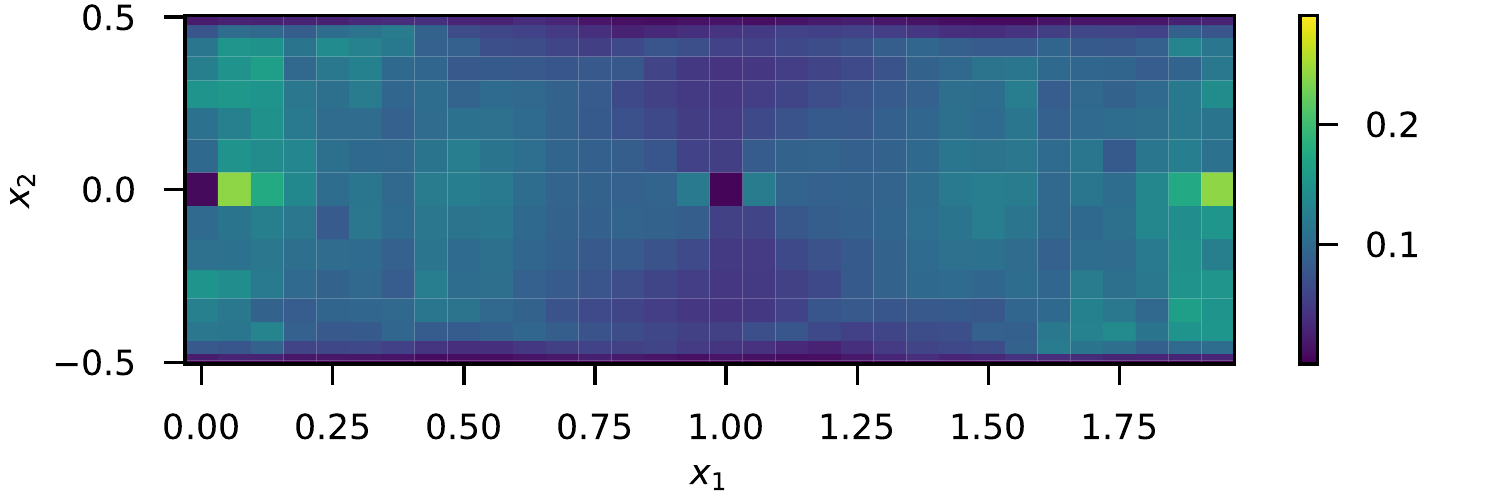}%
    \label{fig:sta:vel-stddev-v}
  }
  \caption{Standard deviations of the time series of true velocity field:
    \protect\subref{fig:sta:vel-stddev-u} $\upsilon_1$-component, \protect\subref{fig:sta:vel-stddev-v} $\upsilon_2$-component.%
    \label{fig:sta:vel-stddev} }
\end{figure}

This example is a continuation of the last example in
Section~\ref{sc:numerics.spatial}
where we reconstructed the velocity field at $t=0.5$ in the Kelvin--Helmholtz
instability. We now aim to reconstruct the velocity flow at any point in the time
interval $t \in [0,20]$, following the method in Section~\ref{sc:compress.sensing}.
The time series resolution is $0.1$, yielding $T=201$
time snapshots. Figure
\ref{fig:sta:vel-stddev} depicts the standard deviations in both components of the
velocity field over time.
The $\upsilon_1$-component of the velocity
is significantly more variable than the $\upsilon_2$ component of the velocity.
Owing to the formation of secondary vortices around the Kelvin--Helmholtz
instability, the velocity field evolves more dynamically in the
regions adjacent to the central vortical structure, specifically, on the left and right boundaries of the domain.
Conversely, in the central vortical structure, there is no significant variations
over time.

\begin{figure}[!htb]
\centering
\includegraphics[width=1.8in]{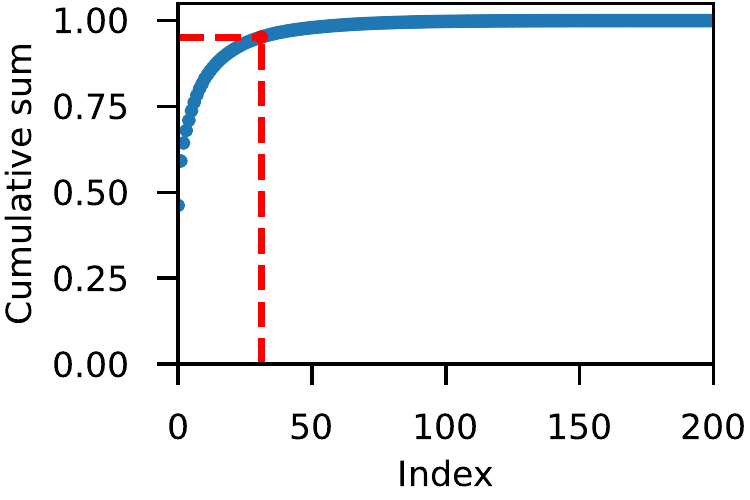}
\caption{Cumulative sum of squares of the singular values of data matrix $u$
  containing the velocity time series at $201$ time moments collected
  at $100$ locations in the
  simulation of the two-dimensional
  Kelvin--Helmholtz instability.\label{fig:sta:sum-sigma-sq}}
\end{figure}

The data matrix~$\bs{u}$ is sampled at $100$ randomly uniformly distributed
locations on the uniform grid at $T=201$ time points. Therefore the matrix size is $201 \times 200$ (see Section\ref{sc:compress.sensing}).
Using \eqref{eq:time_compression_approx}, we compute the truncated SVD of the data matrix. To understand the importance of each mode encompassed in $\bs{u}$,
we compute the cumulative sum of squares of
the singular values (energy).
The cumulative sum versus the singular-value index is plotted in Figure \ref{fig:sta:sum-sigma-sq}.
As expected, the first few singular values represent most of the energy.
Among the $200$ modes, $100$ modes
explain more than $99\,\%$ of the data variability. Moreover, this plot
demonstrates that $95\,\%$ of the explained variance in the data is captured
by the low-rank approximation of rank $31$.

The compressed sensing temporal approximation is given by
\eqref{eq:time_compression_approx}, with $\{ \bs{e}_k \}_{k=1}^K$ corresponding
to the $K$ first right-singular vectors of $\bs{u}$. Based on the cumulative sum analysis, we set $K=31$ and train $32$ models (the additional model captures the mean of the time series).
In all models, we use $\epsilon=10^{-5}$ and $k=1.5$. In a grid search, these values were
found to minimize the prediction error defined by expression \eqref{eq:sta:error} below.

After training the models, we evaluate their accuracy on the
test data $\bs u_{\text{test}}$,
disjoint from the training data $\bs{u}$.
The test data is sampled on a uniform $32 \times 16$ spatial grid,
giving $P_{\mathrm{test}} = 512$ test locations at
the same time points as the training data.

\begin{figure}[htbp]
    \centering
    \includegraphics[width=0.6\textwidth]{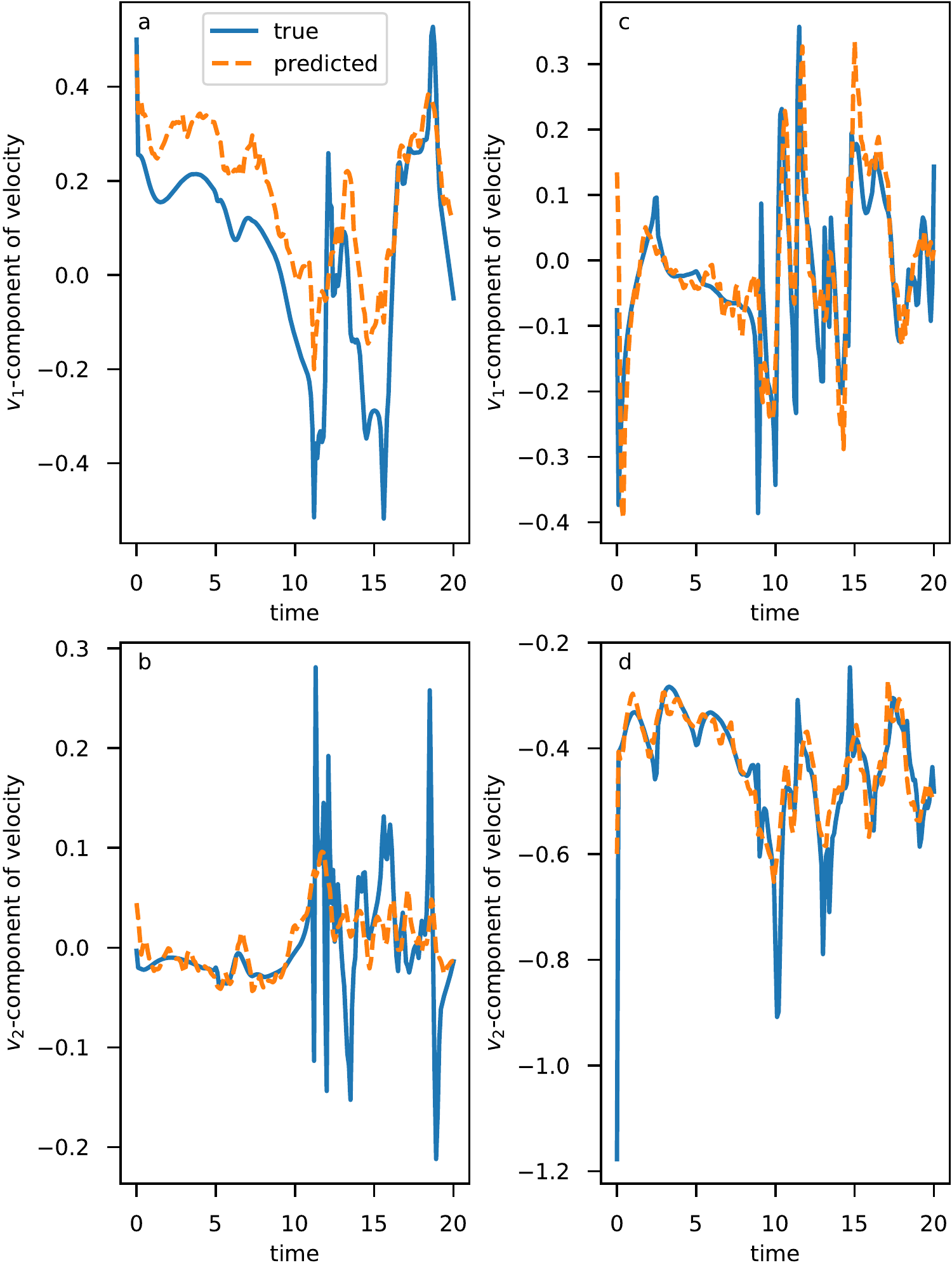}
    \caption{True and predicted time series of the velocity components at different locations:
      (left to right, top) at location $(0.1875, -0.4614)$; (left to right, bottom) at location $(0.4375, 0.0015)$.%
    }%
    \label{fig:sta:example-approx}
  \end{figure}

Figure \ref{fig:sta:example-approx} shows two examples of the test time series
along with the reconstructed times series of both velocity field components
taken at the locations $(0.1875, -0.4614)$ and $(0.4375, 0.0015)$.
Although the predicted values follow the trends of the true time series, the spikes in the true time series are not captured in
the reconstruction.

\begin{figure}[htbp]
  \captionsetup[subfloat]{farskip=0pt,captionskip=1pt}
  \centering
  \subfloat[]{%
    \includegraphics[width=0.6\linewidth]{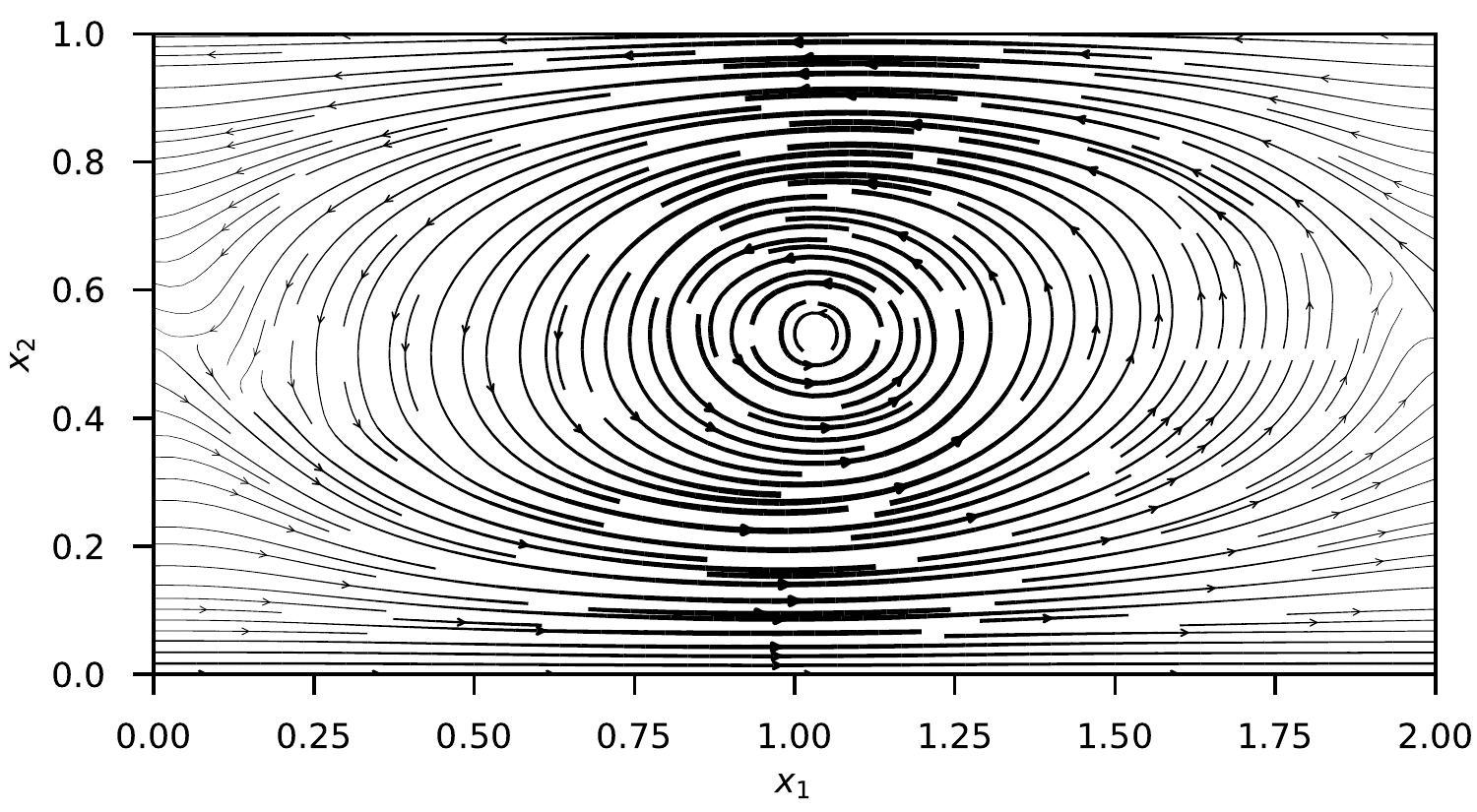}%
    \label{fig:true-field-t_1}
  }\hfill
  \subfloat[]{%
    \includegraphics[width=0.6\linewidth]{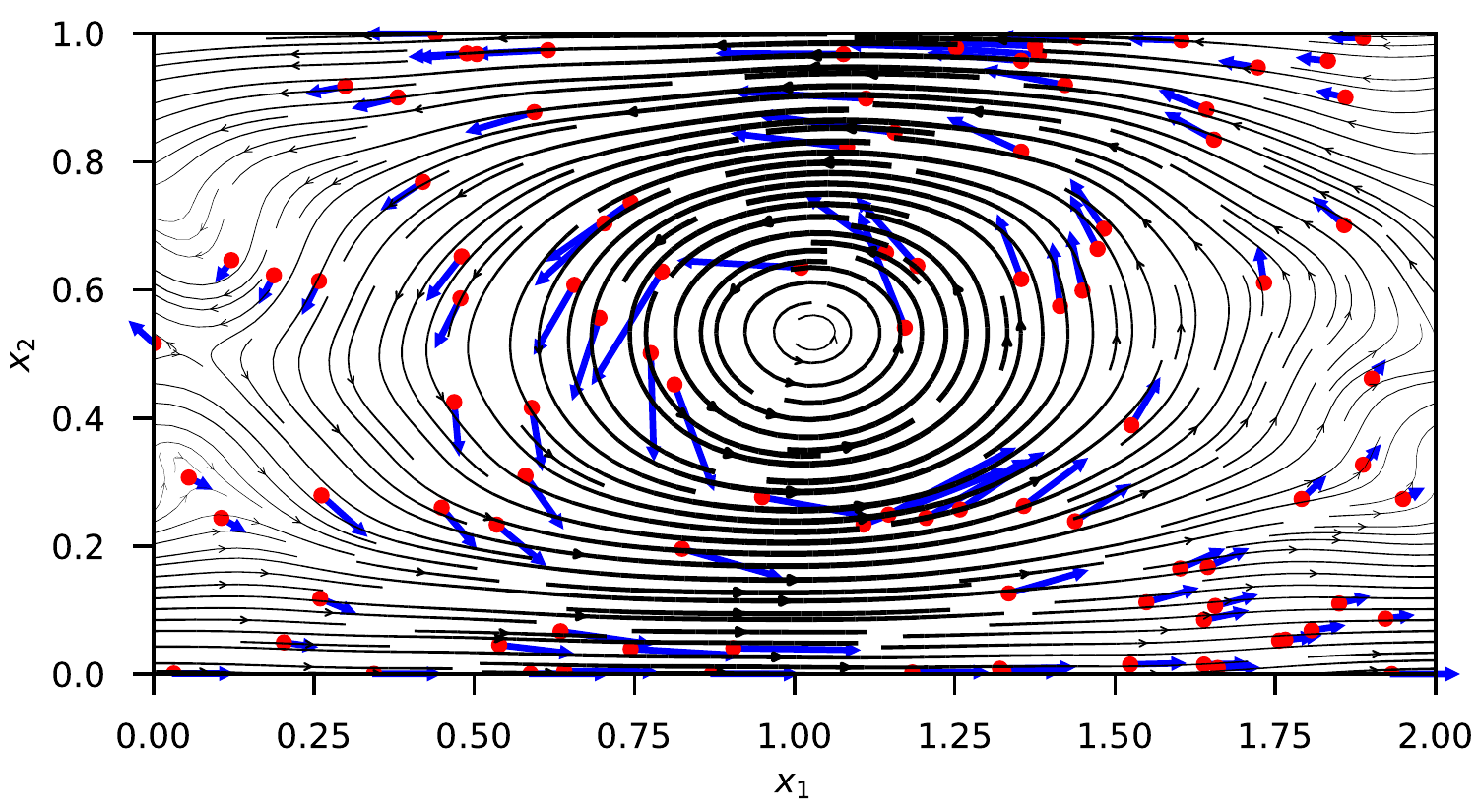}%
    \label{fig:pred-field-t_1}
  }
  \caption{%
    Streamplots of the velocity field at time $1.0$:
    (A)~true field; (B)~predicted field.%
    \label{fig:true-and-pred-field-t=1.0}
  }
\end{figure}

\begin{figure}[htbp]
  \captionsetup[subfloat]{farskip=0pt,captionskip=1pt}
  \centering
  \subfloat[]{%
    \includegraphics[width=0.6\textwidth]{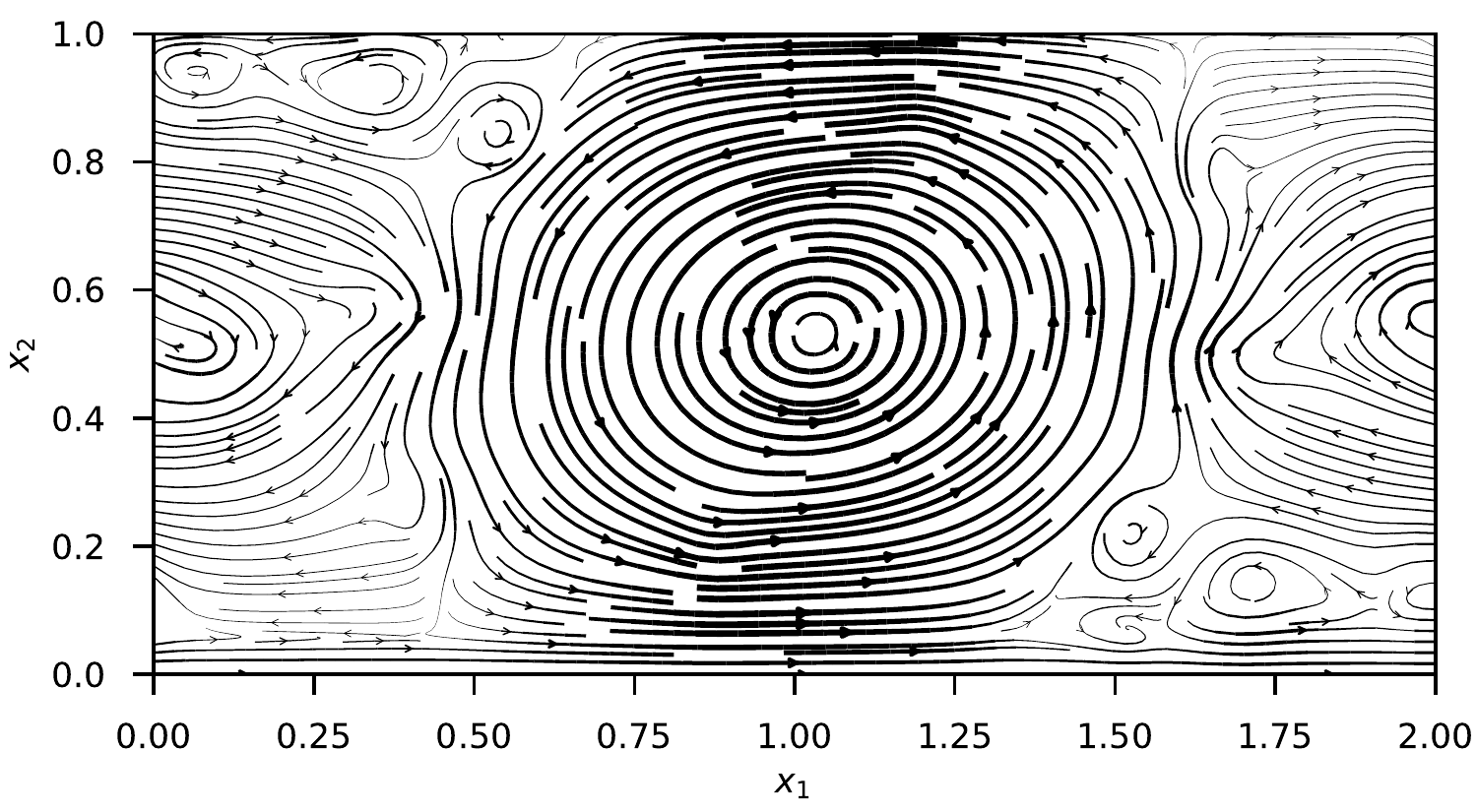}%
    \label{fig:true-field-t_10}
  }\hfill
  \subfloat[]{%
    \includegraphics[width=0.6\textwidth]{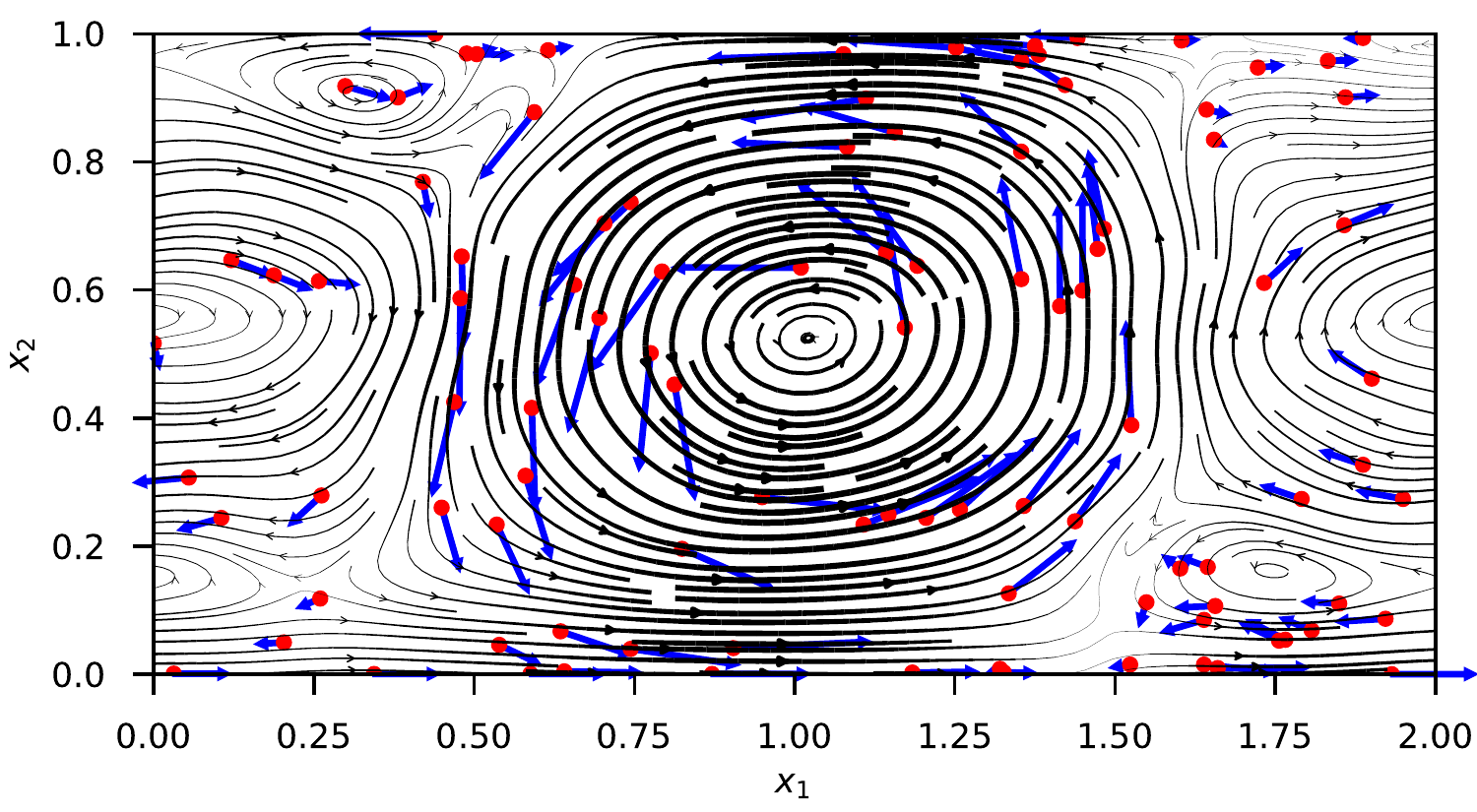}%
    \label{fig:pred-field-t_10}
  }
  \caption{%
    Streamplots of the velocity field at time $10.0$:
    (A)~true field; (B)~predicted field.%
    \label{fig:true-and-pred-field-t_10}
  }
\end{figure}

Figure \ref{fig:true-and-pred-field-t=1.0} and \ref{fig:true-and-pred-field-t_10} plot the streamlines of the true and reconstructed velocity fields at times $t=1.0$ and $t=10.0$, respectively.
Both figures well represent the true fields, although the number of measurements is small with a significant SVD truncation of the data matrix $\bs u$. However, in the center of the domain, the predicted field is somewhat less intense than the true field.

\begin{figure}[htbp]
    \centering
    \includegraphics[width=0.6\textwidth]{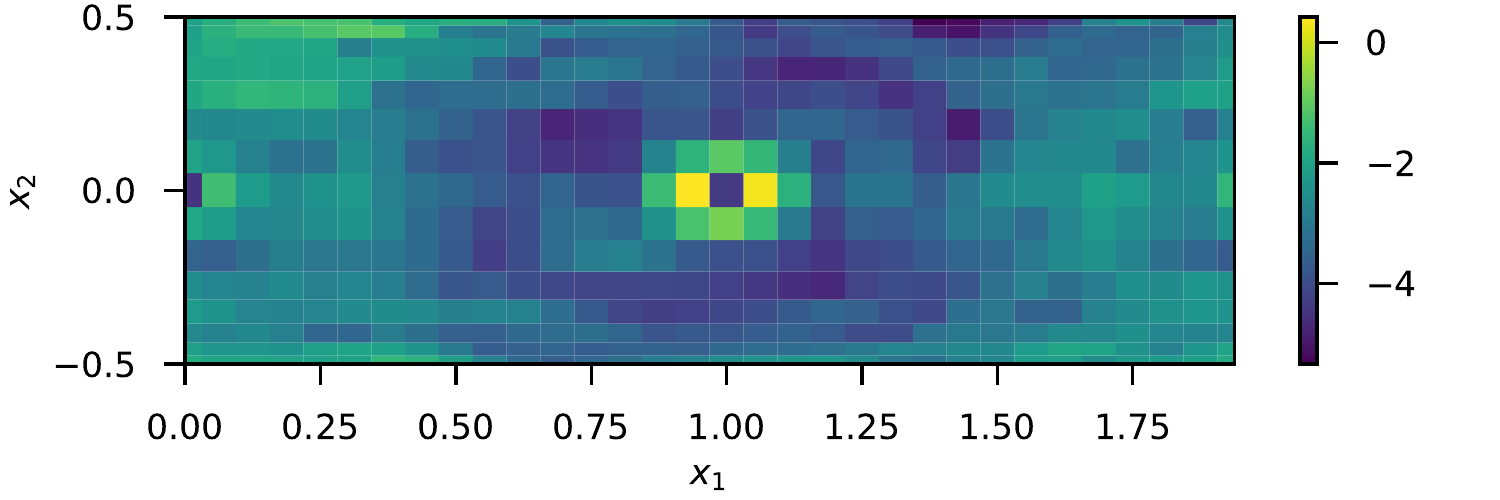}
    \caption{Heatmap of the prediction error defined by~\eqref{eq:sta:error}.}%
    \label{fig:sta:error-heatmap-logscale}
\end{figure}

To assess the quality of the spatiotemporal approximation, we compute the
pointwise normalized mean squared error on a test data set, distinct from the
training data $\bs{u}$ used to parametrize the model.
This error at point $\bs{x}_p$ is computed as
\begin{equation}
e_p = \log
    \frac{%
    \frac1T \sum_{m=1}^T
      \|\bs{\upsilon}_{\cl{I}}^{\mathrm{opt}}(\bs{x}_p,t_m) - {\bs{u}_{\mathrm{test}}}_{m,p}\|^2
  }{%
    \frac{1}{P_{\mathrm{test}}} \sum_{i=1}^{P_{\mathrm{test}}}
    \frac1T \sum_{m=1}^T \|{\bs{u}_{\mathrm{test}}}_{m,i}\|^2
  },
\label{eq:sta:error}
\end{equation}
where we have followed the notation of Section \ref{sc:compress.sensing},
and $\|\cdot \|$ denotes the Euclidean norm in $\R^n$.
The choice of the normalizing factor in the denominator (averaging over all
test locations and time) is selected to avoid division by zero.
This error may be interpreted as a quality comparison of the
proposed model and a model that predicts a constant averaged field. On the logarithmic scale, a positive error denotes that the proposed model is worse (at that spatial point) than the constant model. Figure \ref{fig:sta:error-heatmap-logscale} shows the heatmap of the error
computed by \eqref{eq:sta:error} at all $P_{\mathrm{test}}$ locations.
The largest prediction errors occur at the center of the
domain where the central vortical structure is located.
This occurrs because the predicted velocities near the center of the domain are smaller than the true values, as observed in Figures \ref{fig:true-and-pred-field-t=1.0} and \ref{fig:true-and-pred-field-t_10}.

Overall, this example shows that by combining time compression with the \SFdf\ model that approximates the spatial data, we can economically and accurately reconstruct time-evolving velocity fields.

%% file: section-6.tex
%auto-ignore
\section{Conclusions}
\label{sc:conclusions}

We proposed a Sparse Fourier divergence-free method (\SFdf) based on a discrete $L^2$ projection. We coupled our method, a spatial approximation, with a temporal approximation based on the truncated SVD. This results in the combination of supervised learning in space and unsupervised learning in time. In this physical-informed type of statistical learning framework, we adaptively build a sparse Fourier set of basis functions and their coefficients by solving a sequence of minimization problems. The sparse Fourier set of basis functions is augmented greedily in each optimization problem. We regularize our minimization problems with the seminorm of the fractional Sobolev space in a Tikhonov fashion. The spatiotemporal approximation is then performed by coupling our \SFdf\ spatial approximation with truncated SVD for the temporal approximation.

The physics-informed regularized supervised learning is powerful enough to reconstruct vector fields from even very sparse data. Moreover, the temporal approximation implies that the run-time of the method is independent of the number of measurements in time, except for the SVD decomposition. Hence, the method is most efficient when the measurement set is sparse in space and dense in time.

To assess the accuracy of our method, we reconstruct incompressible flows from velocity measurements in various numerical examples. The spatial and temporal approximations well agreed with the true velocity fields, as verified in various numerical experiments with a high degree of accuracy.

\section{Acknowledgments}

This work was partially supported by the KAUST Office of Sponsored Research (OSR) under Award numbers URF$/1/2281-01-01$, URF$/1/2584-01-01$ in the KAUST Competitive Research Grants Program Round 8, the Alexander von Humboldt Foundation, and Coordination for the Improvement of Higher Education Personnel (CAPES).

Last but not least, we want to thank Prof. Jesper Oppelstrup for providing us with valuable ideas and constructive comments.

%% file: appendix.tex
%auto-ignore
\section{Sobolev embedding}\label{ap:sobolev.embedding}

From the embedding Sobolev theorem, with $k>l$ and $1\le{p}<q<\infty$ and
\begin{equation}
\dfrac{1}{q}=\dfrac{1}{p}-\dfrac{k-l}{n},
\end{equation}
we have that the ${W}^{l,q}(\cl{D})$ continuously embeds ${W}^{k,p}(\cl{D})$, that is,
\begin{equation}
{W}^{k,p}(\cl{D})\subseteq{W}^{l,q}(\cl{D}).
\end{equation}
Moreover, when $k>n/p$, we have that $\bs{\upsilon}\in{W}^{k,p}(\cl{D})$ also belongs to a H\"{o}lder space, namely $\bs{\upsilon}\in{C}^{k-\left[\frac{n}{p}\right]-1,\gamma}(\cl{D})$, where $\cl{D}\subset\bb{R}^n$ and
\begin{equation}
\gamma=
\begin{cases}
1+\left[\frac{n}{p}\right]-\frac{n}{p}&\frac{n}{p}\notin\bb{Z},\\
\text{any element in }(0,1)&\frac{n}{p}\in\bb{Z}.
\end{cases}
\end{equation}
Finally, we have the following estimate
\begin{equation}
\|\bs{\upsilon}\|_{C^{k-\left[\frac{n}{p}\right]-1,\gamma}(\cl{D})}\leq A \|\bs{\upsilon}\|_{W^{k,p}(\cl{D})},
\end{equation}
where $A$ does not depend on $\bs{\upsilon}$. Interest readers are referred to \cite{Eva98}.

\section{Wirtinger calculus}\label{ap:wirtinger}

This Appendix presents the relevant mathematical treatment for computing the gradient of our residual \eqref{eq:optimality}, which is a real function depending on complex variables.

Let $\bb{A}\subset\bb{C}$ be an open set and $f(z)=u(z)+\jmath{v}(z)$ be holomorphic (analytic) in the open set $\bb{U}\subseteq\bb{A}$, where $\jmath$ is the imaginary unit, $u(z),v(z)\in\bb{R}$, and $z\coloneqq{x}+\jmath{y}\in\bb{U}$. If $f$ is holomorphic in the open set $\bb{U}\subseteq\bb{A}$, then, $f$ is differentiable at $z_0$ for all $z_0\in\bb{U}$. Letting $\hat{f}(x,y)\coloneqq{f}(z)$, $\hat{u}(x,y)\coloneqq{u}(z)$, and $\hat{v}(x,y)\coloneqq{v}(z)$, then differentiability at $z$ implies that
\begin{equation}
\dd{f(z)}{z}=\lim_{\Delta{z}\to0}\dfrac{f(z+\Delta{z})-f(z)}{\Delta{z}}=\lim_{\substack{\Delta{x}\to0\\\Delta{y}\to0}}\dfrac{\hat{f}(x+\Delta{x},y+\Delta{y})-\hat{f}(x,y)}{\Delta{x}+\jmath\Delta{y}},
\end{equation}
exists. Moreover, with $\Delta{z}=\Delta{x}+\jmath\Delta{y}$ and considering two cases: ($\Delta{x}=0,\Delta{y}\to0$) and ($\Delta{x}\to0,\Delta{y}=0$), we are led to the Cauchy--Riemann equations
\begin{equation}\label{eq:cauchy.riemann}
\dfrac{\partial\hat{u}(x,y)}{\partial{x}}=\dfrac{\partial\hat{v}(x,y)}{\partial{y}}\qquad\text{and}\qquad\dfrac{\partial\hat{u}(x,y)}{\partial{y}}=-\dfrac{\partial\hat{v}(x,y)}{\partial{x}}.
\end{equation}
A necessary condition for $f$ being holomorphic in $\bb{U}$ is that the Cauchy--Riemann equations are satisfied.

Total differential of the bivariate function $\hat{f}(x,y)$ associated with the univariate differential of $f(z)$ is given by
\begin{equation}\label{eq:total.diff.f.hat}
\di{\hat{f}}=\dfrac{\partial\hat{u}(x,y)}{\partial{x}}\di{x}+\jmath\dfrac{\partial\hat{v}(x,y)}{\partial{x}}\di{x}+\dfrac{\partial\hat{u}(x,y)}{\partial{y}}\di{y}+\jmath\dfrac{\partial\hat{v}(x,y)}{\partial{y}}\di{y},
\end{equation}
with
\begin{equation}\label{eq:diff}
\di{x}=\fr{1}{2}(\di{z}+\di{z}^\ast)\qquad\text{and}\qquad\di{y}=\fr{1}{2\jmath}(\di{z}-\di{z}^\ast).
\end{equation}
Thus, the total differential \eqref{eq:total.diff.f.hat} with expressions \eqref{eq:diff} becomes
\begin{align}\label{eq:total.diff.f.hat.rearranged}
\di{\hat{f}}=&{}\dfrac{1}{2}\Bigg[\dfrac{\partial\hat{u}(x,y)}{\partial{x}}+\dfrac{\partial\hat{v}(x,y)}{\partial{y}}+\jmath\Bigg(\dfrac{\partial\hat{v}(x,y)}{\partial{x}}-\dfrac{\partial\hat{u}(x,y)}{\partial{y}}\Bigg)\Bigg]\di{z}\nonumber\\[4pt]
&+\dfrac{1}{2}\Bigg[\dfrac{\partial\hat{u}(x,y)}{\partial{x}}-\dfrac{\partial\hat{v}(x,y)}{\partial{y}}+\jmath\Bigg(\dfrac{\partial\hat{v}(x,y)}{\partial{x}}+\dfrac{\partial\hat{u}(x,y)}{\partial{y}}\Bigg)\Bigg]\di{z}^\ast.
\end{align}
Alternatively, in view of \eqref{eq:total.diff.f.hat.rearranged}, we have
\begin{equation}\label{eq:total.diff.f}
\di{f}=\dfrac{\partial{f}(z)}{\partial{z}}\di{z}+\dfrac{\partial{f}(z)}{\partial{z}^\ast}\di{z}^\ast,
\end{equation}
where the Wirtinger differential operators are
\begin{equation}\label{eq:wirtinger.diff.operator}
\dfrac{\partial}{\partial{z}}\coloneqq\frac{1}{2}\Bigg[\dfrac{\partial}{\partial{x}}-\jmath\dfrac{\partial}{\partial{y}}\Bigg]\qquad\text{and}\qquad\dfrac{\partial}{\partial{z}^\ast}\coloneqq\frac{1}{2}\Bigg[\dfrac{\partial}{\partial{x}}+\jmath\dfrac{\partial}{\partial{y}}\Bigg].
\end{equation}
Also, note that
\begin{equation}\label{eq:conj.derivative}
\bigg(\dfrac{\partial{f}(z)}{\partial{z}}\bigg)^\ast=\dfrac{\partial{f^\ast}(z)}{\partial{z}^\ast}.
\end{equation}

In dealing with real-valued functions, $f\colon{z}\in\bb{U}\subseteq\bb{C}\to{f}(z)\in\bb{R}$, we have that the Cauchy--Riemann equations \eqref{eq:cauchy.riemann} are no longer satisfied unless $f$ is trivial (constant) with respect to its arguments. Consequently, $f$ is non analytic in the general cases. Thus, $f(z)$ becomes $\hat{f}(x,y)=\hat{u}(x,y)$ and expressions \eqref{eq:total.diff.f.hat} and \eqref{eq:total.diff.f.hat.rearranged} specialize to
\begin{align}\label{eq:total.differential.f.hat}
\di{\hat{f}}&=\dfrac{\partial\hat{u}(x,y)}{\partial{x}}\di{x}+\dfrac{\partial\hat{u}(x,y)}{\partial{y}}\di{y},\nonumber\\[4pt]
&=\dfrac{1}{2}\Bigg[\dfrac{\partial\hat{u}(x,y)}{\partial{x}}-\jmath\dfrac{\partial\hat{u}(x,y)}{\partial{y}}\Bigg]\di{z}+\dfrac{1}{2}\Bigg[\dfrac{\partial\hat{u}(x,y)}{\partial{x}}+\jmath\dfrac{\partial\hat{u}(x,y)}{\partial{y}}\Bigg]\di{z}^\ast.
\end{align}
Moreover, from \eqref{eq:total.diff.f.hat.rearranged} with \eqref{eq:wirtinger.diff.operator}, we have that
\begin{align}\label{eq:df.dz}
\dfrac{\partial{f}(z)}{\partial{z}}\di{z}=\dfrac{1}{2}\Bigg[\dfrac{\partial\hat{u}(x,y)}{\partial{x}}-\jmath\dfrac{\partial\hat{u}(x,y)}{\partial{y}}\Bigg](\!\di{x}+\jmath\di{y}),
\end{align}
and
\begin{align}\label{eq:df.dzast}
\dfrac{\partial{f}(z)}{\partial{z}^\ast}\di{z}^\ast=\dfrac{1}{2}\Bigg[\dfrac{\partial\hat{u}(x,y)}{\partial{x}}+\jmath\dfrac{\partial\hat{u}(x,y)}{\partial{y}}\Bigg](\!\di{x}-\jmath\di{y}).
\end{align}
Also, note that for a real-valued function $f$, expression \eqref{eq:conj.derivative} reduces to
\begin{equation}\label{eq:conj.derivative.real}
\bigg(\dfrac{\partial{f}(z)}{\partial{z}}\bigg)^\ast=\dfrac{\partial{f}(z)}{\partial{z}^\ast}.
\end{equation}
Next, \eqref{eq:total.differential.f.hat} can be written as
\begin{equation}\label{eq:total.diff.real.f.hat}
\di{\hat{f}}=\nabla\hat{f}\cdot(\!\di{x},\di{y}),
\end{equation}
where
\begin{equation}
\nabla\hat{f}\coloneqq\bigg(\dfrac{\partial\hat{u}(x,y)}{\partial{x}},\dfrac{\partial\hat{u}(x,y)}{\partial{y}}\bigg).
\end{equation}
Combining \eqref{eq:total.differential.f.hat}, \eqref{eq:df.dz}, and emulating the same reasoning for $\partial{f}/\partial{z}^\ast$, we obtain the following relation
\begin{equation}\label{eq:total.diff.real.f}
\di{f}=2\,\mathfrak{R}\bigg\{\dfrac{\partial{f}(z)}{\partial{z}}\di{z}\bigg\}=2\,\mathfrak{R}\bigg\{\dfrac{\partial{f}(z)}{\partial{z}^\ast}\di{z}^\ast\bigg\},
\end{equation}
Last, from \eqref{eq:total.diff.real.f.hat}, \eqref{eq:total.diff.real.f}, and noting that $\di{f}(z)=\di{\hat{f}}(x,y)$, the steepest ascent direction of a real-valued function $f$ is given by
\begin{equation}\label{eq:gradient.wirtinger}
\nabla{\hat{f}}\cdot(\!\di{x},\di{y})=2\,\mathfrak{R}\bigg\{\dfrac{\partial{f}(z)}{\partial{z}}\di{z}\bigg\}=2\,\mathfrak{R}\bigg\{\dfrac{\partial{f}(z)}{\partial{z}^\ast}\di{z}^\ast\bigg\}.
\end{equation}
Additional details are given in \cite{Sob12}.